\documentclass[aps,pra,reprint,twocolumn]{revtex4-2}
\usepackage{soul}
\usepackage{color}
\usepackage{graphicx}
\usepackage{enumitem}
\usepackage{amsfonts}
\usepackage{amssymb}
\usepackage{amsmath,latexsym,amsthm}
\usepackage{physics}
\usepackage{bm}
\usepackage[dvipsnames]{xcolor}
\usepackage[breaklinks=true,colorlinks=true,linkcolor=teal,urlcolor=teal,citecolor=teal]{hyperref}
\usepackage[bottom]{footmisc}
\begin{document}

\title{Signatures of a gravitational quantum vacuum on dynamics of massive particles}
\author{Aaron R. Malcolm}
\email{Aaron.Malcolm@warwick.ac.uk}
\altaffiliation{Authors contributed equally.}
\author{Zhi-Wei Wang}
\email{zhiweiwang.phy@gmail.com}
\altaffiliation{Authors contributed equally.}
\author{B. Sharmila}
\email{Sharmila.Balamurugan@warwick.ac.uk}
\author{Animesh Datta}
\email{Animesh.Datta@warwick.ac.uk}
\affiliation{Department of Physics, University of Warwick, Coventry CV4 7AL, United Kingdom.}

\begin{abstract}
We study the interaction of two massive particles with a quantised gravitational field in its vacuum state using two different position observables: 
(i) a frame-dependent coordinate separation and 
(ii) a frame-independent geodesic separation.
For free particles, (i) leads to purely unitary dynamics  but (ii) leads to dissipation.
For two particles coupled through a linear spring, (i) and (ii) lead to different cut-off dependences in the frequency shift harmonic ladder of the differential motional mode. 
Our findings highlight the observable consequences of choosing different position observables at the interface of quantum mechanics and gravity.

\end{abstract}

\maketitle

\noindent

The quest for the unification of general relativity and quantum mechanics has been one of the most profound challenges in theoretical physics.
Efforts to develop a coherent theoretical framework have spurred various approaches, including string theory \cite{Green2012}, loop quantum gravity \cite{Rovelli2008}, and the study of entanglement entropy in the context of the holographic principle \cite{Ryu2006, VanRaamsdonk2010}. 
Central to this endeavour is the unravelling of the quantum nature of gravity and its interaction with matter, a challenge dating back to Bronstein's pioneering work in 1936 \cite{bronstein1936quantentheorie}. This remains an active area of research studied by various low-energy approaches~\cite{Breuer2009,Anastopoulos2013,Blencowe2013,Parikh2021,Torovs2020,Sen2024,Chen2024}. These works underscore the necessity for a deeper understanding of gravity's behaviour at the quantum level.
 
To that end, several studies have analysed the decoherence effects induced by gravitational fields through different models \cite{Anastopoulos2013,Blencowe2013,Breuer2009,Torovs2020,Sen2024}.
One popular model developed by Anastopolous, Blencowe and Hu (ABH) \cite{Anastopoulos2013,Blencowe2013} considers a scalar matter field interacting with quantised, weak gravitational modes resulting in dissipation in the energy basis. 
Coupling the stress-energy tensor directly to the metric~\cite{Anastopoulos2013,Blencowe2013} results in a potential term which is modified by the metric perturbations. 
Another model treats the gravitational field as an external, classical field in the non-relativistic limit before quantising it~\cite{Speliotopoulos, Parikh2021, Torovs2020, Sen2024, Kanno2020}.
This results in a potential term which is not coupled to the metric. Instead only the separation between particles is modified by the metric perturbations. 

In this article, we find these two models to lead to different outcomes in certain physical scenarios.
In particular, we study the effect of a quantised gravitational field in its vacuum state on two massive particles.
We study their dynamics for two different position observables, namely
(i) a frame-dependent coordinate separation $\hat{x}$ 
and (ii) a frame-independent geodesic separation $\hat{\xi}$ between the two particles. 
For these, we evaluate the expectation and variance of the position. 



We first study the case of two free, unbound particles, and find our results to be in agreement with the ABH models for free particles~\cite{Anastopoulos2013} only when using the frame-dependent coordinate separation $\hat{x}$.
We then study the case of two particles coupled through a linear spring, and find a signature of amplitude damping as opposed to the phase damping that is found from applying our analysis to the Hamiltonian reported in Ref.~\cite{Blencowe2013}.
We also identify additional quantum signatures of the gravitational vacuum, such as a frequency shift and the generation of quantum coherence. 
Notably, the dependence of the frequency shift of the high-frequency cut-off illustrates another consequence of the choice of the two different position observables.

We believe the frame-independent $\hat{\xi}$ to be the physically relevant observable predicting the correct dynamics. However, the scalar field approaches \cite{Anastopoulos2013,Blencowe2013} used $\hat{x}$ as their position observable. 
We show that the latter choice leads to different observable consequences both for free particles and those coupled by a linear spring.

As opposed to models using scalar matter fields, other works such as those by Sen \textit{et al.} \cite{Sen2024} and Toro\v{s} \textit{et al.} \cite{Torovs2020} describe the matter as test particles as we do. 
Specifically, Sen \textit{et al.} analyse a gravitational wave detector placed inside a harmonic trap that interacts with quantised gravitational waves \cite{Sen2024}. They calculate the transition probability for the system to move from an initial to a final state, showing that resonant absorption agrees with semi-classical predictions.
For a similar harmonic system, Toro\v{s} \textit{et al.} demonstrate how the rate of the dissipation agrees with the classical decay one would expect from an accelerating mass \cite{Torovs2020}. 


\begin{figure}[ht]
\centering
\includegraphics[width=0.47\textwidth]{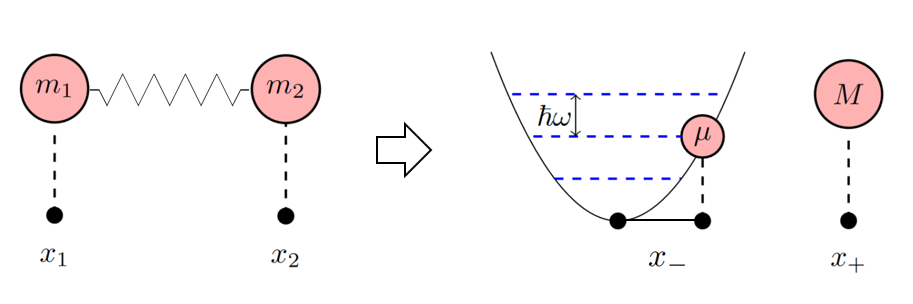}
\caption{The system consists of two test particles with masses $m_{1}$ and $m_{2}$ located at positions $x_{1}$ and $x_{2}$, respectively. They are coupled through a linear spring interaction. We then transform variables and consider the centre of mass variables $x_{+}$ and $x_{-}$, with total mass $M$ and reduced mass $\mu$. The system then becomes two decoupled particles. A single free particle of mass $M$ and a particle in a harmonic potential of mass $\mu$.}
\label{fig:osc}
\end{figure}
\begin{figure}[ht]
\centering
\includegraphics[width=0.47\textwidth]{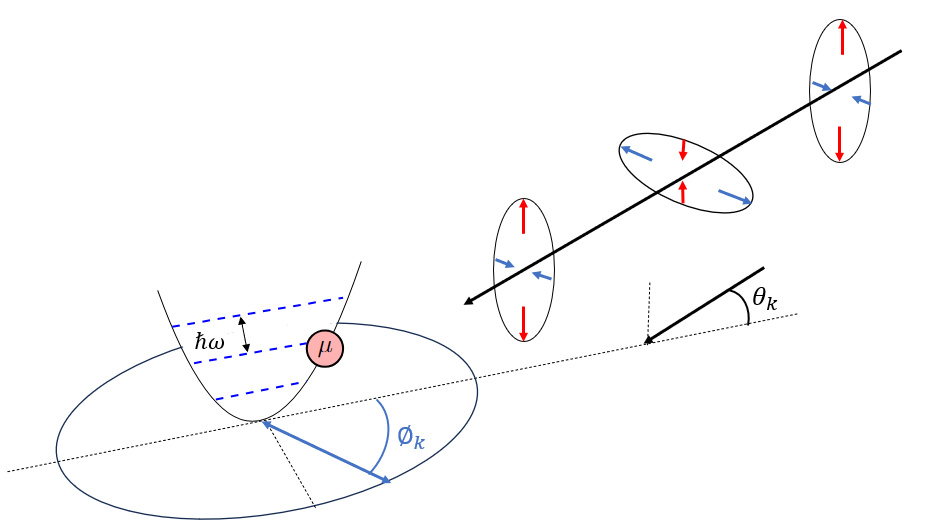}
\caption{Schematic of a single-mode gravitational wave with frequency $\Omega_{\mathbf{k}}$ interacting with a particle of mass $\mu$ in a harmonic oscillator potential with frequency $\omega$. The axis of the oscillator is at an angle $\theta_{\mathbf{k}}$ from the direction of propagation and angle $\phi_{\mathbf{k}}$ from the axis of the polarisation, denoted by the blue arrows. This figure illustrates one mode of the gravitational field; in our analysis, we consider the field to include modes of all frequencies and orientations.}
\label{fig:detectiongw}
\end{figure}

\section{Model}
We model the gravitational field as a quantum field in the vacuum state, following the approach outlined in both Parikh \textit{et al.} \cite{Parikh2021} and Kanno \textit{et al.} \cite{Kanno2020}. 
This treats the gravitational field as a small perturbation of the Minkowski metric, represented as \( g_{\mu\nu} = \eta_{\mu\nu} + h_{\mu\nu} \), where $|h_{\mu \nu}| \ll 1$. Here, \( h_{\mu\nu} \) denotes a small perturbation, facilitating a linear approximation in our treatment of the field. To simplify the complexity inherent in Einstein's field equations, we make use of the gauge freedom inherent in these equations and adopt the synchronous gauge and the de Donder gauge. This approach effectively isolates the unphysical gauge freedoms. Subsequent application of the Transverse-Traceless (TT) gauge enables us to extract the physically meaningful wave solutions from these perturbations. This results in the following action for the gravitational field,  up to second order in $h_{\mu \nu}$ (see section A1 in the Supplementary Material for further details)
\begin{equation}
    \label{eq:ehactiontt}
    S_{\textsc{eh}} = - \frac{\rho_{\textsc{f}} c^{2}}{4} \int \text{d}^{4} x \, \partial_{\mu} h_{i j} \partial^{\mu} h^{i j},
\end{equation}
where $\rho_{\textsc{f}} = c^{2} /8 \pi G $.

We aim to model the response of a particle confined in a trap of a harmonic oscillator potential with frequency $\omega$, to the presence of quantum vacuum fluctuations of the gravitational field. To achieve this, we first consider two particles of mass $m_{1}$ and $m_{2}$ coupled through a linear spring interaction. The coordinate positions of the particles are labelled $x_{1}$ and $x_{2}$ as shown in Fig. \ref{fig:osc}. Since our primary interest is the gravitational field's effect on the separation between the two masses, we transform the positions into the centre of mass variables $x_{+}$ and $x_{-}$ (see Fig. \ref{fig:osc}) where
\begin{align}
    \label{eq:comx}
    x_{+} &= \frac{m_{1} x_{1} + m_{2} x_{2}}{m_{1} + m_{2}}, \\
    x_{-} &= x_{1} - x_{2}.
\end{align}
In section A2 of the Supplementary Material, we show that the centre of mass motion $x_{+}$ takes the form of a single, free particle and will therefore follow a geodesic path. As $x_{+}$ is only a single point, there can be no frame-independent observable that can be measured. On the other hand, the differential mode $x_{-}$ describes the frame-dependent coordinate separation between the two masses, effectively modelled by a reduced mass $\mu$. This reduced mass system does not follow a geodesic path because it is bound in a harmonic oscillator. Therefore, we focus only on the differential mode motion $x_{-}$ which takes the form of a particle, with mass $\mu$ in a harmonic oscillator potential of frequency $\omega$, where
\begin{align}
    \label{eq:mu}
    \mu &= \frac{m_{1} m_{2}}{m_{1} + m_{2}}.
\end{align}
As we have shown that the linear spring description of the system is equivalent to a single particle in a harmonic potential in the differential mode $x_-$, 
we will now refer to the system as a harmonic potential rather than a linear spring.
We also denote $x \equiv x_{-}$ for ease of notation.
It is important to note that relative position $x$ is an arbitrary coordinate and is not frame-independent. 
We therefore transform this coordinate into the more physically relevant geodesic separation, $\xi$, between the particles
\begin{equation}
    \label{eq:geosep}
    \xi^{i} = x^{i} + \frac{1}{2} \delta^{i j} h_{j k} \xi^{k},
\end{equation}
where it has been assumed that the metric perturbation, $h_{j k }$, is constant over the separation between the particles.

In order to move to a quantum treatment, the perturbation is then decomposed into its Fourier modes
\begin{align}
    \label{eq:hexp}
    h_{ij}(t, \mathbf{X}) = \frac{1}{V} \sum_{\mathbf{k}, s} q_{\mathbf{k}, s} e^{i \mathbf{k} \cdot \mathbf{X}} \epsilon^{s}_{ij}(\mathbf{k}),
\end{align}
where $q_{\mathbf{k}, s}$ is the amplitude of each mode, $\epsilon^{s}_{ij}(\mathbf{k})$ is the polarisation tensor associated with each mode, $V$ is some normalization volume, $\mathbf{X}$ is the position in some arbitrary coordinate frame and $\mathbf{k}$ is the conjugate momentum of $\mathbf{X}$ such that each mode has frequency $ \Omega_{\mathbf{k}} = c \mathbf{k}$. In keeping with the assumption that the metric perturbation is constant over the separation of the particles, we invoke the long-wavelength approximation such that $e^{i \mathbf{k} \cdot \mathbf{X}} \approx 1$.

From Eqs.~(\ref{eq:ehactiontt}) and (\ref{eq:hexp}) we can see that the gravitational vacuum is modelled by many low-amplitude gravitational waves with frequencies $\Omega_{\mathbf{k}}$. Our system thus comprises of a single particle of reduced mass $\mu$ in a harmonic oscillator potential interacting with these gravitational waves as depicted in Fig.~\ref{fig:detectiongw}.

We choose to consider the modes of the field to be only in the ``+" polarisation state such that
\begin{align}
\epsilon^{+}_{ij} = \begin{pmatrix}
0&0&0&0\\
0&1&0&0\\
0&0&-1&0\\
0&0&0&0
\end{pmatrix}.
\end{align}
The summation over $\mathbf{k}$ encompasses all modes of the field, $k=0$ to $k=\infty$. 
To avoid a divergence, we impose a high-frequency cut-off, $\Omega_{\textsc{max}}$ which sets the maximum energy scale at which we expect this model to be valid.

We then promote the dynamical variables $x$, $p_{x}$, $\xi$, $p_{\xi}$, $q_\mathbf{k}$ and $p_\mathbf{k}$ to quantum operators where $\hat{p}_{x}$, $\hat{p}_{\xi}$ and $\hat{p}_\mathbf{k}$ are the canonical momenta of $\hat{x}$, $\hat{\xi}$ and $\hat{q}_\mathbf{k}$ respectively. They obey the commutation relations $ [\hat{x}, \hat{p}_{x}] = [\hat{\xi}, \hat{p}_{\xi}]  =i \hbar$, and $[\hat{q}_{\mathbf{k}}, \hat{p}_{\mathbf{k}'}] = i \hbar \delta_{\mathbf{k},\mathbf{k}'}$.

The interaction of the masses with the gravitational field leads to two Hamiltonians for $\hat{x}$ and $\hat{\xi}$ respectively (see section A2 of the Supplementary Material for full derivation.)
\begin{align}
    \label{eq:Hamx}
    \hat{H}_{x} &= \frac{\hat{p}_{x}^{2}}{2 \mu} + \frac{1}{2} \mu \omega^{2} \hat{x}^{2} - \frac{1}{V} \sum_{k} \left[  \left( \frac{\hat{p}_{x}^{2}}{2m} - \frac{m \omega^{2} \hat{x}^{2}}{2} \right) \hat{q}_{\mathbf{k}} \right] \nonumber \\
    &+ \sum_{\mathbf{k}} \left( \frac{V \hat{p}_{\mathbf{k}}^{2}}{2 \rho_{\textsc{f}}} + \frac{\rho_{\textsc{f}}}{2V} \Omega_{\mathbf{k}}^{2} \hat{q}_{\mathbf{k}}^{2} \right),\\
    \label{eq:Hamxi}
    \hat{H}_{\xi} &= \frac{\hat{p}_{\xi}^{2}}{2 \mu} + \frac{1}{2} \mu \omega^{2} \hat{\xi}^{2} + \frac{\hat{\xi} \hat{p}_{\xi} + \hat{p}_{\xi} \hat{\xi}}{4 \rho_{\textsc{f}}}  \sum_{\mathbf{k}} \hat{p}_{\mathbf{k}} \nonumber \\
    &+ \sum_{\mathbf{k}} \left( \frac{V \hat{p}_{\mathbf{k}}^{2}}{2 \rho_{\textsc{f}}} + \frac{\rho_{\textsc{f}}}{2V} \Omega_{\mathbf{k}}^{2} \hat{q}_{\mathbf{k}}^{2} \right),
\end{align}
In deriving the Eqs.~(\ref{eq:Hamx}) and (\ref{eq:Hamxi}), we invoke both the long-wavelength approximation and the non-relativistic limit. We also assume that the gravitational self interaction is negligible ($A(\xi) \approx 1 $ in section A of the Supplementary Material). From comparison of Eqs.~(\ref{eq:Hamx}) and (\ref{eq:Hamxi}), we see that the interaction of the gravitational field with the masses will lead to different dynamics of the position variables $\hat{x}$ and $\hat{\xi}$.

We study the dynamics of the masses by taking an open quantum system approach. We define the differential motional quantum state of the masses $\hat{\sigma}(t)$ at an instant of time $t$ is given by a master equation, obtained by tracing out the field modes of the joint state $\hat{\rho}(t)$ of the particle and the field subsystems at each instant, $\hat{\sigma}(t) = \text{Tr}_{\textsc{f}}[\hat{\rho}(t)]$.

The master equation describing the evolution of $\hat{\sigma}$ is obtained using the Born-Markov approximations listed below:
\begin{enumerate}[label=(\roman*)]
\setcounter{enumi}{0}
    \item \label{Ass:tscale} There exists a separation of timescales between matter and field subsystems;
    \item \label{Ass:weakInt} Weak interaction between the matter and field is assumed. The master equation is therefore found to be second-order in the interaction;
    \item \label{Ass:eqbm} The field is assumed to be at thermodynamic equilibrium. Therefore, the density matrix of the field can be considered a statistical mixture of the eigenstates of the field Hamiltonian.
\end{enumerate}

Being interested in the signatures of quantum vacuum fluctuations, we derive the master equation for the evolution of the quantum state $\hat{\sigma}$ of the masses under both Hamiltonian $\hat{H}_{x}$ and $\hat{H}_{\xi}$, as given in Eq.~(\ref{eq:Hamx}) and Eq.~(\ref{eq:Hamxi}), using Assumptions \ref{Ass:tscale}--\ref{Ass:eqbm}. In this derivation, the field is considered to be in the vacuum state. The detailed steps of the derivation are outlined in section C of the Supplementary Material.

Before presenting these two master equations in the context of a single reduced mass in a harmonic potential, we consider a similar analysis in the context of two free particles, obtained by setting $\omega=0$.

\section{Free Particle}
We first consider no interaction between the particles by setting, $\omega = 0$. We derive the master equation in terms of both $\hat{x}$ and $\hat{\xi}$ (see section B of the Supplementary Material). In the $\hat{x}$ variable
\begin{align}
    \label{eq:mastereqx}
    \frac{\text{d} \hat{\sigma}}{\text{d} t} = - \frac{i}{\hbar} \left[\frac{\hat{p}_{x}^{2}}{2\mu}, \hat{\sigma} \right] - \frac{i}{\hbar}  \Delta_{x} \left[\left( \frac{\hat{p}_{x}^{2}}{2 \mu} \right)^{2}, \hat{\sigma} \right],
\end{align}
where
\begin{align}
    \Delta_{x} = \frac{32}{15 \pi} \frac{t_{\textsc{p}}^{2} \Omega_{\textsc{max}}}{\hbar},
\end{align}
and $t_{\textsc{p}}$ is the Planck time.

The effect of the quantum gravitational field is to introduce a higher order kinetic energy shift, $\Delta_{x}.$ 
This is in agreement with the shift reported in Ref.~\cite[Eq.~(53)]{Anastopoulos2013} up to a numerical prefactor, where $\Theta \rightarrow 0$, the gravitational self interaction has been neglected and $\hbar = c = 1$ in the latter refefence. From Eq.~(\ref{eq:mastereqx}) the mean and variance of $\hat{x}$ and $\hat{p}_{x}$ can be calculated
\begin{align}
    \label{eq:eqomx}
    \langle \hat{x} \rangle (t) &= \left( \frac{1}{\mu} \langle \hat{p}_{x} \rangle (0) +  \frac{\Delta_{x}}{\mu^{2}} \langle \hat{p}^{3}_{x} \rangle (0) \right) t + \langle \hat{x} \rangle (0) , \\
    \langle \hat{x}^{2} \rangle (t) &= \left( \frac{\langle \hat{p}_{x}^{2} \rangle (0)}{\mu^{2}} + \frac{2 \Delta_{x}}{\mu^{3}} \langle \hat{p}_{x}^{4} \rangle (0) + \frac{\Delta_{x}^{2}}{\mu^{4}} \langle \hat{p}_{x}^{6} \rangle (0) \right) t^{2} \nonumber \\
    &+ \left( \frac{\langle \hat{x} \hat{p}_{x} + \hat{p}_{x} \hat{x} \rangle (0)}{\mu} + \frac{\Delta_{x}}{\mu^{2}} \langle \hat{p}_{x}^{3} \hat{x} + \hat{x} \hat{p}_{x}^{3} \rangle (0) \right) t + \langle \hat{x}^{2} \rangle (0), \\
    \langle \hat{p}_{x}^{n} \rangle (t) &= \langle \hat{p}_{x}^{n} \rangle (0),
\end{align}
for any positive integer $n$.

We compare the above to the master equation in terms of $\hat{\xi}$
\begin{align}
    \label{eq:mastereqxi}
    \frac{\text{d}\hat{\sigma}(t)}{\text{d}t} &= - \frac{i}{\hbar} \left(1 - \Delta_{\xi} \right) \left[\frac{\hat{p}_{\xi}^{2}}{2\mu}, \hat{\sigma} \right] \nonumber \\ 
    &+ \gamma_{\xi} \left( \hat{p}^{2}_{\xi} \hat{\sigma} \hat{p}_{\xi} \hat{\xi} + \hat{\xi} \hat{p}_{\xi} \hat{\sigma} \hat{p}^{2}_{\xi} - \hat{\sigma} \hat{p}^{3}_{\xi} \hat{\xi} - \hat{\xi} \hat{p}^{3}_{\xi} \hat{\sigma} \right),
\end{align}
where
\begin{align}
    \Delta_{\xi} &= \frac{8}{15 \pi} t^{2}_{\textsc{p}} \Omega_{\text{max}}^{2}, \\
    \gamma_{\xi} &= \frac{\Delta_{\xi}}{2 \hbar^{2} \mu}.
\end{align}
In addition to a shift in the kinetic energy of the particles, $\Delta_{\xi}$, there is a non-unitary dissipation term, $\gamma_{\xi}$. The shift to the kinetic energy allows a renormalised mass, $\mu_{\xi}$, to be defined, 
\begin{align}
    \mu_{\xi} = \mu \left(1 - \Delta_{\xi} \right)^{-1}.
\end{align}
This was not possible in the coordinate separation where the shift was to the square of the kinetic energy.
Finding the mean and variance of $\hat{\xi}$ and $\hat{p}_{\xi}$
\begin{align}
    \label{eq:expxiFree1}
    \langle \hat{\xi} \rangle(t) &= \frac{\langle \hat{p}_{\xi} \rangle(0)}{\mu_{\xi}}t + \langle \hat{\xi} \rangle (0), \\
    \langle \hat{\xi}^{2} \rangle (t) &= \frac{\langle \hat{p}_{\xi}^{2} \rangle (0)}{\mu_{\xi}^{2}} \left(1 - \Delta_{\xi} \right) t^{2} \nonumber \\
    &+ \frac{\langle \hat{\xi} \hat{p}_{\xi} + \hat{p}_{x} \hat{\xi} \rangle (0)}{\mu_{\xi}} \left(1 + \Delta_{\xi} \right) t + \langle \hat{\xi}^{2} \rangle (0), \\
    \langle p^{n}_{\xi} \rangle (t) &= \langle p^{n}_{\xi} \rangle (0),
\end{align}
for any positive integer $n$.

We see from comparison of the master equations Eq.~(\ref{eq:mastereqx}) and Eq.~(\ref{eq:mastereqxi}) that the resulting dynamics of the $\hat{x}$ and $\hat{\xi}$ will differ. Specifically, the coordinate separation will only show a unitary change in the dynamics through a shift in the square of the kinetic energy. This is in contrast to the geodesic separation, which presents not only a shift in the kinetic energy, but also a non-unitary dissipation term. 

One consequence of this can be seen through comparison of the variances $\hat{x}^{2}$ and $\hat{\xi}^{2}$. We see that the variances will increase at different rates for the coordinate separation and geodesic separation. 
As the geodesic separation is frame-independent, we expect the dynamics of the free particles it predicts to be physically relevant.
 This could be confirmed by observing the rate at which the variance of the free particles increases. However, we note that effects of the gravitational vacuum are at the Planck scale. 

\section{Harmonic Oscillator}
We now resume considering a single reduced mass $\mu$ in a harmonic potential, i.e., $\omega \neq 0$. As the masses are now in a harmonic potential (Eqs. (\ref{eq:Hamx}) and (\ref{eq:Hamxi})), we introduce the standard ladder operators ($\hat{b}_{x}$, $\hat{b}^{\dagger}_{x}$) and ($\hat{b}_{\xi}$, $\hat{b}^{\dagger}_{\xi}$) for the $\hat{x}$ and $\hat{\xi}$ variable respectively. We present the master equation for each model in the limit of the rotating wave approximation (Full master equations beyond this limit can be found in section C of the Supplementary Material ) For $\hat{x}$ the master equation is
\begin{align}
    \label{eq:mastereqhox}
    \frac{\text{d} \hat{\sigma}}{\text{d} t} &= -i \left(\omega - \delta_{-} \right)  \left[ \hat{b}_{x}^{\dagger} \hat{b}_{x}, \hat{\sigma} \right] \nonumber - i \delta_{-} \left[ \left( \hat{b}_{x}^{\dagger} \hat{b}_{x} \right)^{2}, \hat{\sigma}  \right] \nonumber \\
    &+ \Gamma \left( \hat{b}_{x}^{2} \hat{\sigma} \hat{b}_{x}^{\dagger 2} - \frac{1}{2} \hat{b}_{x}^{\dagger 2} \hat{b}_{x}^{2} \hat{\sigma} - \frac{1}{2} \hat{\sigma} \hat{b}_{x}^{\dagger 2} \hat{b}_{x}^{2} \right).
\end{align}
where
\begin{align}
    \label{eq:gamma}
    \Gamma &= \frac{32}{15}\frac{G \hbar \omega^{3}}{c^{5}}, \\
    \delta_{-} &= \frac{\Gamma}{2 \pi \omega} \left(- \omega \ln{\left| \frac{ \Omega_{\textsc{max}} - 2 \omega}{\omega}  \right| } - \frac{\Omega_{\textsc{max}}}{2} \right).
\end{align}
From Eq.~(\ref{eq:mastereqhox}) we find two signatures of the gravitational vacuum fluctuations: 
shifts in the energy levels of the oscillator given by Eq. (24) and dissipation, given by Eq. (23).
The first and second terms of the master equation describe a shift in the energy levels of the harmonic oscillator, or equivalently, a shift in transition frequency between these energy levels. The first term shows an energy-level-independent shift to the frequency between any two adjacent energy levels $\delta \omega^{(1)} = -\delta_{-} $. This is similar to the frequency shift obtained for a harmonically trapped charged particle in the presence of vacuum fluctuations of an electromagnetic field \cite{EMPaper}.
The second term outlines an energy-level-dependent shift, $\delta \omega^{(2)} = \left(\delta_{-} \right) \left(2n + 1 \right)$ at $n$-energy-level of the oscillator. This dependence on the energy level $n$, is not seen in the electromagnetic case and is a result of the non-linearity of the gravitational interaction. We have computed and corroborated the shift using two techniques, namely, the time-dependent (Eq. C34 in section C of the Supplementary Material) and the time-independent perturbation techniques (Eq. D22 in section D of the Supplementary Material).

The constant $\delta_{-}$ depends on the UV cut-off $\Omega_{\text{max}}$. However, as argued by Bethe \cite{Bethe} the mass renormalisation of the free particle should be accounted for. The regularisation of the energy shift of a particle could be achieved by subtracting the energy shift of a free particle. This is done for the full master equation first before taking the RWA (see Eq.~(\ref{eq:deltarenormx}) in Appendix~\ref{app:freex} for more details). Following this procedure we find renormalised constants $\delta_{-}^{\textsc{(r)}}$ to be the true, physical constants of the master equation Eq.~(\ref{eq:mastereqhox}).
\begin{align}
    \label{eq:deltar}
    \delta_{-}^{\textsc{(r)}} &= -\frac{\Gamma}{ 2\pi } \ln{\left| \frac{ \Omega_{\textsc{max}} - 2 \omega}{\omega}  \right| }.
\end{align}

The third term in Eq.~(\ref{eq:mastereqhox}) describes dissipation of the matter system that matches the one presented in Ref.~\cite{Torovs2020} with decay rate, $\Gamma$. In the long time limit, the dissipation in Eq.~(\ref{eq:mastereqhox}) will result in a steady state that consists of a mixture of the ground state and first excited state. As explained in Ref.~\cite{Torovs2020}, this occurs as a result of decay from the first excited state to the ground state being forbidden by the two-boson decay process.

We now comapare the dynamics of $\hat{x}$ to that of $\hat{\xi}$ once again. The resulting master equation is
\begin{align}
    \label{eq:mastereqhoxi}
    \frac{\text{d} \hat{\sigma}}{\text{d} t} &= -i \left(\omega - \Delta_{-}  \right)  \left[ \hat{b}_{\xi}^{\dagger} \hat{b}_{\xi}, \hat{\sigma} \right] - i\Delta_{-} \left[ \left( \hat{b}_{\xi}^{\dagger} \hat{b}_{\xi} \right)^{2}, \hat{\sigma}  \right] \nonumber \\
    &+ \Gamma \left( \hat{b}_{\xi}^{2} \sigma \hat{b}_{\xi}^{\dagger 2} - \frac{1}{2} \hat{b}_{\xi}^{\dagger 2} \hat{b}_{\xi}^{2} \hat{\sigma} - \frac{1}{2} \hat{\sigma} \hat{b}_{\xi}^{\dagger 2} \hat{b}_{\xi}^{2} \right).
\end{align}
where
\begin{align}
    \Delta_{-} &= \frac{\Gamma}{2 \pi} \left(- \ln{\left| \frac{ \Omega_{\textsc{max}} - 2 \omega}{\omega} \right|} - \frac{\Omega_{\textsc{max}}}{2 \omega} - \frac{\Omega^{2}_{\textsc{max}}}{8 \omega^{2}} - \frac{\Omega^{3}_{\textsc{max}}}{24 \omega^{3}} \right),
\end{align}
and $\Gamma$ is the same as that defined Eq.~(\ref{eq:gamma}).
As before, renormalising the constant $\Delta_{-}$ through the same procedure outlined earlier (see Eq. B97 in section B2 of the Supplementary Material for more details), results in
\begin{align}
	\label{eq:deltarXi}
    \Delta^{(\textsc{r})}_{-} &= \frac{\Gamma}{2 \pi} \left(- \ln{\left| \frac{ \Omega_{\textsc{max}} - 2 \omega}{\omega}    \right|} - \frac{\Omega_{\textsc{max}}}{2 \omega} - \frac{\Omega^{3}_{\textsc{max}}}{24 \omega^{3}} \right).
\end{align}
Comparison of Eq.~(\ref{eq:mastereqhoxi}) with Eq.~(\ref{eq:mastereqhox}) reveals that the mechanism of dissipation is the same for the particle when measuring in either $\hat{x}$ or $\hat{\xi}$ with the same decay rate $\Gamma$. In both variables, the dissipation occurs through amplitude damping in which the Lindbladian superoperator in the master equations, $\mathcal{L}_{\textsc{amp}}$, is of the form
\begin{align}
    \label{eq:Lamp}
    \mathcal{L}_{\textsc{amp}} [\hat{\sigma}] = \left( \hat{b}_{i}^{2} \sigma \hat{b}_{i}^{\dagger 2} - \frac{1}{2} \hat{b}_{i}^{\dagger 2} \hat{b}_{i}^{2} \hat{\sigma} - \frac{1}{2} \hat{\sigma} \hat{b}_{i}^{\dagger 2} \hat{b}_{i}^{2} \right),
\end{align}
where $i = \{ x, \xi \}$.

The ABH model predicts interaction of a harmonic oscillator with the gravitational vacuum through the energy basis~\cite{Blencowe2013}. Applying the same open quantum system analysis as before to the Hamiltonian reported in Ref.~\cite{Blencowe2013} (where the position is taken to be the coordinate separation, $\hat{x}$) results in a phase damped Lindbladian term, $\mathcal{L}_{\textsc{pha}}$,
\begin{align}
    \label{eq:Lpha}
    \mathcal{L}_{\textsc{pha}} [\hat{\sigma}] = \left( \hat{n}_{x} \sigma \hat{n}_{x} - \frac{1}{2} \hat{n}_{x}^{2} \hat{\sigma} - \frac{1}{2} \hat{\sigma} \hat{n}_{x}^{2} \right),
\end{align}
where $\hat{n}_{x} = \hat{b}_{x}^{\dagger} \hat{b}_{x}$.

Unlike the free particles, it can be seen from comparison of Eq.~(\ref{eq:Lamp}) and Eq.~(\ref{eq:Lpha}), that transforming from $\hat{\xi}$ to $\hat{x}$ does not yield a master equation that is in agreement with the ABH model. Comparison is only possible in the case of free particles due to the difference in the modelling of the potential. Specifically, in the ABH model the system is a modelled as a scalar matter field on a curved space-time. This results in a potential term which itself changes due to the metric perturbation $h_{ij}$. This is physically different from the way we model the potential, in agreement with Refs.~\cite{Speliotopoulos, Torovs2020, Sen2024, Parikh2021, Kanno2020}, in which the system is taken in the non-relativistic limit first. The gravitational field is also taken to be comprised of classical gravitational waves which are then subsequently quantised. This results in a model in which only the separation between particles changes due to the metric perturbations and not the harmonic potential itself.

Although the mechanism of dissipation between $\hat{x}$ and $\hat{\xi}$ is the same (Eq.~(\ref{eq:mastereqhoxi}) with Eq.~(\ref{eq:mastereqhox}) respectively), comparing the frequency shifts reveals that the two types of position variables depend differently on the high-frequency cut-off $\Omega_{\textsc{max}}$. In Eq. (\ref{eq:deltar}), we see that the frequency shift $\delta_{-}^{(\textsc{r})}$ has only logarithmic dependence on the cut-off. This is similar to the dependence found in the case of electromagnetic vacuum fluctuations as reported in \cite{EMPaper}. However, modelling with the frame-independent observable, the frequency shift $\Delta_{-}^{(\textsc{r})}$ in Eq. (\ref{eq:deltarXi}) is extremely sensitive to the cut-off with cubic dependence.

Even in the limit of the rotating wave approximation, the difference in measuring the particles with either the coordinate separation $\hat{x}$ or the geodesic separation $\hat{\xi}$ can be seen. However, we expect that the rotating-wave approximation is not appropriate for the vacuum state of the field, which contains an infinite range of modes, both near and far from resonance with the potential frequency. Going beyond the rotating-wave approximation leads to additional signatures of the vacuum such as the generation of coherences (see Eqs.~C34 and~C69 in section C of the Supplementary Material). The generation of coherences is itself not unique to the gravitational interaction and is also found in the electromagnetic interaction \cite{EMPaper}. As the terms in the master equation corresponding to the coherence shift are also dependent on $\delta_{\pm}$ and $\Delta_{\pm}$, transforming from the coordinate separation to the geodesic separation will again result in different dependence to the high-frequency cut-off. 

\section{Conclusion}
We find that the dynamics of the two types of position variables, the coordinate separation $\hat{x}$ and the geodesic separation $\hat{\xi}$ are distinctly different for two masses, free or harmonically bound, interacting with quantised gravitational modes. In the case of the coordinate separation, $\hat{x}$, we find that interaction with the gravitational field results in purely unitary dynamics with a shift to the square of the kinetic energy. This is in contrast to the geodesic separation, $\hat{\xi}$, which leads to not only a shift in the energy, but a dissipation term.

We also find in the case of the harmonic oscillator, transforming between the position observables has no effect on the mechanism of dissipation, it is only the UV cut-off dependence of the frequency shift that changes. Modelling with the former, as in \cite{Anastopoulos2013, Blencowe2013}, leads to a frequency shift of the oscillator which only has logarithmic dependence to the high-frequency cut-off $\Omega_{\textsc{max}}$, similar to that found due to vacuum fluctuations of the electromagnetic field. In contrast, modelling with the latter, as in \cite{Parikh2021,Torovs2020, Sen2024, Kanno2020} , results in a frequency shift that is extremely sensitive to the high-frequency cut-off with cubic dependence.

We note that going beyond the rotating-wave approximation leads to other signatures such as the generation of coherences in the state of the particle. While this is not a unique feature of the gravitational field--it also occurs in vacuum fluctuations of the electromagnetic field--there are notable differences. Specifically, unlike the EM field, the gravitational field imposes restrictions on the types of coherence that can be generated.

As $\hat{\xi}$ is the frame-independent variable, we expect the resulting dynamics from modelling with $\hat{\xi}$ to be correct. This is in contrast to the models of \cite{Anastopoulos2013} and \cite{Blencowe2013} which model the position with the frame-dependent $\hat{x}$. The differing dynamics in both cases highlights the necessity of finding the appropriate definition of position, not only for vacuum phenomena, but for further modelling of non-vacuum phenomena and strong field interactions.

\textit{Acknowledgements:} This work has been funded, in part, by the UK STFC “Quantum Technologies for Fundamental Physics” programme (Grant Numbers ST/T006404/1, ST/W006308/1 and ST/Y004493/1).

\newpage
\onecolumngrid
\bibliography{refs}

\begin{thebibliography}{18}%
\makeatletter
\providecommand \@ifxundefined [1]{%
 \@ifx{#1\undefined}
}%
\providecommand \@ifnum [1]{%
 \ifnum #1\expandafter \@firstoftwo
 \else \expandafter \@secondoftwo
 \fi
}%
\providecommand \@ifx [1]{%
 \ifx #1\expandafter \@firstoftwo
 \else \expandafter \@secondoftwo
 \fi
}%
\providecommand \natexlab [1]{#1}%
\providecommand \enquote  [1]{``#1''}%
\providecommand \bibnamefont  [1]{#1}%
\providecommand \bibfnamefont [1]{#1}%
\providecommand \citenamefont [1]{#1}%
\providecommand \href@noop [0]{\@secondoftwo}%
\providecommand \href [0]{\begingroup \@sanitize@url \@href}%
\providecommand \@href[1]{\@@startlink{#1}\@@href}%
\providecommand \@@href[1]{\endgroup#1\@@endlink}%
\providecommand \@sanitize@url [0]{\catcode `\\12\catcode `\$12\catcode `\&12\catcode `\#12\catcode `\^12\catcode `\_12\catcode `\%12\relax}%
\providecommand \@@startlink[1]{}%
\providecommand \@@endlink[0]{}%
\providecommand \url  [0]{\begingroup\@sanitize@url \@url }%
\providecommand \@url [1]{\endgroup\@href {#1}{\urlprefix }}%
\providecommand \urlprefix  [0]{URL }%
\providecommand \Eprint [0]{\href }%
\providecommand \doibase [0]{https://doi.org/}%
\providecommand \selectlanguage [0]{\@gobble}%
\providecommand \bibinfo  [0]{\@secondoftwo}%
\providecommand \bibfield  [0]{\@secondoftwo}%
\providecommand \translation [1]{[#1]}%
\providecommand \BibitemOpen [0]{}%
\providecommand \bibitemStop [0]{}%
\providecommand \bibitemNoStop [0]{.\EOS\space}%
\providecommand \EOS [0]{\spacefactor3000\relax}%
\providecommand \BibitemShut  [1]{\csname bibitem#1\endcsname}%
\let\auto@bib@innerbib\@empty
\bibitem [{\citenamefont {Green}\ \emph {et~al.}(2012)\citenamefont {Green}, \citenamefont {Schwarz},\ and\ \citenamefont {Witten}}]{Green2012}%
  \BibitemOpen
  \bibfield  {author} {\bibinfo {author} {\bibfnamefont {M.~B.}\ \bibnamefont {Green}}, \bibinfo {author} {\bibfnamefont {J.~H.}\ \bibnamefont {Schwarz}},\ and\ \bibinfo {author} {\bibfnamefont {E.}~\bibnamefont {Witten}},\ }\href {https://doi.org/10.1017/cbo9781139248570} {\emph {\bibinfo {title} {Superstring Theory: 25th Anniversary Edition}}}\ (\bibinfo  {publisher} {Cambridge University Press},\ \bibinfo {year} {2012})\BibitemShut {NoStop}%
\bibitem [{\citenamefont {Rovelli}(2008)}]{Rovelli2008}%
  \BibitemOpen
  \bibfield  {author} {\bibinfo {author} {\bibfnamefont {C.}~\bibnamefont {Rovelli}},\ }\href {https://doi.org/10.12942/lrr-2008-5} {\bibfield  {journal} {\bibinfo  {journal} {Living reviews in relativity}\ }\textbf {\bibinfo {volume} {11}},\ \bibinfo {pages} {1} (\bibinfo {year} {2008})}\BibitemShut {NoStop}%
\bibitem [{\citenamefont {Ryu}\ and\ \citenamefont {Takayanagi}(2006)}]{Ryu2006}%
  \BibitemOpen
  \bibfield  {author} {\bibinfo {author} {\bibfnamefont {S.}~\bibnamefont {Ryu}}\ and\ \bibinfo {author} {\bibfnamefont {T.}~\bibnamefont {Takayanagi}},\ }\href {https://doi.org/10.1103/PhysRevLett.96.181602} {\bibfield  {journal} {\bibinfo  {journal} {Phys. Rev. Lett.}\ }\textbf {\bibinfo {volume} {96}},\ \bibinfo {pages} {181602} (\bibinfo {year} {2006})}\BibitemShut {NoStop}%
\bibitem [{\citenamefont {Van~Raamsdonk}(2010)}]{VanRaamsdonk2010}%
  \BibitemOpen
  \bibfield  {author} {\bibinfo {author} {\bibfnamefont {M.}~\bibnamefont {Van~Raamsdonk}},\ }\href {https://doi.org/10.1142/S0218271810018529} {\bibfield  {journal} {\bibinfo  {journal} {International Journal of Modern Physics D}\ }\textbf {\bibinfo {volume} {19}},\ \bibinfo {pages} {2429} (\bibinfo {year} {2010})}\BibitemShut {NoStop}%
\bibitem [{\citenamefont {Bronstein}(1936)}]{bronstein1936quantentheorie}%
  \BibitemOpen
  \bibfield  {author} {\bibinfo {author} {\bibfnamefont {M.~P.}\ \bibnamefont {Bronstein}},\ }\href {https://doi.org/0.34663/9783945561317-22} {\bibfield  {journal} {\bibinfo  {journal} {Phys. Z. Sowjetunion}\ }\textbf {\bibinfo {volume} {9}},\ \bibinfo {pages} {140} (\bibinfo {year} {1936})}\BibitemShut {NoStop}%
\bibitem [{\citenamefont {Breuer}\ \emph {et~al.}(2009)\citenamefont {Breuer}, \citenamefont {Göklü},\ and\ \citenamefont {Lämmerzahl}}]{Breuer2009}%
  \BibitemOpen
  \bibfield  {author} {\bibinfo {author} {\bibfnamefont {H.-P.}\ \bibnamefont {Breuer}}, \bibinfo {author} {\bibfnamefont {E.}~\bibnamefont {Göklü}},\ and\ \bibinfo {author} {\bibfnamefont {C.}~\bibnamefont {Lämmerzahl}},\ }\href {https://doi.org/10.1088/0264-9381/26/10/105012} {\bibfield  {journal} {\bibinfo  {journal} {Classical and Quantum Gravity}\ }\textbf {\bibinfo {volume} {26}},\ \bibinfo {pages} {105012} (\bibinfo {year} {2009})}\BibitemShut {NoStop}%
\bibitem [{\citenamefont {Anastopoulos}\ and\ \citenamefont {Hu}(2013)}]{Anastopoulos2013}%
  \BibitemOpen
  \bibfield  {author} {\bibinfo {author} {\bibfnamefont {C.}~\bibnamefont {Anastopoulos}}\ and\ \bibinfo {author} {\bibfnamefont {B.~L.}\ \bibnamefont {Hu}},\ }\href {https://doi.org/10.1088/0264-9381/30/16/165007} {\bibfield  {journal} {\bibinfo  {journal} {Classical and Quantum Gravity}\ }\textbf {\bibinfo {volume} {30}},\ \bibinfo {pages} {165007} (\bibinfo {year} {2013})}\BibitemShut {NoStop}%
\bibitem [{\citenamefont {Blencowe}(2013)}]{Blencowe2013}%
  \BibitemOpen
  \bibfield  {author} {\bibinfo {author} {\bibfnamefont {M.~P.}\ \bibnamefont {Blencowe}},\ }\bibfield  {journal} {\bibinfo  {journal} {Physical Review Letters}\ }\textbf {\bibinfo {volume} {111}},\ \href {https://doi.org/10.1103/PhysRevLett.111.021302} {10.1103/PhysRevLett.111.021302} (\bibinfo {year} {2013})\BibitemShut {NoStop}%
\bibitem [{\citenamefont {Parikh}\ \emph {et~al.}(2021)\citenamefont {Parikh}, \citenamefont {Wilczek},\ and\ \citenamefont {Zahariade}}]{Parikh2021}%
  \BibitemOpen
  \bibfield  {author} {\bibinfo {author} {\bibfnamefont {M.}~\bibnamefont {Parikh}}, \bibinfo {author} {\bibfnamefont {F.}~\bibnamefont {Wilczek}},\ and\ \bibinfo {author} {\bibfnamefont {G.}~\bibnamefont {Zahariade}},\ }\href {https://doi.org/10.1103/PhysRevD.104.046021} {\bibfield  {journal} {\bibinfo  {journal} {Phys. Rev. D}\ }\textbf {\bibinfo {volume} {104}},\ \bibinfo {pages} {046021} (\bibinfo {year} {2021})}\BibitemShut {NoStop}%
\bibitem [{\citenamefont {Toroš}\ \emph {et~al.}(2024)\citenamefont {Toroš}, \citenamefont {Mazumdar},\ and\ \citenamefont {Bose}}]{Torovs2020}%
  \BibitemOpen
  \bibfield  {author} {\bibinfo {author} {\bibfnamefont {M.}~\bibnamefont {Toroš}}, \bibinfo {author} {\bibfnamefont {A.}~\bibnamefont {Mazumdar}},\ and\ \bibinfo {author} {\bibfnamefont {S.}~\bibnamefont {Bose}},\ }\bibfield  {journal} {\bibinfo  {journal} {Physical Review D}\ }\textbf {\bibinfo {volume} {109}},\ \href {https://doi.org/10.1103/PhysRevD.109.084050} {10.1103/PhysRevD.109.084050} (\bibinfo {year} {2024})\BibitemShut {NoStop}%
\bibitem [{\citenamefont {Sen}\ \emph {et~al.}(2024)\citenamefont {Sen}, \citenamefont {Gangopadhyay},\ and\ \citenamefont {Bhattacharyya}}]{Sen2024}%
  \BibitemOpen
  \bibfield  {author} {\bibinfo {author} {\bibfnamefont {S.}~\bibnamefont {Sen}}, \bibinfo {author} {\bibfnamefont {S.}~\bibnamefont {Gangopadhyay}},\ and\ \bibinfo {author} {\bibfnamefont {S.}~\bibnamefont {Bhattacharyya}},\ }\href {https://doi.org/10.1103/PhysRevD.110.026008} {\bibfield  {journal} {\bibinfo  {journal} {Phys. Rev. D}\ }\textbf {\bibinfo {volume} {110}},\ \bibinfo {pages} {026008} (\bibinfo {year} {2024})}\BibitemShut {NoStop}%
\bibitem [{\citenamefont {Chen}\ and\ \citenamefont {Giacomini}(2024)}]{Chen2024}%
  \BibitemOpen
  \bibfield  {author} {\bibinfo {author} {\bibfnamefont {L.-Q.}\ \bibnamefont {Chen}}\ and\ \bibinfo {author} {\bibfnamefont {F.}~\bibnamefont {Giacomini}},\ }\href@noop {} {\bibinfo {title} {Quantum effects in gravity beyond the newton potential from a delocalised quantum source}} (\bibinfo {year} {2024}),\ \Eprint {https://arxiv.org/abs/2402.10288} {arXiv:2402.10288 [quant-ph]} \BibitemShut {NoStop}%
\bibitem [{\citenamefont {Speliotopoulos}(1995)}]{Speliotopoulos}%
  \BibitemOpen
  \bibfield  {author} {\bibinfo {author} {\bibfnamefont {A.~D.}\ \bibnamefont {Speliotopoulos}},\ }\href {https://doi.org/10.1103/physrevd.51.1701} {\bibfield  {journal} {\bibinfo  {journal} {Physical Review D}\ }\textbf {\bibinfo {volume} {51}},\ \bibinfo {pages} {1701–1709} (\bibinfo {year} {1995})}\BibitemShut {NoStop}%
\bibitem [{\citenamefont {Kanno}\ \emph {et~al.}(2021)\citenamefont {Kanno}, \citenamefont {Soda},\ and\ \citenamefont {Tokuda}}]{Kanno2020}%
  \BibitemOpen
  \bibfield  {author} {\bibinfo {author} {\bibfnamefont {S.}~\bibnamefont {Kanno}}, \bibinfo {author} {\bibfnamefont {J.}~\bibnamefont {Soda}},\ and\ \bibinfo {author} {\bibfnamefont {J.}~\bibnamefont {Tokuda}},\ }\href {https://doi.org/10.1103/PhysRevD.103.044017} {\bibfield  {journal} {\bibinfo  {journal} {Phys. Rev. D}\ }\textbf {\bibinfo {volume} {103}},\ \bibinfo {pages} {044017} (\bibinfo {year} {2021})}\BibitemShut {NoStop}%
\bibitem [{\citenamefont {Malcolm}\ \emph {et~al.}(2024)\citenamefont {Malcolm}, \citenamefont {Sharmila}, \citenamefont {Wang},\ and\ \citenamefont {Datta}}]{EMPaper}%
  \BibitemOpen
  \bibfield  {author} {\bibinfo {author} {\bibfnamefont {A.}~\bibnamefont {Malcolm}}, \bibinfo {author} {\bibfnamefont {B.}~\bibnamefont {Sharmila}}, \bibinfo {author} {\bibfnamefont {Z.}~\bibnamefont {Wang}},\ and\ \bibinfo {author} {\bibfnamefont {A.}~\bibnamefont {Datta}},\ }\bibfield  {journal} {\bibinfo  {journal} {Quantum Science and Technology}\ }\href {https://doi.org/10.1088/2058-9565/ad8eef} {10.1088/2058-9565/ad8eef} (\bibinfo {year} {2024})\BibitemShut {NoStop}%
\bibitem [{\citenamefont {Bethe}(1947)}]{Bethe}%
  \BibitemOpen
  \bibfield  {author} {\bibinfo {author} {\bibfnamefont {H.~A.}\ \bibnamefont {Bethe}},\ }\href {https://doi.org/10.1103/PhysRev.72.339} {\bibfield  {journal} {\bibinfo  {journal} {Phys. Rev.}\ }\textbf {\bibinfo {volume} {72}},\ \bibinfo {pages} {339} (\bibinfo {year} {1947})}\BibitemShut {NoStop}%
\bibitem [{\citenamefont {Cohen-Tannoudji}\ \emph {et~al.}(2004)\citenamefont {Cohen-Tannoudji}, \citenamefont {Dupont-Roc},\ and\ \citenamefont {Grynberg}}]{Cohen}%
  \BibitemOpen
  \bibfield  {author} {\bibinfo {author} {\bibfnamefont {C.}~\bibnamefont {Cohen-Tannoudji}}, \bibinfo {author} {\bibfnamefont {J.}~\bibnamefont {Dupont-Roc}},\ and\ \bibinfo {author} {\bibfnamefont {G.}~\bibnamefont {Grynberg}},\ }\href {https://doi.org/10.1002/9783527617197} {\emph {\bibinfo {title} {Atom - Photon Interactions: Basic Processes and Applications}}}\ (\bibinfo  {publisher} {WILEY-VCH Verlag GmbH \& Co. KGaA},\ \bibinfo {address} {Weinheim},\ \bibinfo {year} {2004})\ Chap.\ \bibinfo {chapter} {Complement B.IV.3}, pp.\ \bibinfo {pages} {326 -- 329}\BibitemShut {NoStop}%
\bibitem [{\citenamefont {Welton}(1948)}]{Welton}%
  \BibitemOpen
  \bibfield  {author} {\bibinfo {author} {\bibfnamefont {T.~A.}\ \bibnamefont {Welton}},\ }\href {https://doi.org/10.1103/PhysRev.74.1157} {\bibfield  {journal} {\bibinfo  {journal} {Phys. Rev.}\ }\textbf {\bibinfo {volume} {74}},\ \bibinfo {pages} {1157} (\bibinfo {year} {1948})}\BibitemShut {NoStop}%
\end{thebibliography}%

\onecolumngrid
\appendix

\section{\label{app:Classical} Classical action and Hamiltonian}

\subsection{ \label{app:EHAction} The Einstein-Hilbert action}

The quantum vacuum of the gravitational field is modelled by the Einstein-Hilbert action in the weak field limit
\begin{equation}
    S_{EH} = \frac{c^{4}}{16 \pi G} \int \text{d}^{4}x \sqrt{-\text{det}(\mathbf{g})} R , 
\end{equation}
where $R$ is the Ricci scalar and $g_{\mu \nu}$ is the metric of the spacetime.
In the weak-field limit $g_{\mu \nu} = \eta_{\mu \nu} + h_{\mu \nu}$, where $\eta_{\mu \nu}$ is the Minkowski metric, $\text{diag}(-1, 1, 1, 1)$, and $| h_{\mu \nu} | \ll 1$.

To second order in $h_{\mu \nu}$, the Einstein-Hilbert action becomes
\begin{equation}
    S_{EH} = \frac{c^{4}}{32 \pi G} \int \text{d}^{4}x  \left( h^{\mu \nu} \partial_{\mu} \partial_{\nu} h + h \, \partial_{\mu} \partial_{\nu} h^{\mu \nu} + h^{\mu \nu} \square h_{\mu \nu} 
    - h \square h - h^{\mu \nu} \partial_{\mu} \partial^{\rho} h_{\rho \nu} - h^{\mu \nu} \partial_{\nu} \partial^{\rho} h_{\rho \mu} \right) ,
    \label{eq:ehactionlin}
\end{equation}
where we have omitted the linear terms in $h_{\mu \nu}$ as they do not affect the dynamics.

The perturbation $h_{\mu \nu}$ has 10 independent components. Making use of the gauge freedom of the Einstein-Hilbert action we impose both the de Donder gauge
\begin{align}
    \partial^{\mu} h_{\mu \nu} = \frac{1}{2} \partial_{\nu} h = 0,
\end{align}
and the transverse-traceless (TT) gauge,
\begin{align}
    \partial^{\mu} h_{\mu \nu} &= 0, \\
    h_{0 \mu} &= 0, \\
    h &= 0.
\end{align}
Together, these gauge choices naturally describe the solution of the Einstein-Hilbert action as gravitational waves. The perturbation metric now has only 2 independent components
\begin{align}
    h_{\mu \nu} =   \begin{pmatrix}
                    0 & 0 & 0 & 0 \\
                    0 & h_{11} & h_{12} & 0 \\
                    0 & h_{12} & -h_{11} & 0 \\
                    0 & 0 & 0 & 0
                    \end{pmatrix}
\end{align}
where we have taken the direction of propagation of gravitational waves to be in the $\mathbf{z}$ direction.
Finally, we can expand the metric perturbation into Fourier modes as follows
\begin{align}
    \label{eq:hexp2}
    h_{ij}(t, \mathbf{X}) = \frac{1}{V} \sum_{\mathbf{k}, s} q_{\mathbf{k}, s} e^{i \mathbf{k} \cdot \mathbf{X}} \epsilon^{s}_{ij}(\mathbf{k}),
\end{align}
where $q_{\mathbf{k}}$ is the amplitude of each mode $\mathbf{k}$ and $\epsilon^{s}_{ij}(\mathbf{k})$ is the polarisation tensor. 

\subsection{\label{app:paction} Modelling interaction of matter with gravitational field}

Having modelled the gravitational field as a collection of linearized gravitational waves, we now look to model the interaction of the field with two interacting massive particles in a quadratic potential. We consider the following Lagrangian,
\begin{align}
    L = \frac{1}{2} m_{1} \dot{x_{1}}^{2} + \frac{1}{2} m_{2} \dot{x_{2}}^{2} - \frac{q}{2} \left( x_{1} - x_{2} \right)^{2},
\end{align}
where $q$ is the strength of interaction. Moving into the centre of mass motion variables
\begin{align}
    x_{+} &= \frac{m_{1} x_{1} + m_{2} x_{2}}{m_{1} + m_{2}}, \\
    x_{-} &= x_{1} - x_{2}, \\
    M &= m_{1} + m_{2}, \\
    \mu &= \frac{m_{1} m_{2}}{ m_{1} + m_{2}}.
\end{align}
The Lagrangian then takes the form
\begin{align}
    L = \frac{1}{2} M \dot{x}^{2}_{+} + \frac{1}{2} \mu \dot{x}^{2}_{-} - \frac{1}{2} \mu \omega^{2} x_{-}^{2},
\end{align}
where $\omega^{2} = q/ \mu$.

It can be seen that the centre of mass motion $x_{+}$ is that of a single free particle. We follow a similar argument to that made in \cite{Torovs2020} in that the motion of a single free particle will follow a geodesic and therefore, has no frame invariant observable that can be measured. We therefore only consider the differential mode motion $x_{-}$.

The separation between the particles, $x_{-}$, depends on the chosen reference frame. We therefore transform the coordinate position $x_{-}$ into the geodesic separation $\xi$ with the following transformation
\begin{equation}
    \xi^{i} = x_{-}^{i} + \frac{1}{2} \delta^{i j} h_{j k} x_{-}^{k}.
\end{equation}
We have assumed that the metric perturbation $h_{ij}$ is constant across the separation between the two masses.

In the non-relativistic limit, the behaviour of a linearized GR wave on a particle in a potential $\Phi (\xi)$ in a flat Minkowski space-time is modelled by the Lagrangian
\begin{align}
    \label{eq:Lgrwave2}
    L = \frac{1}{2}\mu \dot{\xi}^{2} - \mu \Gamma^{i}_{0 j} \dot{\xi}_{i} \xi^{j} - \frac{1}{2} \mu \omega^{2} \xi^{2}.
\end{align}
where $\xi$ is the geodesic separation of the test particle from the origin of our coordinate system. The derivation of (\ref{eq:Lgrwave2}) is found in \cite{Speliotopoulos} where the TT-guage is chosen and two main assumptions have been made. The first is that the velocity of the particle is low enough to be in the non-relativistic limit $v \ll c$. The second is the long-wavelength approximation.

In the TT-gauge,
\begin{equation}
    \Gamma^{i}_{0 j} = \frac{1}{2}\dot{h}^{i}_{j},
\end{equation}
where the dot indicates the derivative with respect to the coordinate time $t$.

Using the long wavelength approximation, the metric expansion Eq.~6 can be simplified
\begin{equation}
    \dot{h}^{i}_{j} = \frac{1}{V} \sum_{\mathbf{k} , s} \dot{q}_{\mathbf{k} , s} (t) \; \epsilon^{i}_{j} (\mathbf{k}, s)
.
\end{equation}
We consider only one polarisation such that $\epsilon^{i}_{j} = \delta^{1}_{1}$ i.e. the "$+$" polarisation. The Lagrangian is then
\begin{equation}
    L = \frac{1}{2}\mu \dot{\xi}^{2} - \frac{1}{2} \mu \omega^{2} \xi^{2}  - \frac{\mu}{2V} \sum_{\mathbf{k}} \dot{q}_{\mathbf{k}} (t) \dot{\xi} \xi .
\end{equation}
Adding the field Lagrangian derived from the linearised Einstein-Hilbert action in the TT-gauge the full Lagrangian is then
\begin{equation}
    L = \frac{1}{2}\mu \dot{\xi}^{2} - \frac{1}{2} \mu \omega^{2} \xi^{2} - \frac{\mu}{2V} \sum_{\mathbf{k}} \dot{q}_{\mathbf{k}} (t) \dot{\xi} \xi + \frac{\rho_{\textsc{f}}}{2 V} \sum_{\mathbf{k}} (\dot{q}_{\mathbf{k}}^{2} - \Omega_{\mathbf{k}}^{2} q_{\mathbf{k}}^{2}),
\end{equation}
where $\Omega_{k} = c k$ and
\begin{equation}
    \rho_{\textsc{f}} = \frac{c^{2}}{8 \pi G}.
\end{equation}
The Hamiltonian is then
\begin{eqnarray}
    H = \left( \frac{p_{\xi}^{2}}{2 \mu} + \frac{\xi p_{\xi}}{2 \rho_{\textsc{f}} } \sum_{\mathbf{k}} p_{\mathbf{k}} +  \sum_{\mathbf{k}} \frac{V p_{\mathbf{k}}^{2}}{2 \rho_{\textsc{f}}} \right) A (\xi) + \frac{1}{2} \mu \omega^{2} \xi^{2} + \frac{1}{V}\sum_{\mathbf{k}}  \frac{\rho_{\textsc{f}}}{2} \Omega_{\mathbf{k}}^{2} q_{\mathbf{k}}^{2},
\end{eqnarray}
where
\begin{equation}
    A(\xi) = \left(1 - \frac{\mu \xi^{2}}{4 \rho_{\textsc{f}} V} \sum_{\mathbf{k}} 1
    \right)^{-1},
\end{equation}
and $p_{\xi}$,$p_{\mathbf{k}}$ are the canonical momenta of $\xi$ and $q_{\mathbf{k}}$ respectively.

We make the assumption that $A(\xi) \approx 1$. The Hamiltonian is then
\begin{equation}
    H = \frac{p_{\xi}^{2}}{2 \mu} + \frac{1}{2} \mu \omega^{2} \xi^{2} + \frac{\xi p_{\xi}}{2  \rho_{\textsc{f}}}  \sum_{\mathbf{k}} p_{\mathbf{k}} + \sum_{\mathbf{k}} \left( \frac{V p_{\mathbf{k}}^{2}}{2 \rho_{\textsc{f}}} + \frac{\rho_{\textsc{f}}}{2V} \Omega_{\mathbf{k}}^{2} q_{\mathbf{k}}^{2}\right).
\end{equation}
The interaction term is then made symmetric in $\xi$ and $p_{\xi}$
\begin{eqnarray}
    \label{eq:Ham}
    H = \frac{p_{\xi}^{2}}{2 \mu} + \frac{1}{2} \mu \omega^{2} \xi^{2} + \frac{\xi p_{\xi} + p_{\xi} \xi}{4 \rho_{\textsc{f}}}  \sum_{\mathbf{k}} p_{\mathbf{k}} + \sum_{\mathbf{k}} \left( \frac{V p_{\mathbf{k}}^{2}}{2 \rho_{\textsc{f}}} + \frac{\rho_{\textsc{f}}}{2V} \Omega_{\mathbf{k}}^{2} q_{\mathbf{k}}^{2} \right).
    \label{OsH}
\end{eqnarray}
In order to compare to the dynamics of the $x$ variable, we transform $\xi$ back into $x$ to find the Hamiltonian
\begin{align}
    \label{eq:Ham2}
    H = \frac{p_{x}^{2}}{2 \mu} + \frac{1}{2} \mu \omega^{2} x_{-}^{2} - \frac{1}{V} \sum_{k} \left[  \left( \frac{p_{x}^{2}}{2\mu} - \frac{\mu \omega^{2} x_{-}^{2}}{2} \right) q_{\mathbf{k}} \right] + \sum_{\mathbf{k}} \left( \frac{V p_{\mathbf{k}}^{2}}{2 \rho_{\textsc{f}}} + \frac{\rho_{\textsc{f}}}{2V} \Omega_{\mathbf{k}}^{2} q_{\mathbf{k}}^{2} \right),
\end{align}
where $p_{x}$ is the conjugate momentum of $x$.
In this paper, we also study the dynamics of two free particles by taking, $\omega = 0$. 

\section{\label{app:freeparticle} Deriving the master equation for free particle}

In this section, we detail the steps involved in obtaining the master equation for the free particle interacting with the gravitational field. We do this for both position variables $x$ and $\xi$. 

\subsection{\label{app:freex} Coordinate separation}

We start from the classical Hamiltonian \ref{eq:Ham2} where we take $\omega = 0$ for the free particle and the subscript has been dropped such that $x_{-} = x$. This Hamiltonian can be partitioned into particle, interaction and field terms as follows:
\begin{align}
    \label{eq:Hamp}
    H_{\textsc{p}, x} &= \frac{p_{x}^{2}}{2\mu}, \\
    \label{eq:Hami}
    H_{\textsc{i}, x} &= -\frac{ p^{2}_{x}}{2 \mu} \frac{1}{V} \sum_{\mathbf{k}} q_{\mathbf{k}}, \\
    \label{eq:Hamf}
    H_{\textsc{f}, x} &= \sum_{\mathbf{k}} \left( \frac{Vp_{\mathbf{k}}^{2}}{2 \rho_{\textsc{f}}} + \frac{\rho_{\textsc{f}}}{2V} \Omega_{\mathbf{k}}^{2} q_{\mathbf{k}}^{2} \right).
\end{align}
The dynamical variables are promoted to operators with the corresponding commutation relations imposed:
\begin{align}
    [\hat{x}, \hat{p}_{x}] = i \hbar, \qquad [\hat{q}_{k}, \hat{p}_{k}] =i \hbar.
\end{align}
The field constant is $\rho_{\textsc{f}}$, where 
\begin{align}
    \rho_{\textsc{f}} = \frac{c^{2} }{8 \pi G}.
\end{align}
As the field Hamiltonian \ref{eq:Hamf} is a collection of harmonic oscillators, the following ladder operators can be defined 

\begin{align}
   \hat{a}_{k} &= \sqrt{\frac{\rho_{\textsc{f}} \Omega_{k}}{2 V \hbar}} \left(\hat{q}_{k} + \frac{i V}{ \rho_{\textsc{f}} \Omega_{k}} \hat{p}_{k} \right), \\
\hat{a}^{\dagger}_{k} &= \sqrt{\frac{ \rho_{\textsc{f}} \Omega_{k}}{2 V \hbar}} \left(\hat{q}_{k} - \frac{i V}{\rho_{\textsc{f}} \Omega_{k}} \hat{p}_{k} \right)
.
\end{align}
The field Hamiltonian is then 
\begin{align}
\hat{H}_{\textsc{f}, x} = \hbar \Omega_{k} \left( \hat{a}^{\dagger}_{k} \hat{a}_{k} + \frac{1}{2} \right),
\end{align}
and interaction Hamiltonian 
\begin{align}
    \hat{H}_{\textsc{i}, x} = \sum_{k} f^{(1)}_{x, k} \hat{p}^{2}_{x} (\hat{a}_{k} + \hat{a}^{\dagger}_{k}),
\end{align}
where
\begin{align}
   f^{(1)}_{x, k} = \sqrt{\frac{4 \pi G \hbar }{ V c^{2} \mu^{2} \Omega_{k}}} \sin^{2}{(\theta_{\mathbf{k}})}\cos{(2 \phi_{\mathbf{k}})},
\end{align}
and $\theta_{\mathbf{k}}$, $\phi_{\mathbf{k}}$ are polar angles denoted in Fig 2.

We want to study the dynamics of the particle subsystem using its time-evolved reduced density matrix $ {\sigma}(t) = \text{Tr}_{\textsc{f}}[  {\rho}(t) ]$ where $ {\rho}(t)$ denotes the density matrix of the full particle-field system. The density matrix corresponding to the field $ {\sigma}_{\textsc{f}} (t) = \text{Tr}_{\textsc{p}}[  {\rho}(t) ]$. We work in the interaction picture denoted by the tilde on operators
\begin{align}
\tilde{a}_{k}(t) &= e^{-i \Omega_{k} t} \hat{a}_{k}, \\
 \\
\tilde{p}_{x}(t) &= \hat{p}_{x},
\end{align}
obtained using
\begin{align}
    \widetilde{\mathcal{X}}(t) = e^{\frac{i}{\hbar}\left(H_{\textsc{p}} + H_{\textsc{f}} \right)t} \hat{\mathcal{X}} (0) e^{-\frac{i}{\hbar}\left(H_{\textsc{p}} + H_{\textsc{f}} \right)t},
\end{align}
for any operator $\hat{\mathcal{X}}$.
Using the assumptions (1) - (3), the Liouville equation to second-order in the interaction strength is presented~\cite{Cohen}, 
\begin{align}
    \label{eq:evoeq1}
    \frac{\Delta \widetilde{\sigma}(t)}{\Delta t} = -\frac{1}{\hbar^{2}} \frac{1}{\Delta t} \int^{\infty}_{0} d(t_{1} - t_{2}) \int_{t}^{t + \Delta t} d t_{1} \text{Tr}_{\textsc{f}} [ \widetilde{H}_{\textsc{i}} ( t_{1} ), [ \widetilde{H}_{\textsc{i}} ( t_{2} ) , \widetilde{\sigma} (t) \otimes \widetilde{\sigma}_{\textsc{f}} ]].
\end{align}
This equation can be expressed in terms of the defined quantity $R$
\begin{align}
\nonumber \frac{\Delta \sigma}{\Delta t} &= -\frac{1}{h^{2}} \frac{1}{\Delta t} \int^{\infty}_{0} d(t_{1} - t_{2}) \int_{t}^{t + \Delta t} d t_{1}\\ &\text{Tr}_{\textsc{f}} [ p^{2}_{x} R ( t_{1} ), [ p^{2}_{x} R^{\dagger} ( t_{2} ) , \sigma (t) \otimes \sigma_{\textsc{f}} ]] + \text{h.c},
\end{align}
where
\begin{align}
    R(t) &= \sum_{k} f^{(1)}_{x, k} a_{k}(t),
\end{align}
and all operators are in the interaction picture but the tilde has been dropped.
Evaluating the first commutator.
\begin{equation}
    \label{eq:comm1}
    \frac{\Delta \sigma}{\Delta t} = -\frac{1}{h^{2}} \frac{1}{\Delta t} \int^{\infty}_{0} d(t_{1} - t_{2}) \int_{t}^{t + \Delta t} d t_{1} \left(\text{Tr}_{\textsc{f}} \left[ p^{2}_{x} R(t_{1}), \left( p^{2}_{x} \sigma \otimes R^{\dagger}(t_{2}) \sigma_{\textsc{f}} - \sigma p^{2}_{x} \otimes \sigma_{\textsc{f}} R^{\dagger}(t_{2}) \right) \right] \right)
\end{equation}
Evaluating the commutator in \ref{eq:comm1} and making use of the invariance of cyclic permutations of trace we obtain
\begin{align}
\nonumber \frac{\Delta \sigma}{\Delta t} &= -\frac{1}{h^{2}} \frac{1}{\Delta t} \int^{\infty}_{0} \text{d} \tau \int_{t}^{t + \Delta t} d t_{1} \sum_{k} \\ 
\Big[&  \left( p^{4}_{x} \sigma - p^{2}_{x} \sigma p^{2}_{x} \right) \langle R (t_{1}) R^{\dagger} (t_{2}) \rangle_{\textsc{f}} \nonumber \\
&+ \left(\sigma p^{4}_{x} - p^{2}_{x} \sigma p^{2}_{x} \right) \langle R^{\dagger} (t_{2}) R (t_{1}) \rangle_{\textsc{f}} + h.c. \Big],
\end{align}
where $\tau = t_{1} - t_{2}$.
Evaluating the expectation values of $\langle R(t_{1})R^{\dagger}(t_2)\rangle_{\textsc{f}}$
\begin{align}
\langle R (t_{1}) R^{\dagger} (t_{2}) \rangle_{\textsc{f}} &= \sum_{k} |f_{x, k}|^{2} ( \langle n_{k} \rangle + 1 ) e^{-i \Omega_{k} (t_{1} - t_{2})}, \\
\langle R^{\dagger} (t_{2}) R (t_{1}) \rangle_{\textsc{f}} &= \sum_{k} |f_{x, k}|^{2} \langle n_{k} \rangle e^{-i \Omega_{k} (t_{1} - t_{2}) } .
\end{align}
The master equation becomes
\begin{align}
\nonumber \frac{\Delta \sigma}{\Delta t} &= -\frac{1}{h^{2}} \frac{1}{\Delta t} \int^{\infty}_{0} \text{d} \tau \int_{t}^{t + \Delta t} d t_{1} \sum_{k} \\ 
\Big[&  \left( p^{4}_{x} \sigma - p^{2}_{x} \sigma p^{2}_{x} \right) |f_{x, k}|^{2} (\langle n_{k} \rangle + 1) e^{-i\Omega_{k} \tau} \nonumber \\
&+ \left(\sigma p^{4}_{x} - p^{2}_{x} \sigma p^{2}_{x} \right) |f_{x, k}|^{2} \langle n_{k} \rangle e^{-i\Omega_{k} \tau} + h.c. \Big],
\end{align}
Integrating over $t_{1}$ and taking the limit $\Delta t \rightarrow0$
\begin{align}
\nonumber \frac{\Delta \sigma}{\Delta t} &= -\frac{1}{h^{2}} \int^{\infty}_{0} \text{d} \tau \sum_{k} \\ 
\Big[&  \left( p^{4}_{x} \sigma - p^{2}_{x} \sigma p^{2}_{x} \right) |f_{x, k}|^{2} (\langle n_{k} \rangle + 1) e^{-i\Omega_{k} \tau} \nonumber \\
&+ \left(\sigma p^{4}_{x} - p^{2}_{x} \sigma p^{2}_{x} \right) |f_{x, k}|^{2} \langle n_{k} \rangle e^{-i\Omega_{k} \tau} + h.c. \Big],
\end{align}
The integral over $\tau$ can be found using
\begin{align}
\int_{0}^{\infty} e^{-i \Omega_{k} \tau} \text{d} \tau &= \frac{-i}{\Omega_{k}} + \pi\delta(\Omega_{k}),
\end{align}
and defining the following useful constants
\begin{align}
    \delta_{x} &= \sum_{k} \frac{|f_{k}|^{2}}{\Omega_{k} \hbar^{2}}, \\
    \delta_{x}' &= \sum_{k} \frac{|f_{k}|^{2}}{\Omega_{k} \hbar^{2}} \langle n_{k} \rangle, \\
    \Gamma_{1} &= \sum_{k} \frac{|f_{x, k}|^{2}}{ \hbar^{2}} \pi \delta(\Omega_{k}), \\
    \Gamma'_{1} &= \sum_{k} \frac{|f_{x, k}|^{2}}{ \hbar^{2}} \pi \delta(\Omega_{k})\langle n_{k} \rangle.
\end{align}
noting that $\Gamma_{1} = \Gamma'_{1} = 0$. 
Moving back into the Schr{\"o}dinger picture
\begin{align}
\nonumber \frac{\Delta \hat{\sigma}}{\Delta t} &=  \frac{i}{\hbar} \left[\hat{\sigma}, \hat{H}_{\textsc{p}, x} \right] \nonumber \\
&-\Big[  \left( \hat{p}^{4}_{x} \hat{\sigma} - \hat{p}^{2}_{x} \hat{\sigma} \hat{p}^{2}_{x} \right) \left( i\delta_{x} + i\delta'_{x} \right) \nonumber \\
&+ \left(\hat{\sigma} \hat{p}^{4}_{x} - \hat{p}^{2}_{x} \hat{\sigma} \hat{p}^{2}_{x} \right) i\delta'_{x} + h.c. \Big],
\end{align}
\\
We aim to study the effect of the gravitational vacuum state on our particle. We therefore set $\langle n_{k} \rangle = 0$ and therefore $\delta'_{x} = 0$
\begin{align}
\frac{\text{d}\hat{\sigma}}{\text{d}t} &= \frac{i}{\hbar} \left[\hat{\sigma}, \hat{H}_{\textsc{p}, x} \right] - i \delta_{x} \left[\hat{p}^{4}_{x} , \hat{\sigma} \right].
\end{align}
Evaluating the constant
\begin{align}
    \delta_{x} &= \frac{G }{2 \pi^{2} \hbar \mu^{2} c^{5}} \int \text{d}^{3}\mathbf{k} \; \frac{1}{\Omega_{k}} \; \sin^{4}{(\theta_{\mathbf{k}})} \cos^{2}{(\phi_\mathbf{k})} \nonumber \\  
    &= \frac{G }{2 \pi^{2} \hbar \mu^{2} c^{5}} \int_{0}^{\Omega_{\textsc{max}}} \text{d} \Omega_{k} \; \int_{0}^{\pi} \text{d} \theta \; \sin^{5}{(\theta_{\mathbf{k}})}\int_{0}^{2 \pi} \text{d} \phi \; \cos^{2}{(\phi_\mathbf{k})} \nonumber \\
    &= \frac{8 }{15 \pi}  \frac{G }{\hbar \mu^{2} c^{5}} \Omega_{\text{max}} \nonumber \\
    &= \frac{8}{15 \pi} \frac{t^{2}_{\textsc{p}} \Omega_{\text{max}}}{\mu^{2} \hbar^{2}}.
\end{align}

From the master equation, the dynamics of expectation values can be found,
\begin{align}
    \frac{\text{d}}{\text{d}t} \langle \hat{p}_{x}^{a} \hat{x}^{b} \rangle &= \frac{b}{\mu} \langle \hat{p}_{x}^{a+1} \hat{x}^{b-1} \rangle + \frac{i \hbar b (b-1) }{2\mu} \langle \hat{p}_{x}^{a} \hat{x}^{b-2} \rangle + 4 \hbar \delta_{x} b \langle \hat{p}_{x}^{a+3} \hat{x}^{b-1} \rangle + 6 i \hbar^{2} \delta_{x} b (b-1) \langle \hat{p}_{x}^{a+2} \hat{x}^{b-2} \rangle \nonumber \\
    &- 4\hbar^{3} \delta_{x} b (b-1) (b-2) \langle \hat{p}_{x}^{a+1} \hat{x}^{b-3} \rangle - i \hbar^{4} \delta_{x} b (b-1) (b-2) (b-3) \langle \hat{p}_{x}^{a} \hat{x}^{b-4} \rangle, 
\end{align}
where $a$, $b$ are positive integers.

The mean and variance of $\hat{x}$ and $\hat{p}_{x}$ can then be calculated
\begin{align}
    \label{eq:eomx}
    \langle \hat{x} \rangle (t) &= \left( \frac{1}{\mu} \langle \hat{p}_{x} \rangle (0) + 4 \hbar \delta_{x} \langle \hat{p}^{3}_{x} \rangle (0) \right) t , \\
    \langle \hat{x}^{2} \rangle (t) &= \left( \frac{\langle \hat{p}_{x}^{2} \rangle (0)}{\mu^{2}} + \frac{8 \hbar \delta_{x}}{\mu} \langle \hat{p}_{x}^{4} \rangle (0) + 16 \hbar^{2} \delta_{x}^{2} \langle \hat{p}_{x}^{6} \rangle (0) \right) t^{2} + \left( \frac{\langle \hat{x} \hat{p}_{x} + \hat{p}_{x} \hat{x} \rangle }{\mu} + 4 \hbar \delta_{x} \langle \hat{p}_{x}^{3} \hat{x} + \hat{x} \hat{p}_{x}^{3} \rangle \right) t + \langle \hat{x}^{2} \rangle (0), \\
    \langle \hat{p}^{n} \rangle (t) &= \langle \hat{p}^{n} \rangle (0),
\end{align}
for any positive integer $n$.

In order to regularise the results in the case of the harmonically trapped particle, the master equation (\ref{eq:mastereqd}) in that case is stated here and is also rewritten as,
\begin{align}
\frac{\text{d} \hat{\sigma}}{\text{d} t} &= -i \left(\omega - \delta_{-} -  3\delta_{+} \right)  \left[ \hat{b}_{x}^{\dagger} \hat{b}_{x}, \hat{\sigma} \right] \nonumber \\
&+ \frac{i}{\hbar} \frac{ \delta_{+} -  \delta_{-}}{\hbar \omega^{2}} \left[ \left( \hat{H}_{\textsc{p}}^{2} \right)^{2}, \hat{\sigma}  \right] \nonumber \\
&+ \Gamma \left( \hat{b}_{x}^{2} \hat{\sigma} \hat{b}_{x}^{\dagger 2} - \frac{1}{2} \hat{b}_{x}^{\dagger 2} \hat{b}_{x}^{2} \hat{\sigma} - \frac{1}{2} \hat{\sigma} \hat{b}_{x}^{\dagger 2} \hat{b}_{x}^{2} \right) \nonumber \\
&-  i \delta_{+} \left( \hat{b}^{\dagger 2} \hat{\sigma} \hat{b}_{x}^{\dagger 2} - \hat{b}_{x}^{\dagger 4} \hat{\sigma} + \hat{\sigma} \hat{b}_{x}^{4} - \hat{b}_{x}^{2} \hat{\sigma} \hat{b}_{x}^{2} \right) \nonumber \\
&-  \left(i \delta_{-} + \frac{\Gamma}{2} \right)\left( \hat{b}_{x}^{4} \hat{\sigma} - \hat{b}_{x}^{2} \hat{\sigma} \hat{b}_{x}^{2} \right) \nonumber \\
&-  \left(i \delta_{-} - \frac{\Gamma}{2} \right) \left( \hat{b}_{x}^{\dagger 2} \hat{\sigma} \hat{b}_{x}^{\dagger 2} - \hat{\sigma} \hat{b}_{x}^{\dagger 4} \right).
\end{align}
The second line corresponds to a shift in the square of the energy $\delta E$, similar to that found in the free particle, 
\begin{align}
    \delta E = \frac{\delta_{+} - \delta_{-}}{\hbar \omega^{2}}.
\end{align}
Using the definition of the parameters $\delta_{\pm}$, we see that the shift has two terms: one linear in $\Omega_{\mathbf{k}}$ and one logarithmic in $\Omega_{\mathbf{k}}$. Considering the linear term $\delta E_{\text{lin}}$ in the shift $\delta E$, 
\begin{align}
    \delta E_{\text{lin}} = \frac{16 G}{15 \pi c^{2}} \Omega_{\mathbf{k}}
\end{align}
Rewriting the master equation for the free particle,
\begin{align}
\frac{\text{d}\sigma}{\text{d}t} &= \frac{i}{\hbar} \left[\sigma, H_{P} \right] - \frac{i}{\hbar}  \delta_{\textsc{fp}} \left[ \hat{H}^{2}_{\textsc{p}} , \sigma \right],
\end{align}
where
\begin{align}
    \delta E_{\textsc{fp}} =4 \mu^{2} \hbar \delta_{x}.
\end{align}
It is interesting to note that 
\begin{align}
   \frac{\delta E_{\textsc{fp}}}{2} =  \delta E_{\text{lin}}.
\end{align}
This can be explained as follows.
Bethe argues that the regularisation of the energy shift in the trapped case is achieved by subtracting the energy shift of a free particle, choosing the kinetic energy of that free particle to be the same as the average kinetic energy of the trapped system~\cite{Bethe}. As our particle is in a harmonic trap, its average kinetic energy is just half the total energy of the system. We, therefore, see that the $\delta E_{\text{lin}}$ is the term that needs to be subtracted out to regularise the results. 
The shift $\delta E$ in the case of the harmonic oscillator is left solely with the logarithmic dependence in $\Omega_{\mathbf{k}}$. Alternatively, this can also be achieved by introducing the appropriate counter-terms to the Hamiltonian as described in \cite{Welton}. This gives the renormalised constants,
\begin{align}
    \label{eq:deltarenormx}
    \delta_{\pm}^{(\textsc{r})} = -\frac{\Gamma}{2 \pi} \ln{\left| \Omega_{\textsc{max}} \pm 2 \omega  \right|}.
\end{align}

\subsection{\label{app:freexi} Geodesic separation}

We start from the classical Hamiltonian \ref{eq:Ham} where we take $\omega = 0$ for the free particle. This Hamiltonian can be partitioned into particle, interaction and field terms as follows:
\begin{align}
    \label{eq:Hamp2b}
    H_{\textsc{p}} &= \frac{p_{\xi}^{2}}{2\mu}, \\
    \label{eq:Hami2b}
    H_{\textsc{i}} &= \frac{ \xi p_{\xi} + p_{\xi} \xi}{4 \rho_{\textsc{f}}} \sum_{\mathbf{k}} p_{\mathbf{k}}, \\
    \label{eq:Hamf2b}
    H_{\textsc{f}} &= \sum_{\mathbf{k}} \left( \frac{Vp_{\mathbf{k}}^{2}}{2 \rho_{\textsc{f}}} + \frac{\rho_{\textsc{f}}}{2V} \Omega_{\mathbf{k}}^{2} q_{\mathbf{k}}^{2} \right).
\end{align}
The dynamical variables are promoted to operators with the corresponding commutation relations imposed:
\begin{align}
    [\hat{\xi}, \hat{p}_{\xi}] = i \hbar, \qquad [\hat{q}_{k}, \hat{p}_{k}] =i \hbar.
\end{align}
The field constant is $\rho_{\textsc{f}}$, where 
\begin{align}
    \rho_{\textsc{f}} = \frac{c^{2} }{8 \pi G}.
\end{align}
As the field Hamiltonian \ref{eq:Hamf2b} is a collection of harmonic oscillators, the following ladder operators can be defined 

\begin{align}
   \hat{a}_{k} &= \sqrt{\frac{\rho_{\textsc{f}} \Omega_{k}}{2 V \hbar}} \left(\hat{q}_{k} + \frac{i V}{ \rho_{\textsc{f}} \Omega_{k}} \hat{p}_{k} \right), \\
    \hat{a}^{\dagger}_{k} &= \sqrt{\frac{ \rho_{\textsc{f}} \Omega_{k}}{2 V \hbar}} \left(\hat{q}_{k} - \frac{i V}{\rho_{\textsc{f}} \Omega_{k}} \hat{p}_{k} \right).
\end{align}
The field Hamiltonian is then 
\begin{align}
\hat{H}_{\textsc{f}, \xi} = \hbar \Omega_{k} \left( \hat{a}^{\dagger}_{k} \hat{a}_{k} + \frac{1}{2} \right),
\end{align}
and interaction Hamiltonian 
\begin{align}
    \hat{H}_{\textsc{i}, \xi} = \sum_{k} f_{k} (\hat{\xi} \hat{p}_{\xi} + \hat{p}_{\xi} \hat{\xi}) (a_{k} - a^{\dagger}_{k}),
\end{align}
where
\begin{align}
   f_{\xi, k} = -i \sqrt{\frac{\pi G \hbar \Omega_{k} }{ V c^{2}}} \sin^{2}{(\theta_{\mathbf{k}})}\cos{(2 \phi_{\mathbf{k}})},
\end{align}
and $\theta_{\mathbf{k}}$, $\phi_{\mathbf{k}}$ are polar angles denoted in Fig. 1.

We want to study the dynamics of the particle subsystem using its time-evolved reduced density matrix $ {\sigma}(t) = \text{Tr}_{\textsc{f}}[  {\rho}(t) ]$ where $ {\rho}(t)$ denotes the density matrix of the full particle-field system. The density matrix corresponding to the field $ {\sigma}_{\textsc{f}} (t) = \text{Tr}_{\textsc{p}}[  {\rho}(t) ]$. We work in the interaction picture denoted by the tilde on operators
\begin{align}
\tilde{a}_{k}(t) &= e^{-i \Omega_{k} t} \hat{a}_{k}, \\
\tilde{\xi}(t) &= \hat{\xi} + \frac{t}{\mu} \hat{p}_{\xi},
 \\
\tilde{p}_{\xi}(t) &= \hat{p}_{\xi},
\end{align}
obtained using
\begin{align}
    \widetilde{\mathcal{X}}(t) = e^{\frac{i}{\hbar}\left(H_{\textsc{p}} + H_{\textsc{f}} \right)t} \hat{\mathcal{X}} (0) e^{-\frac{i}{\hbar}\left(H_{\textsc{p}} + H_{\textsc{f}} \right)t},
\end{align}
for any operator $\hat{\mathcal{X}}$.
Using the assumptions (1) - (3), the Liouville equation to second-order in the interaction strength is presented\cite{Cohen}, 
\begin{align}
    \label{eq:evoeq2}
    \frac{\Delta \widetilde{\sigma}(t)}{\Delta t} = -\frac{1}{\hbar^{2}} \frac{1}{\Delta t} \int^{\infty}_{0} d(t_{1} - t_{2}) \int_{t}^{t + \Delta t} d t_{1} \text{Tr}_{\textsc{f}} [ \widetilde{H}_{\textsc{i}} ( t_{1} ), [ \widetilde{H}_{\textsc{i}} ( t_{2} ) , \widetilde{\sigma} (t) \otimes \widetilde{\sigma}_{\textsc{f}} ]].
\end{align}
This equation can be expressed in terms of the defined quantities $R$ and $\mathcal{O}$
\begin{align}
\nonumber \frac{\Delta \sigma}{\Delta t} &= -\frac{1}{h^{2}} \frac{1}{\Delta t} \int^{\infty}_{0} d(t_{1} - t_{2}) \int_{t}^{t + \Delta t} d t_{1}\\ &\text{Tr}_{\textsc{f}} [ \mathcal{O} (t_{1}) R ( t_{1} ), [ \mathcal{O} ( t_{2} ) R^{\dagger} ( t_{2} ) , \sigma (t) \otimes \sigma_{\textsc{f}} ]] + \text{h.c},
\end{align}
where
\begin{align}
    R(t) &= \sum_{k} f_{\xi, k} a_{k}(t), \\
    \mathcal{O}(t) &= (\xi p_{\xi} + p_{\xi} \xi - \frac{2 t}{\mu}p^{2}_{\xi}),
\end{align}
and all operators are in the interaction picture but the tilde has been dropped.
Evaluating the first commutator.
\begin{equation}
    \label{eq:comm2}
    \frac{\Delta \sigma}{\Delta t} = -\frac{1}{h^{2}} \frac{1}{\Delta t} \int^{\infty}_{0} d(t_{1} - t_{2}) \int_{t}^{t + \Delta t} d t_{1} \left(\text{Tr}_{\textsc{f}} \left[ \mathcal{O}_{1}, \mathcal{O}_{2} \right] \right)
\end{equation}
where
\begin{align}
\mathcal{O}_{1} &= \mathcal{O}(t_{1}) R(t_{1}), \\
\mathcal{O}_{2} &= \mathcal{O}(t_{2}) \sigma(t) \otimes R^{\dagger} \sigma_{\textsc{f}} - \sigma(t) \mathcal{O}(t_{2}) \otimes \sigma_{\textsc{f}} R^{\dagger}.
\end{align}
Evaluating the commutator in \ref{eq:comm1} and making use of the invariance of cyclic permutations of trace we obtain
\begin{align}
\nonumber \frac{\Delta \sigma}{\Delta t} &= -\frac{1}{h^{2}} \frac{1}{\Delta t} \int^{\infty}_{0} \text{d} \tau \int_{t}^{t + \Delta t} d t_{1} \sum_{k} \\ 
\Big[&  \left(\mathcal{O}' \mathcal{O}' \sigma - \mathcal{O}' \sigma \mathcal{O}' \right) \langle R (t_{1}) R^{\dagger} (t_{2}) \rangle_{\textsc{f}} \nonumber \\
&+ \left(\sigma \mathcal{O}'\mathcal{O}' - \mathcal{O}' \sigma \mathcal{O}' \right) \langle R^{\dagger} (t_{2}) R (t_{1}) \rangle_{\textsc{f}} \nonumber \\
&- \left( p^{2} \xi p \sigma + p^{3} \xi \sigma - \xi p \sigma p^{2} - p \xi \sigma p^{2}\right)  \frac{2 t_{1}}{\mu} \langle R (t_{1}) R^{\dagger} (t_{2}) \rangle_{\textsc{f}}  \nonumber \\
&- \left( \xi p^{3} \sigma + p \xi p^{2} \sigma - p^{2} \sigma \xi p -p^{2} \sigma p \xi \right) \frac{2 (t_{1} - \tau) }{\mu} \langle R (t_{1}) R^{\dagger} (t_{2}) \rangle_{\textsc{f}}  \nonumber \\
&+ ( p^{4}  \sigma - p^{2} \sigma p^{2} ) \frac{4 t_{1}(t_{1} - \tau)}{m^{2} } \langle R (t_{1}) R^{\dagger} (t_{2}) \rangle_{\textsc{f}}  \nonumber \\
&- \left( \sigma \xi p^{3} + \sigma p \xi p^{2}  - p^{2} \sigma \xi p  - p^{2} \sigma p \xi \right) \langle R^{\dagger} (t_{2}) R (t_{1}) \rangle_{\textsc{f}} \nonumber \\
&- \left( \sigma p^{2} \xi p  + \sigma p^{3} \xi - \xi p \sigma p^{2} - p \xi \sigma p^{2} \right) \langle R^{\dagger} (t_{2}) R (t_{1}) \rangle_{\textsc{f}} \nonumber \\
&+ (\sigma p^{4} -p^{2} \sigma p^{2}) \langle R^{\dagger} (t_{2}) R (t_{1}) \rangle_{\textsc{f}} + h.c. \Big],
\end{align}
where $\tau = t_{1} - t_{2}$ and $\mathcal{O}' = \xi p + p \xi$.
Evaluating the expectation values of $\langle R(t_{1})R^{\dagger}(t_2)\rangle_{\textsc{f}}$
\begin{align}
\langle R (t_{1}) R^{\dagger} (t_{2}) \rangle_{\textsc{f}} &= \sum_{k} |f_{k}|^{2} ( \langle n_{k} \rangle + 1 ) e^{-i \Omega_{k} (t_{1} - t_{2})}, \\
\langle R^{\dagger} (t_{2}) R (t_{1}) \rangle_{\textsc{f}} &= \sum_{k} |f_{k}|^{2} \langle n_{k} \rangle e^{-i \Omega_{k} (t_{1} - t_{2}) } .
\end{align}
The master equation becomes
\begin{align}
\nonumber \frac{\Delta \sigma}{\Delta t} &= -\frac{1}{h^{2}} \frac{1}{\Delta t} \int^{\infty}_{0} \text{d} \tau \int_{t}^{t + \Delta t} d t_{1} \sum_{k} \\ 
\Big[&  \left(\mathcal{O}' \mathcal{O}' \sigma - \mathcal{O}' \sigma \mathcal{O}' \right) |f_{\xi, k}|^{2} ( \langle n_{k} \rangle + 1 ) e^{  - i \omega_{k} \tau } \nonumber \\
&+ \left(\sigma \mathcal{O}'\mathcal{O}' - \mathcal{O}' \sigma \mathcal{O}' \right) |f_{\xi, k}|^{2} \langle n_{k} \rangle e^{  - i \omega_{k} \tau } \nonumber \\
&- \left( p^{2} \xi p \sigma + p^{3} \xi \sigma - \xi p \sigma p^{2} - p \xi \sigma p^{2}\right) |f_{\xi, k}|^{2} ( \langle n_{k} \rangle + 1 ) \frac{2 t_{1}}{\mu} e^{  - i \omega_{k} \tau} \nonumber \\
&- \left( \xi p^{3} \sigma + p \xi p^{2} \sigma - p^{2} \sigma \xi p -p^{2} \sigma p \xi \right) |f_{\xi, k}|^{2} \left( \langle  n_{k} \rangle + 1\right) \frac{2 (t_{1} - \tau) }{\mu} e^{  - i \omega_{k} \tau} \nonumber \\
&+ ( p^{4}  \sigma - p^{2} \sigma p^{2} ) |f_{k}|^{2} \langle \left( n_{k} \rangle + 1\right) \frac{4 t_{1}(t_{1} - \tau)}{\mu^{2} } e^{  - i \omega_{k} \tau } \nonumber \\
&- \left( \sigma \xi p^{3} + \sigma p \xi p^{2}  - p^{2} \sigma \xi p  - p^{2} \sigma p \xi \right) |f_{\xi, k}|^{2}  \langle n_{k} \rangle  \frac{2 t_{1}}{\mu } e^{  - i \omega_{k} \tau} \nonumber \\
&- \left( \sigma p^{2} \xi p  + \sigma p^{3} \xi - \xi p \sigma p^{2} - p \xi \sigma p^{2} \right) |f_{\xi, k}|^{2} \langle  n_{k} \rangle  \frac{2 (t_{1} - \tau)}{\mu } e^{  - i \omega_{k} \tau} \nonumber \\
&+ (\sigma p^{4} -p^{2} \sigma p^{2}) |f_{\xi, k}|^{2} \langle n_{k} \rangle \frac{4 t_{1}(t_{1} - \tau)}{\mu^{2}} e^{  - i \omega_{k} \tau } + h.c. \Big].
\end{align}
Integrating over $t_{1}$
\begin{align}
\nonumber \frac{\Delta \sigma}{\Delta t} &= -\frac{1}{\hbar^{2}} \frac{1}{\Delta t} \int^{\infty}_{0} \text{d} \tau \sum_{k} \\ 
\Big[&  \left(\mathcal{O}' \mathcal{O}' \sigma - \mathcal{O}' \sigma \mathcal{O}' \right) |f_{\xi, k}|^{2} ( \langle n_{k} \rangle + 1 ) \Delta t e^{- i \Omega_{k} \tau } \nonumber \\
&+ \left(\sigma \mathcal{O}'\mathcal{O}' - \mathcal{O}' \sigma \mathcal{O}' \right) |f_{\xi, k}|^{2} \langle n_{k} \rangle \Delta t e^{- i \Omega_{k} \tau } \nonumber \\
&- \left( p^{2} \xi p \sigma + p^{3} \xi \sigma - \xi p \sigma p^{2} - p \xi \sigma p^{2}\right) |f_{\xi, k}|^{2} ( \langle n_{k} \rangle + 1 ) \frac{(\Delta t^{2} + 2 t \Delta t)}{\mu} e^{- i \Omega_{k} \tau} \nonumber \\
&- \left( \xi p^{3} \sigma + p \xi p^{2} \sigma - p^{2} \sigma \xi p -p^{2} \sigma p \xi \right) |f_{\xi, k}|^{2} \left( \langle  n_{k} \rangle + 1\right) \frac{(\Delta t^{2} + 2 t \Delta t - \tau \Delta t)}{\mu} e^{- i \Omega_{k} \tau} \nonumber \\
&+ ( p^{4}  \sigma - p^{2} \sigma p^{2} ) |f_{\xi, k}|^{2} \langle \left( n_{k} \rangle + 1\right) \frac{4(\Delta t^{3} + 3 t^{2} \Delta t + 3  t \Delta t^{2} ) - 6\tau(\Delta t^{2} + 2 t\Delta t )}{3\mu^{2} } e^{- i \Omega_{k} \tau } \nonumber \\
&- \left( \sigma \xi p^{3} + \sigma p \xi p^{2}  - p^{2} \sigma \xi p  - p^{2} \sigma p \xi \right) |f_{\xi, k}|^{2}  \langle n_{k} \rangle  \frac{(\Delta t^{2} + 2 t \Delta t)}{m} e^{- i \Omega_{k} \tau} \nonumber \\
&- \left( \sigma p^{2} \xi p  + \sigma p^{3} \xi - \xi p \sigma p^{2} - p \xi \sigma p^{2} \right) |f_{\xi, k}|^{2} \langle  n_{k} \rangle  \frac{(\Delta t^{2} + 2 t \Delta t - \tau \Delta t)}{\mu} e^{- i \Omega_{k} \tau} \nonumber \\
&+ (\sigma p^{4} -p^{2} \sigma p^{2}) |f_{\xi, k}|^{2} \langle n_{k} \rangle \frac{4(\Delta t^{3} + 3 t^{2} \Delta t + 3  t \Delta t^{2} ) - 6\tau(\Delta t^{2} + 2 t\Delta t )}{3\mu^{2} } e^{- i \Omega_{k} \tau } + h.c. \Big]
,
\end{align}
As $\Delta t \rightarrow 0$
\begin{align}
\nonumber \frac{\text{d}\sigma}{\text{d}t} &= -\frac{1}{\hbar^{2}} \int^{\infty}_{0} \text{d} \tau \sum_{k} \\ 
\Big[&  \left(\mathcal{O}' \mathcal{O}' \sigma - \mathcal{O}' \sigma \mathcal{O}' \right) |f_{\xi, k}|^{2} ( \langle n_{k} \rangle + 1 ) e^{- i \Omega_{k} \tau } \nonumber \\
&+ \left(\sigma \mathcal{O}'\mathcal{O}' - \mathcal{O}' \sigma \mathcal{O}' \right) |f_{\xi, k}|^{2} \langle n_{k} \rangle e^{- i \Omega_{k} \tau } \nonumber \\
&- \left( p^{2} \xi p \sigma + p^{3} \xi \sigma - \xi p \sigma p^{2} - p \xi \sigma p^{2}\right) |f_{\xi, k}|^{2} ( \langle n_{k} \rangle + 1 ) \frac{2 t}{\mu} e^{- i \Omega_{k} \tau} \nonumber \\
&- \left( \xi p^{3} \sigma + p \xi p^{2} \sigma - p^{2} \sigma \xi p -p^{2} \sigma p \xi \right) |f_{\xi, k}|^{2} \left( \langle  n_{k} \rangle + 1\right) \frac{2 (t - \tau)}{\mu} e^{- i \Omega_{k} \tau} \nonumber \\
&+ ( p^{4}  \sigma - p^{2} \sigma p^{2} ) |f_{k}|^{2} \langle \left( n_{k} \rangle + 1\right) \frac{4( t^{2} - \tau t )}{m^{2} } e^{- i \Omega_{k} \tau } \nonumber \\
&- \left( \sigma \xi p^{3} + \sigma p \xi p^{2}  - p^{2} \sigma \xi p  - p^{2} \sigma p \xi \right) |f_{\xi, k}|^{2}  \langle n_{k} \rangle  \frac{2 t}{\mu} e^{- i \Omega_{k} \tau} \nonumber \\
&- \left( \sigma p^{2} \xi p  + \sigma p^{3} \xi - \xi p \sigma p^{2} - p \xi \sigma p^{2} \right) |f_{\xi, k}|^{2} \langle  n_{k} \rangle  \frac{2 (t - \tau)}{\mu} e^{- i \Omega_{k} \tau} \nonumber \\
&+ (\sigma p^{4} -p^{2} \sigma p^{2}) |f_{\xi, k}|^{2} \langle n_{k} \rangle \frac{4( t^{2} - \tau t )}{m^{2} } e^{- i \Omega_{k} \tau } + h.c. \Big]
,
\end{align}
The integral over $\tau$ can be found using
\begin{align}
\int_{0}^{\infty} e^{-i \Omega_{k} \tau} \text{d} \tau &= \frac{-i}{\Omega_{k}} + \pi\delta(\Omega_{k}), \nonumber
 \\
\int_{0}^{\infty} \tau e^{-i \Omega_{k} \tau} \text{d} \tau &= -\frac{1}{\Omega_{k}^{2}} + i \pi \frac{\text{d}}{\text{d}\Omega_{k}} \delta (\Omega_{k})\
\end{align}
and defining the following useful constants
\begin{align}
    \Delta_{1} &= \sum_{k} \frac{|f_{\xi, k}|^{2}}{\Omega_{k} \hbar^{2}}, \\
\Delta_{1}' &= \sum_{k} \frac{|f_{\xi, k}|^{2}}{\Omega_{k} \hbar^{2}} \langle n_{k} \rangle, \\
\Gamma_{1} &= \sum_{k} \frac{|f_{\xi, k}|^{2}}{ \hbar^{2}} \pi \delta(\Omega_{k}), \\
\Gamma'_{1} &= \sum_{k} \frac{|f_{\xi, k}|^{2}}{ \hbar^{2}} \pi \delta(\Omega_{k})\langle n_{k} \rangle, \\
\Delta_{2} &= \sum_{k} \frac{2|f_{\xi, k}|^{2}}{ \mu\hbar^{2} \Omega_{k}^{2}}, \\
\Delta'_{2} &= \sum_{k} \frac{2|f_{\xi, k}|^{2}}{ \mu\hbar^{2} \Omega_{k}^{2}} \langle n_{k} \rangle, \\
\Gamma_{2} &= \frac{i 2 \pi |f_{\xi, k}|^{2}}{\mu \hbar^{2}} \frac{\text{d}}{\text{d}\Omega_{k}} \delta(\Omega_{k}), \\
\Gamma'_{2} &= \frac{i 2 \pi |f_{\xi, k}|^{2}}{\mu \hbar^{2}} \frac{\text{d}}{\text{d}\Omega_{k}} \delta(\Omega_{k}) \langle n_{k} \rangle
.
\end{align}
noting that $\Gamma_{1} = \Gamma'_{1} = \Gamma_{2} = \Gamma'_{2} = 0$. 
Moving back into the Schrodinger picture
\begin{align}
\frac{\text{d}\sigma}{\text{d}t} &= \frac{i}{\hbar} \left[\sigma, H_{\textsc{p}} \right] \nonumber \\
&- \left( (\xi p + p \xi) \sigma (\xi p + p \xi) - (\xi p + p \xi)^{2} \sigma\right) \left( \Delta_{1}' + \Delta_{1} \right) \nonumber \\
&- \left(\sigma (\xi p + p \xi)^{2} - (\xi p + p \xi) \sigma (\xi p + p \xi)   \right) \Delta_{1}' \nonumber \\
&- \left( \xi p^{3} \sigma + p \xi p^{2} \sigma - p^{2} \sigma \xi p - p^{2} \sigma p \xi \right) \left( \Delta_{2} + \Delta_{2}' \right) \nonumber \\ 
&- \left( \sigma p^{2} \xi p + \sigma p^{3} \xi - \xi p \sigma p^{2} - p \xi \sigma p^{2}  \right) \Delta_{2}' + h.c.
\end{align}
\\
We aim to study the effect of the gravitational vacuum state on our particle. We therefore set $\langle n_{k} \rangle = 0$ and therefore $\Delta'_{1} = \Delta'_{2} = 0$
\begin{align}
\frac{\text{d}\sigma}{\text{d}t} &= \frac{i}{\hbar} \left[\sigma, H_{P} \right] \nonumber \\
&- \Delta_{1} \left( (\xi p + p \xi) \sigma (\xi p + p \xi) - (\xi p + p \xi)^{2} \sigma\right) \nonumber \\
&- \Delta_{2} \left( \xi p^{3} \sigma + p \xi p^{2} \sigma - p^{2} \sigma \xi p - p^{2} \sigma p \xi \right) \nonumber \\ 
&+ h.c.
\end{align}
Evaluating the constants
\begin{align}
    \Delta_{1} &= \frac{G}{8 \pi^{2} \hbar c^{2}} \int \text{d}^{3}\mathbf{k} \; \sin^{4}{(\theta_{\mathbf{k}})} \cos^{2}{(\phi_\mathbf{k})} \nonumber \\  
    &= \frac{G}{8 \pi^{2} \hbar c^{5}} \int_{0}^{\Omega_{\textsc{max}}} \text{d} \Omega_{k} \; \Omega^{2}_{k} \int_{0}^{\pi} \text{d} \theta \; \sin^{5}{(\theta_{\mathbf{k}})}\int_{0}^{2 \pi} \text{d} \phi \; \cos^{2}{(\phi_\mathbf{k})} \nonumber \\
    &= \frac{2 G }{45 \pi c^{5} \hbar} \Omega_{\text{max}}^{3}, \\
    \Delta_{2} &= \frac{G}{4 \pi^{2} \hbar c^{2} m} \int \text{d}^{3}\mathbf{k} \; \frac{1}{\Omega_{\mathbf{k}}} \sin^{4}{(\theta_{\mathbf{k}})} \cos^{2}{(\phi_\mathbf{k})} \nonumber \\  
    &= \frac{G}{4 \pi^{2} \hbar c^{5} m} \int_{0}^{\Omega_{\textsc{max}}} \text{d} \Omega_{k} \; \Omega_{k} \int_{0}^{\pi} \text{d} \theta \; \sin^{5}{(\theta_{\mathbf{k}})}\int_{0}^{2 \pi} \text{d} \phi \; \cos^{2}{(\phi_\mathbf{k})} \nonumber \\
    &= \frac{2G}{15 \pi \hbar c^{5} \mu} \Omega_{\text{max}}^{2}.
\end{align}
Comparing the magnitude of terms in the second line $\Delta_{1} p^{2} \xi^{2} \sigma$ with those in the first line $p^{2} \sigma / 2 \mu \hbar$
\begin{align}
    2 \mu \hbar \Delta_{1} \xi^{2} = \frac{\mu \xi^{2}}{90 \pi^{2} \rho_{\textsc{f}}} k_{\text{max}}^{3} = \frac{\mu \xi^{2}}{90 V \pi^{2} \rho_{\textsc{f}}} \sum_{k} 1.
\end{align}
Noting that the magnitude of the ratio of these two terms is smaller than the term neglected in assuming $A(\xi) = 1$, the second line is neglected from the master equation.
\begin{align}
\frac{\text{d}\sigma}{\text{d}t} = \frac{i}{\hbar} \left[\sigma, H_{P} \right] - 2 \Delta_{\xi} \left[p \xi , \left[p^{2} , \sigma \right] \right],
\end{align}
where $\Delta_{2}$ has been relabelled $\Delta_{\xi}$,
or alternatively
\begin{align}
    \label{eq:fpmastereq}
    \frac{\text{d}\sigma}{\text{d}t} &= -\frac{i}{\hbar} \left[ \hat{H}_{P}, \hat{\sigma} \right] \left(1 - 4 \mu \hbar^{2} \Delta_{\xi} \right) \nonumber \\
    &- \Delta_{\xi} \left( \hat{\xi} \hat{p}^{3}_{\xi} \sigma - \hat{\xi} \hat{p}_{\xi} \sigma \hat{p}^{2}_{\xi} - \hat{p}^{2}_{\xi} \sigma \hat{p}_{\xi} \hat{\xi} + \sigma \hat{p}^{3}_{\xi} \hat{\xi} \right).
\end{align}

From the master equation, the dynamics of expectation values can be found,
\begin{align}
    \frac{\text{d}}{\text{d}t} \langle \hat{p}_{\xi}^{a} \hat{\xi}^{b} \rangle &= \left( \frac{b}{\mu} + 2 \hbar \Delta_{\xi} b (b-a) \right) \langle \hat{p}_{\xi}^{a+1} \hat{\xi}^{b-1} \rangle + \left( \frac{i \hbar b (b-1) }{2\mu} + i \hbar^{3} \Delta_{\xi} b (b-1) (b-a) \right) \langle \hat{p}_{\xi}^{a} \hat{\xi}^{b-2} \rangle,
\end{align}
where $a$, $b$ are positive integers.

The mean and variance of $\hat{\xi}$ and $\hat{p}_{\xi}$ can then be calculated
\begin{align}
    \label{eq:eomxi}
    \langle \hat{\xi} \rangle (t) &= \left( \frac{1}{\mu} + 2 \hbar \Delta_{\xi} \right) \langle \hat{p}_{x} \rangle (0), \\
    \langle \hat{\xi}^{2} \rangle (t) &= \left(\frac{1}{\mu^{2}} + \frac{4 \hbar \Delta_{\xi}^{2}}{\mu} + 16 \hbar^{4} \Delta_{\xi}^{2} \right) \langle p^{2}_{\xi} \rangle (0) t^{2} + \langle \hat{\xi} \hat{p}_{\xi} + \hat{p}_{\xi} \hat{\xi} \rangle (0) t + \langle \hat{\xi}^{2} \rangle (0) \\
    \langle \hat{p}_{\xi}^{n} \rangle (t) &= \langle \hat{p}_{\xi}^{n} \rangle (0),
\end{align}
for any positive integer $n$.

The variance can be rewritten in terms of the renormalised mass $\mu_{\xi}$
\begin{align}
    \langle \hat{\xi}^{2} \rangle (t) &= \frac{\langle \hat{p}_{\xi}^{2} \rangle (0)}{\mu_{\xi}^{2}} \left(1 - 4 \hbar^{4} \Delta_{\xi} \right) t^{2} + \frac{\langle \hat{\xi} \hat{p}_{\xi} + \hat{p}_{x} \hat{\xi} \rangle (0)}{\mu_{\xi}} \left(1 + 4 \hbar^{2} \Delta_{\xi} \right) t + \langle \hat{\xi}^{2} \rangle (0),
\end{align}
where
\begin{align}
    \mu_{\xi} = \frac{\mu}{\left( 1 + 2 \hbar \mu \Delta_{\xi} \right)}.
\end{align}

As before in the $\hat{x}$ variable, we must regularise the harmonic oscillator results by subtracting out the corresponding free particle energy shift. It can be seen in Eq.~(\ref{eq:fpmastereq}) that the total energy of the free particle is shifted by the amount $\Delta E_{\textsc{fp}}$ where
\begin{align}
    \Delta E_{\textsc{fp}} = 4 \mu \hbar^{2} \Delta_{\xi} = \frac{8}{15 \pi} t_{\textsc{p}}^{2} \Omega_{\textsc{max}}^{2}.
\end{align}
Again restating and rewriting the master equation for the harmonic oscillator,
\begin{align}
\frac{\text{d} \sigma}{\text{d} t} &= - \frac{i}{\hbar} \left(1 - \frac{\delta_{-} -  3\delta_{+}}{\omega} \right)  \left[ \hat{H}_{\textsc{p}}, \sigma \right] \nonumber \\
&+ i \left( \delta_{+} -  \delta_{-} \right) \left[ \left( \hat{b}^{\dagger} b \right)^{2}, \sigma  \right] \nonumber \\
&+ \Gamma \left( b^{2} \sigma b^{\dagger 2} - \frac{1}{2} b^{\dagger 2} b^{2} \sigma - \frac{1}{2} \sigma b^{\dagger 2} b^{2} \right) \nonumber \\
&-  i \delta_{+} \left( b^{\dagger 2} \sigma b^{\dagger 2} - b^{\dagger 4} \sigma + \sigma b^{4} - b^{2} \sigma b^{2} \right) \nonumber \\
&-  \left(i \delta_{-} + \frac{\Gamma}{2} \right)\left( b^{4} \sigma - b^{2} \sigma b^{2} \right) \nonumber \\
&-  \left(i \delta_{-} - \frac{\Gamma}{2} \right) \left( b^{\dagger 2} \sigma b^{\dagger 2} - \sigma b^{\dagger 4} \right).
\end{align}
The first line corresponds to shift in the energy $\Delta E$,
\begin{align}
    \Delta E = -\frac{\Delta_{-} -  3\Delta_{+}}{\omega}.
\end{align}
Considering only the term in $\Delta_{\pm}$ that depends on the square of the $\Omega_{\mathbf{k}}$
\begin{align}
    \Delta E_{\text{quad}} = \frac{4 G \hbar}{15 c^{5}}
\end{align}
Again we note that
\begin{align}
   \frac{\Delta E_{\textsc{fp}}}{2} =  \Delta E_{\text{quad}}.
\end{align}
It is therefore the term in $\Delta_{\pm}$ that is quadratic in the cut-off $\Omega_{\mathbf{k}}$ that must be subtracted out when renormalising the constants. This gives the renormalised constants,
\begin{align}
    \label{eq:deltarenormxi}
    \Delta_{\pm}^{(\textsc{r})} = \frac{\Gamma}{2 \pi} \left(- \ln{\left| \Omega_{\textsc{max}} \pm 2 \omega  \right|} \pm \frac{\Omega_{\textsc{max}}}{2 \omega} \pm \frac{\Omega^{3}_{\textsc{max}}}{24 \omega^{3}} \right).
\end{align}

\section{\label{app:harmosc} Deriving the master equation for harmonic oscillator potential}

\subsection{\label{app:hox} Coordinate Separation}

We now follow the same approach outlined in Appendix \ref{app:freeparticle} to derive the master equation of a particle confined to a harmonic oscillator potential under the influence of the vacuum state of the gravitational field. We can again partition the Hamiltonian into particle, interaction and field terms
\begin{align}
    \label{eq:Hamp3}
    H_{\textsc{p}} &= \frac{p_{x}^{2}}{2\mu} + \frac{m \omega^{2} x^{2}}{2}, \\
    \label{eq:Hami3}
    H_{\textsc{i}} &= -\left( \frac{ p^{2}_{x}}{2 \mu} - \frac{\mu \omega^{2} x^{2}}{2} \right) \frac{1}{V} \sum_{\mathbf{k}} q_{\mathbf{k}}, \\
    \label{eq:Hamf3}
    H_{\textsc{f}} &= \sum_{\mathbf{k}} \left( \frac{V p_{\mathbf{k}}^{2}}{2 \rho_{\textsc{f}}} + \frac{\rho_{\textsc{f}}}{2 V} \Omega_{\mathbf{k}}^{2} q_{\mathbf{k}}^{2} \right).
\end{align}
The interaction and field Hamiltonians are the same as those defined for the free particle in \ref{app:freex}. The particle Hamiltonian now includes the harmonic potential term.
The dynamical variables are then promoted to operators and the following ladder operators defined
\begin{align}
       \hat{a}_{\mathbf{k}} &= \sqrt{\frac{\rho_{\textsc{f}} \Omega_{\mathbf{k}}}{2 V \hbar }} \left( {\hat{q}}_{\mathbf{k}} + \frac{i V}{\rho_{\textsc{f}} \Omega_{\mathbf{k}}}  {p}_{\mathbf{k}} \right), \\
     {\hat{b}}_{x} &= \sqrt{\frac{\mu \omega}{2 \hbar }} \left(  x + \frac{i}{\mu \omega} {\hat{p}_{x}} \right).
\end{align}
where the respective commutation relations are imposed
\begin{align}
    \left[  \hat{a}_{\mathbf{k}},  {\hat{a}}^{\dagger}_{\mathbf{k}'} \right] &= \delta_{\mathbf{k}, \mathbf{k}'}, \\
    \left[  {\hat{b}}_{x},  {\hat{b}}_{x}^{\dagger} \right] &= 1.
\end{align}
The interaction Hamiltonian can then be written in terms of quantum operators,
\begin{equation}
    \hat{H}_{\textsc{i}, x} = \left(\hat{b}_{x}^{2} + (\hat{b}_{x}^{\dagger})^{2} \right) \sum_{\mathbf{k}} g_{x, \mathbf{k}} \left( \hat{a}_{\mathbf{k}} + \hat{a}^{\dagger}_{\mathbf{k}} \right),
\end{equation}
where
\begin{align}
    \label{eq:gkapp}
    g_{x, \mathbf{k}} = \sqrt{\frac{ 4 \pi G \hbar^{3} \omega^{2} }{V c^{2} \Omega_{\mathbf{k}}}} \sin^{2}{(\theta_{\mathbf{k}})}\cos{(2 \phi_{\mathbf{k}})},
\end{align}
and $\theta$ is the angle between the polarization of the wave and the alignment of the one-dimensional oscillator and a factor of 2 has been included to account for the two independent polarisation modes.

We again want to study the dynamics of the particle subsystem using its time-evolved reduced density matrix $ {\sigma}(t) = \text{Tr}_{\textsc{f}}[  {\rho}(t) ]$ where $ {\rho}(t)$ denotes the density matrix of the full particle-field system. The density matrix corresponding to the field $ {\sigma}_{\textsc{f}} (t) = \text{Tr}_{\textsc{p}}[  {\rho}(t) ]$. We work in the interaction picture denoted by the tilde on operators
\begin{align}
    \widetilde{b}_{x}(t) &= \hat{b}_{x} (0) e^{-i \omega t}, \\
    \widetilde{a}_{\mathbf{k}}(t) &= \hat{a}_{\mathbf{k}}(0) e^{-i \Omega_{\mathbf{k}} t}.
\end{align}
obtained using
\begin{align}
    \widetilde{\mathcal{X}}(t) = e^{\frac{i}{\hbar}\left(H_{\textsc{p}} + H_{\textsc{f}} \right)t} \hat{\mathcal{X}} (0) e^{-\frac{i}{\hbar}\left(H_{\textsc{p}} + H_{\textsc{f}} \right)t},
\end{align}

for any operator $\hat{\mathcal{X}}$.
Using the assumptions (1) - (3), the Liouville equation to second-order in the interaction strength is presented\cite{Cohen}, 
\begin{align}
    \label{eq:evoeq1b}
    \frac{\Delta \widetilde{\sigma}(t)}{\Delta t} = -\frac{1}{\hbar^{2}} \frac{1}{\Delta t} \int^{\infty}_{0} d(t_{1} - t_{2}) \int_{t}^{t + \Delta t} d t_{1} \text{Tr}_{\textsc{f}} [ \widetilde{H}_{\textsc{i}} ( t_{1} ), [ \widetilde{H}_{\textsc{i}} ( t_{2} ) , \widetilde{\sigma} (t) \otimes \widetilde{\sigma}_{\textsc{f}} ]].
\end{align}
This equation can be expressed in terms of the defined quantity $R$
\begin{align}
\nonumber \frac{\Delta \sigma}{\Delta t} &= -\frac{1}{h^{2}} \frac{1}{\Delta t} \int^{\infty}_{0} d(t_{1} - t_{2}) \int_{t}^{t + \Delta t} d t_{1}\\ &\text{Tr}_{\textsc{f}} [ \left(b^{2}(t_{1}) + (b^{\dagger})^{2}(t_{1}) \right) R ( t_{1} ), [ \left((b^{\dagger})^{2}(t_{2}) + b^{2}(t_{2}) \right) R^{\dagger} ( t_{2} ) , \sigma (t) \otimes \sigma_{\textsc{f}} ]] + \text{h.c},
\end{align}
where
\begin{align}
    R(t) &= -\sum_{k} g_{x, k} a_{k}(t),
\end{align}
and all operators are in the interaction picture but the tilde has been dropped for convenience.
Evaluating the first commutator.
\begin{equation}
    \label{eq:comm1b}
    \frac{\Delta \sigma}{\Delta t} = -\frac{1}{h^{2}} \frac{1}{\Delta t} \int^{\infty}_{0} d(t_{1} - t_{2}) \int_{t}^{t + \Delta t} d t_{1} \left(\text{Tr}_{\textsc{f}} \left[ \mathcal{O}_{3}, \mathcal{O}_{4} \right] \right)
\end{equation}
where
\begin{align}
\mathcal{O}_{3} &= \left(b_{x}^{2}(t_{1}) + (b_{x}^{\dagger})^{2}(t_{1}) \right) R(t_{1}), \\
\mathcal{O}_{4} &= \left((b_{x}^{\dagger})^{2}(t_{2}) + b_{x}^{2}(t_{2}) \right) \sigma(t) \otimes R^{\dagger}(t_{2}) \sigma_{\textsc{f}} - \sigma(t) \left( b_{x}^{2}(t_{2}) + (b_{x}^{\dagger})^{2}(t_{2}) \right) \otimes \sigma_{\textsc{f}} R^{\dagger}(t_{2}).
\end{align} 
Evaluating the second commutator
\begin{align}
\frac{\Delta \widetilde{\sigma}}{\Delta t} = &-\frac{1}{\hbar^{2}} \frac{1}{\Delta t} \int^{\infty}_{0} d(t_{1} - t_{2}) \int_{t}^{t + \Delta t} d t_{1} \nonumber \\ 
& \bigg(  \left[ (\widetilde{b}_{x}^{\dagger 2} \widetilde{b}_{x}^{2} \widetilde{\sigma} - \widetilde{b}_{x}^{2} \widetilde{\sigma} \widetilde{b}_{x}^{\dagger 2} ) e^{ 2 i \omega ( t_{1} - t_{2})} \right] \nonumber \\
+ &  \left[ ( \widetilde{b}_{x}^{2} \widetilde{b}_{x}^{\dagger 2} \widetilde{\sigma} - \widetilde{b}_{x}^{\dagger 2} \widetilde{\sigma} \widetilde{b}_{x}^{2} ) e^{ -2 i \omega ( t_{1} - t_{2})}  \right] \nonumber \\
- &  \left[ ( \widetilde{b}_{x}^{\dagger 2} \widetilde{\sigma} \widetilde{b}^{\dagger 2} - \widetilde{b}_{x}^{\dagger 4} \widetilde{\sigma} ) e^{ 2 i \omega ( t_{1} + t_{2})} \right] \nonumber \\
- &  \left[ ( \widetilde{b}^{2} \widetilde{\sigma} \widetilde{b}_{x}^{2} - \widetilde{b}_{x}^{4} \widetilde{\sigma}  ) e^{- 2 i \omega ( t_{1} + t_{2})}  \right] \bigg) \langle R ( t_{1} ) R^{\dagger} ( t_{2} ) \rangle_{\textsc{f}} \nonumber \\
+ & \bigg(  \left[ ( \widetilde{\sigma} \widetilde{b}_{x}^{2} \widetilde{b}^{\dagger 2} - \widetilde{b}_{x}^{\dagger 2} \widetilde{\sigma} \widetilde{b}_{x}^{2} ) e^{ 2 i \omega ( t_{1} - t_{2})}  \right] \nonumber \\
+ &  \left[ ( \widetilde{\sigma} \widetilde{b}_{x}^{\dagger 2} \widetilde{b}_{x}^{2} - \widetilde{b}_{x}^{2} \widetilde{\sigma} \widetilde{b}_{x}^{\dagger 2} ) e^{-2 i \omega ( t_{1} - t_{2})}  \right] \nonumber \\
- &  \left[ ( \widetilde{b}_{x}^{2} \widetilde{\sigma} \widetilde{b}_{x}^{2} - \widetilde{\sigma} \widetilde{b}_{x}^{4} ) e^{- 2 i \omega ( t_{1} + t_{2})} \right] \nonumber \\
- &  \left[ ( \widetilde{b}_{x}^{\dagger 2} \widetilde{\sigma} \widetilde{b}_{x}^{\dagger 2} - \widetilde{\sigma} \widetilde{b}_{x}^{\dagger 4}  ) e^{ 2 i \omega ( t_{1} + t_{2})}  \right] \bigg)  \langle R^{\dagger} ( t_{2} ) R ( t_{1} ) \rangle_{\textsc{f}} \nonumber \\
+ & \text{h.c.}
\end{align}
The partial trace over the field gives
\begin{align}
\langle R (t_{1}) R^{\dagger} (t_{2}) \rangle_{\textsc{f}} &= \sum_{\mathbf{k}} |g_{x, \mathbf{k}}|^{2} ( \langle n_{\mathbf{k}} \rangle + 1 ) e^{ - i \Omega_{\mathbf{k}} ( t_{1} - t_{2} ) }, \\
\langle R^{\dagger} (t_{2}) R (t_{1}) \rangle_{\textsc{f}} &= \sum_{\mathbf{k}} |g_{x, \mathbf{k}}|^{2} \langle n_{\mathbf{k}} \rangle e^{ - i \Omega_{\mathbf{k}} ( t_{1} - t_{2} ) }.
\end{align}
We then have
\begin{align}
\frac{\Delta \widetilde{\sigma}}{\Delta t} = -\frac{1}{\hbar^{2}} \frac{1}{\Delta t} \sum_{k} |g_{\xi, k}|^{2}\int^{\infty}_{0} d(t_{1} - t_{2}) \int_{t}^{t + \Delta t} d t_{1} & \bigg(  \left[ (\widetilde{b}_{x}^{\dagger 2} \widetilde{b}_{x}^{2} \widetilde{\sigma} - \widetilde{b}_{x}^{2} \widetilde{\sigma} \widetilde{b}_{x}^{\dagger 2} ) e^{ 2 i \omega ( t_{1} - t_{2})} \right] \nonumber \\
+ &  \left[ ( \widetilde{b}_{x}^{2} \widetilde{b}_{x}^{\dagger 2} \widetilde{\sigma} - \widetilde{b}_{x}^{\dagger 2} \widetilde{\sigma} \widetilde{b}_{x}^{2} ) e^{ -2 i \omega ( t_{1} - t_{2})}  \right] \nonumber \\
- &  \left[ ( \widetilde{b}_{x}^{\dagger 2} \widetilde{\sigma} \widetilde{b}_{x}^{\dagger 2} - \widetilde{b}_{x}^{\dagger 4} \widetilde{\sigma} ) e^{ 2 i \omega ( t_{1} + t_{2})} \right] \nonumber \\
- &  \left[ ( \widetilde{b}_{x}^{2} \widetilde{\sigma} \widetilde{b}_{x}^{2} - \widetilde{b}_{x}^{4} \widetilde{\sigma}  ) e^{- 2 i \omega ( t_{1} + t_{2})}  \right] \bigg) ( \langle n_{k} \rangle + 1 ) \nonumber \\
+ & \bigg(  \left[ ( \widetilde{\sigma} \widetilde{b}_{x}^{2} \widetilde{b}_{x}^{\dagger 2} - \widetilde{b}_{x}^{\dagger 2} \widetilde{\sigma} \widetilde{b}_{x}^{2} ) e^{ 2 i \omega ( t_{1} - t_{2})}  \right] \nonumber \\
+ &  \left[ ( \widetilde{\sigma} \widetilde{b}_{x}^{\dagger 2} \widetilde{b}_{x}^{2} - \widetilde{b}_{x}^{2} \widetilde{\sigma} \widetilde{b}_{x}^{\dagger 2} ) e^{-2 i \omega ( t_{1} - t_{2})}  \right] \nonumber \\
- &  \left[ ( \widetilde{b}_{x}^{2} \widetilde{\sigma} \widetilde{b}_{x}^{2} - \widetilde{\sigma} \widetilde{b}_{x}^{4} ) e^{- 2 i \omega ( t_{1} + t_{2})} \right] \nonumber \\
- &  \left[ ( \widetilde{b}_{x}^{\dagger 2} \widetilde{\sigma} \widetilde{b}_{x}^{\dagger 2} - \widetilde{\sigma} \widetilde{b}_{x}^{\dagger 4}  ) e^{ 2 i \omega ( t_{1} + t_{2})}  \right] \bigg)  \langle n_{k} \rangle \nonumber \\
+ & \text{h.c.}
\end{align}
Evaluating the integral over $d t_{1}$
\begin{align}
\frac{\Delta \widetilde{\sigma}}{\Delta t} = -\frac{1}{\hbar^{2}} \frac{1}{\Delta t} \sum_{k} |g_{\xi, k}|^{2} \int^{\infty}_{0} d\tau & \bigg(  \left[ (\widetilde{b}_{x}^{\dagger 2} \widetilde{b}_{x}^{2} \widetilde{\sigma} - \widetilde{b}_{x}^{2} \widetilde{\sigma} \widetilde{b}_{x}^{\dagger 2} ) e^{ -i \tau (\Omega_{k} - 2 \omega )} \right] \nonumber \\
+ &  \left[ ( \widetilde{b_{x}}^{2} \widetilde{b}_{x}^{\dagger 2} \widetilde{\sigma} - \widetilde{b}_{x}^{\dagger 2} \widetilde{\sigma} \widetilde{b}_{x}^{2} ) e^{ -i \tau (\Omega_{k} + 2 \omega )}  \right] \nonumber \\
- &  \zeta_{1} \left[ ( \widetilde{b}_{x}^{\dagger 2} \widetilde{\sigma} \widetilde{b}_{x}^{\dagger 2} - \widetilde{b}_{x}^{\dagger 4} \widetilde{\sigma} ) e^{ -i \tau (\Omega_{k} + 2 \omega )} \right] \nonumber \\
- &  \zeta_{2} \left[ ( \widetilde{b}_{x}^{2} \widetilde{\sigma} \widetilde{b}_{x}^{2} - \widetilde{b}_{x}^{4} \widetilde{\sigma} ) e^{-i \tau (\Omega_{k} - 2 \omega )}  \right] \bigg) ( \langle n_{k} \rangle + 1 ) \nonumber \\
+ & \bigg(  \left[ ( \widetilde{\sigma} \widetilde{b}^{2} \widetilde{b}_{x}^{\dagger 2} - \widetilde{b}_{x}^{\dagger 2} \widetilde{\sigma} \widetilde{b}_{x}^{2} ) e^{ -i \tau (\Omega_{k} - 2 \omega )}  \right] \nonumber \\
+ &   \left[ ( \widetilde{\sigma} \widetilde{b}_{x}^{\dagger 2} \widetilde{b}_{x}^{2} - \widetilde{b}^{2} \widetilde{\sigma} \widetilde{b}_{x}^{\dagger 2} ) e^{-i \tau (\Omega_{k} + 2 \omega )}  \right] \nonumber \\
- &  \zeta_{2} \left[ ( \widetilde{b}_{x}^{2} \widetilde{\sigma} \widetilde{b}_{x}^{2} - \widetilde{\sigma} \widetilde{b}_{x}^{4} ) e^{-i \tau (\Omega_{k} - 2 \omega )} \right] \nonumber \\
- &  \zeta_{1} \left[ ( \widetilde{b}_{x}^{\dagger 2} \widetilde{\sigma} \widetilde{b}_{x}^{\dagger 2} - \widetilde{\sigma} \widetilde{b}_{x}^{\dagger 4}  ) e^{ -i \tau (\Omega_{k} + 2 \omega )}  \right] \bigg) \langle n_{k} \rangle \nonumber \\
+ & \text{h.c.}
\end{align}
where $\tau = t_{1} - t_{2}$, and the phase factors $\zeta_{1}$ and $\zeta_{2}$ are defined as
\begin{align}
    \zeta_{1} &= \frac{e^{i 4 \omega t}( e^{4 i \omega \Delta t} - 1 )}{4 i \omega \Delta t}, \\
    \zeta_{2} &= \frac{e^{- i 4 \omega t}( 1 - e^{- 4 i \omega \Delta t} )}{4 i \omega \Delta t}.
\end{align}
Further, when $\Delta t \rightarrow 0$, $\zeta_{1} \rightarrow e^{i 4 \omega t}$ and $\zeta_{2} \rightarrow e^{- i 4 \omega t}$ and the master equation becomes
\begin{align}
\frac{\text{d} \widetilde{\sigma}}{\text{d} t} = -\frac{1}{\hbar^{2}} \sum_{k} |g_{x, k}|^{2} \int^{\infty}_{0} d\tau & \bigg(  \left[ (\widetilde{b}_{x}^{\dagger 2} \widetilde{b}_{x}^{2} \widetilde{\sigma} - \widetilde{b}_{x}^{2} \widetilde{\sigma} \widetilde{b}_{x}^{\dagger 2} ) e^{ -i \tau (\Omega_{k} - 2 \omega )} \right] \nonumber \\
+ &  \left[ ( \widetilde{b}_{x}^{2} \widetilde{b}_{x}^{\dagger 2} \widetilde{\sigma} - \widetilde{b}_{x}^{\dagger 2} \widetilde{\sigma} \widetilde{b}_{x}^{2} ) e^{ -i \tau (\Omega_{k} + 2 \omega )}  \right] \nonumber \\
- &  \zeta_{1} \left[ ( \widetilde{b}_{x}^{\dagger 2} \widetilde{\sigma} \widetilde{b}_{x}^{\dagger 2} - \widetilde{b}_{x}^{\dagger 4} \widetilde{\sigma} ) e^{ -i \tau (\Omega_{k} + 2 \omega )} \right] \nonumber \\
- &  \zeta_{2} \left[ ( \widetilde{b}_{x}^{2} \widetilde{\sigma} \widetilde{b}_{x}^{2} - \widetilde{b}_{x}^{4} \widetilde{\sigma} ) e^{-i \tau (\Omega_{k} - 2 \omega )}  \right] \bigg) ( \langle n_{k} \rangle + 1 ) \nonumber \\
+ & \bigg(  \left[ ( \widetilde{\sigma} \widetilde{b}_{x}^{2} \widetilde{b}_{x}^{\dagger 2} - \widetilde{b}_{x}^{\dagger 2} \widetilde{\sigma} \widetilde{b}_{x}^{2} ) e^{ -i \tau (\Omega_{k} - 2 \omega )}  \right] \nonumber \\
+ &   \left[ ( \widetilde{\sigma} \widetilde{b}^{\dagger 2} \widetilde{b}_{x}^{2} - \widetilde{b}_{x}^{2} \widetilde{\sigma} \widetilde{b}_{x}^{\dagger 2} ) e^{-i \tau (\Omega_{k} + 2 \omega )}  \right] \nonumber \\
- &  \zeta_{2} \left[ ( \widetilde{b}_{x}^{2} \widetilde{\sigma} \widetilde{b}_{x}^{2} - \widetilde{\sigma} \widetilde{b}_{x}^{4} ) e^{-i \tau (\Omega_{k} - 2 \omega )} \right] \nonumber \\
- &  \zeta_{1} \left[ ( \widetilde{b}_{x}^{\dagger 2} \widetilde{\sigma} \widetilde{b}_{x}^{\dagger 2} - \widetilde{\sigma} \widetilde{b}_{x}^{\dagger 4}  ) e^{ -i \tau (\Omega_{k} + 2 \omega )}  \right] \bigg) \langle n_{k} \rangle \nonumber \\
+ & \text{h.c.}
\end{align}
The integral over $\tau$ can be found using
\begin{align}
\int_{0}^{\infty}d\tau e^{-i \tau (\Omega_{k} \pm 2 \omega )} &= - \frac{i}{\Omega_{k} \pm 2 \omega } + \pi \delta (\Omega_{k} \pm 2 \omega ),
\end{align}
and defining
\begin{subequations}
\begin{align}
\label{eq:constants1}
\Gamma_{\pm} &= \sum_{k} \frac{|g_{x, k}|^{2}}{\hbar^{2}} 2 \pi \delta \left( \Omega_{k} \pm 2 \omega \right), \\
\label{eq:constants2}
\Gamma_{\pm}^{'} &= \sum_{k} \frac{|g_{x, k}|^{2}}{\hbar^{2}} 2 \pi \langle n_{k} \rangle \delta \left( \Omega_{k} \pm 2 \omega \right), \\
\label{eq:constants3}
\delta_{\pm} &= \sum_{k} \frac{|g_{x, k}|^{2}}{\hbar^{2}} \frac{1}{2 \omega \pm \Omega_{k}}, \\
\label{eq:constants4}
\delta_{\pm}^{'} &= \sum_{k} \frac{|g_{x, k}|^{2}}{\hbar^{2}} \frac{1}{2 \omega \pm \Omega_{k}} \langle n_{k} \rangle.
\end{align}
\label{eq:constset}
\end{subequations}

Noting that $\Gamma_{+} = \Gamma'_{+} = 0$, the subscript is dropped from $\Gamma_{-}$ such that $\Gamma_{-} = \Gamma$.

In order to evaluate the effective rates of damping, gain and detuning in the system, we need to evaluate the summation over $\mathbf{k}$ in (\ref{eq:constants1}) - (\ref{eq:constants4}). In a typical setup of our interest, the interaction with the reservoir is mainly due to vacuum fluctuations in an infinite range of field modes. So, it is more relevant to consider an integral over $\mathbf{k}$ in place of the summation, as indicated below 
\begin{equation}
\label{eq:ksum2}
\sum_{\mathbf{k}} g_{x, \mathbf{k}} \rightarrow \frac{V}{8 \pi^{3}}\int_{\mathbf{k}} g_{x, \mathbf{k}} d^{3} \mathbf{k}.
\end{equation}
Using this along with Eq. (\ref{eq:gkapp}) in Eq. (\ref{eq:constset}), we get, for instance,
\begin{align}
    \Gamma &= \frac{ G \hbar \omega^{2} }{ c^{5}} \int \text{d}^{3}\mathbf{k} \; \frac{1}{\Omega_{\mathbf{k}}} \; \delta \left(\Omega_{\mathbf{k}} - 2 \omega \right) \sin^{4}{(\theta_{\mathbf{k}})}\cos^{2}{(2 \phi_{\mathbf{k}})} , \nonumber \\
    &= \frac{ G \hbar \omega^{2} }{ c^{5}} \int_{0}^{\Omega_{\textsc{max}}} \text{d}\Omega_{k} \Omega_{k} \delta \left(\Omega_{\mathbf{k}} - 2 \omega \right) \int^{\pi}_{0} \text{d}\theta \sin^{5}{(\theta)} \int_{0}^{2 \pi} \text{d}\phi \cos^{2}{(2 \phi)}, \nonumber \\
    &= \frac{32}{15}\frac{G \hbar \omega^{3}}{c^{5}}.
\end{align}
This exactly matches the decay rate in \cite{Torovs2020}.
\begin{align}
    \delta_{\pm} &=  \frac{ G \hbar^{3} \omega^{2} }{2 \pi c^{5}}  \int \text{d}^{3}\mathbf{k} \frac{1}{\Omega_{\mathbf{k}}}\frac{1}{2 \omega \pm \Omega_{\mathbf{k}}} \sin^{4}{(\theta_{\mathbf{k}})}\cos^{2}{(2 \phi_{\mathbf{k}})}, \nonumber \\
    &=  \frac{ G \hbar^{3} \omega^{2} }{2 \pi c^{5}} \int_{0}^{\Omega_{\textsc{max}}} \text{d} \Omega_{k} \frac{\Omega_{k}}{2 \omega \pm \Omega_{\mathbf{k}}} \int^{\pi}_{0} \text{d}\theta \sin^{5}{(\theta)} \int_{0}^{2 \pi} \text{d}\phi \cos^{2}{(2 \phi)}, \nonumber \\
    &= \frac{\Gamma}{2 \pi \omega} \left(- \omega \ln{\left| \Omega_{\textsc{max}} \pm 2 \omega  \right|} \pm \frac{\Omega_{\textsc{max}}}{2} \right)
\end{align}

As we are interested in the influence of the vacuum state of the gravitational field on our particle, we set $\langle n_{\mathbf{k}} \rangle = 0$
\begin{align}
\frac{\text{d} \widetilde{\sigma}}{\text{d} t} = & (  \left[ (\widetilde{b}_{x}^{2} \widetilde{\sigma} \widetilde{b}_{x}^{\dagger 2} - \widetilde{b}_{x}^{\dagger 2} \widetilde{b}_{x}^{2} \widetilde{\sigma}  ) \left( i \delta_{-} + \Gamma \right) \right] \nonumber \\
+ & \left[ ( \widetilde{b}_{x}^{\dagger 2} \widetilde{\sigma} \widetilde{b}_{x}^{2} - \widetilde{b}_{x}^{2} \widetilde{b}_{x}^{\dagger 2} \widetilde{\sigma}  ) \left( - i \delta_{+} \right)  \right] \nonumber \\
- &  \zeta_{1} \left[ (\widetilde{b}^{\dagger 4} \widetilde{\sigma} - \widetilde{b}_{x}^{\dagger 2} \widetilde{\sigma} \widetilde{b}_{x}^{\dagger 2}  ) \left( - i \delta_{+} \right) \right] \nonumber \\
- &  \zeta_{2} \left[ ( \widetilde{b}_{x}^{4} \widetilde{\sigma} - \widetilde{b}_{x}^{2} \widetilde{\sigma} \widetilde{b}_{x}^{2} ) \left( i \delta_{-} + \Gamma \right) )  \right] ) \nonumber \\
+ & h.c.
\end{align}

Moving back into the Schr{\"o}dinger picture
\begin{align}
\frac{\text{d} \sigma}{\text{d} t} = -&i \omega \hat{b}_{x}^{\dagger} \hat{b}_{x} \sigma \nonumber \\
+& ( \left[ (\hat{b}_{x}^{2} \sigma \hat{b}_{x}^{\dagger 2} - \hat{b}_{x}^{\dagger 2} \hat{b}_{x}^{2} \sigma  ) \left( i \delta_{-} + \Gamma \right) \right] \nonumber \\
+ &  \left[ ( \hat{b}_{x}^{\dagger 2} \sigma \hat{b}_{x}^{2} - \hat{b}_{x}^{2} \hat{b}_{x}^{\dagger 2} \sigma  ) \left( - i \delta_{+} \right)  \right] \nonumber \\
- &  \left[ (\hat{b}_{x}^{\dagger 4} \sigma - \hat{b}_{x}^{\dagger 2} \sigma \hat{b}_{x}^{\dagger 2}  ) \left( - i \delta_{+} \right) \right] \nonumber \\
- &  \left[ ( \hat{b}_{x}^{4} \sigma - \hat{b}_{x}^{2} \sigma \hat{b}_{x}^{2} ) \left( i \delta_{-} + \Gamma \right) )  \right] ) \nonumber \\
+ & h.c.
\end{align}

Including the Hermitian conjugate and rearranging, the final master equation is then
\begin{align}
\label{eq:mastereqd}
\frac{\text{d} \sigma}{\text{d} t} &= -i \left(\omega - \delta_{-} -  3\delta_{+} \right)  \left[ \hat{b}_{x}^{\dagger} \hat{b}_{x}, \sigma \right] \nonumber \\
&+ i \left( \delta_{+} -  \delta_{-} \right) \left[ \left( \hat{b}_{x}^{\dagger} \hat{b}_{x} \right)^{2}, \sigma  \right] \nonumber \\
&+ \Gamma \left( \hat{b}_{x}^{2} \sigma \hat{b}_{x}^{\dagger 2} - \frac{1}{2} \hat{b}_{x}^{\dagger 2} \hat{b}_{x}^{2} \sigma - \frac{1}{2} \sigma \hat{b}_{x}^{\dagger 2} \hat{b}_{x}^{2} \right) \nonumber \\
&-  i \delta_{+} \left( \hat{b}_{x}^{\dagger 2} \sigma \hat{b}_{x}^{\dagger 2} - \hat{b}_{x}^{\dagger 4} \sigma + \sigma \hat{b}_{x}^{4} - \hat{b}_{x}^{2} \sigma \hat{b}_{x}^{2} \right) \nonumber \\
&-  \left(i \delta_{-} + \frac{\Gamma}{2} \right)\left( \hat{b}_{x}^{4} \sigma - \hat{b}_{x}^{2} \sigma \hat{b}_{x}^{2} \right) \nonumber \\
&-  \left(i \delta_{-} - \frac{\Gamma}{2} \right) \left( \hat{b}_{x}^{\dagger 2} \sigma \hat{b}_{x}^{\dagger 2} - \sigma \hat{b}_{x}^{\dagger 4} \right).
\end{align}
In the limit of the rotating-wave approximation,
\begin{align}
\label{eq:mastereqdrwa}
\frac{\text{d} \sigma}{\text{d} t} &= -i \left(\omega - \delta_{-}  \right)  \left[ \hat{b}_{x}^{\dagger} \hat{b}_{x}, \sigma \right] - i \delta_{-} \left[ \left( \hat{b}_{x}^{\dagger} \hat{b}_{x} \right)^{2}, \sigma  \right] \nonumber \\
&+ \Gamma \left( \hat{b}_{x}^{2} \sigma \hat{b}_{x}^{\dagger 2} - \frac{1}{2} \hat{b}_{x}^{\dagger 2} \hat{b}_{x}^{2} \sigma - \frac{1}{2} \sigma \hat{b}_{x}^{\dagger 2} \hat{b}_{x}^{2} \right).
\end{align}

\subsection{\label{app:hoxi} Geodesic Separation}

We now follow the same approach outlined in Appendix \ref{app:freeparticle} to derive the master equation of a particle confined to a harmonic oscillator potential under the influence of the vacuum state of the gravitational field. We can again partition the Hamiltonian into particle, interaction and field terms
\begin{align}
    \label{eq:Hamp3b}
    H_{\textsc{p}, \xi} &= \frac{p_{\xi}^{2}}{2m} + \frac{m \omega^{2} \xi^{2}}{2}, \\
    \label{eq:Hami3b}
    H_{\textsc{i}, \xi} &= \frac{ \xi p_{\xi} + p_{\xi} \xi}{4 \rho_{\textsc{f}}} \sum_{\mathbf{k}} p_{\mathbf{k}}, \\
    \label{eq:Hamf3b}
    H_{\textsc{f}, \xi} &= \sum_{\mathbf{k}} \left( \frac{V p_{\mathbf{k}}^{2}}{2 \rho_{\textsc{f}}} + \frac{\rho_{\textsc{f}}}{2 V} \Omega_{\mathbf{k}}^{2} q_{\mathbf{k}}^{2} \right).
\end{align}
The interaction and field Hamiltonians are the same as those defined for the free particle. The particle Hamiltonian now includes the harmonic potential term.
The dynamical variables are then promoted to operators and the following ladder operators defined
\begin{align}
       \hat{a}_{\mathbf{k}} &= \sqrt{\frac{\rho_{\textsc{f}} \Omega_{\mathbf{k}}}{2 V \hbar }} \left( \hat{q}_{\mathbf{k}} + \frac{i V}{\rho_{\textsc{f}} \Omega_{\mathbf{k}}}  \hat{p}_{\mathbf{k}} \right), \\
     \hat{b}_{\xi} &= \sqrt{\frac{m \omega}{2 \hbar }} \left(  \hat{\xi} + \frac{i}{m \omega} \hat{p}_{\xi} \right).
\end{align}
where the respective commutation relations are imposed
\begin{align}
    \left[  \hat{a}_{\mathbf{k}},  \hat{a}^{\dagger}_{\mathbf{k}'} \right] &= \delta_{\mathbf{k}, \mathbf{k}'}, \\
    \left[  \hat{b},  \hat{b}^{\dagger} \right] &= 1.
\end{align}
The interaction Hamiltonian can then be written in terms of quantum operators,
\begin{equation}
    \hat{H}_{\textsc{i}, \xi} = - \left(\hat{b}_{\xi}^{2} - (\hat{b}_{\xi}^{\dagger})^{2} \right) \sum_{\mathbf{k}} g_{\xi, \mathbf{k}} \left( \hat{a}_{\mathbf{k}} - \hat{a}^{\dagger}_{\mathbf{k}} \right),
\end{equation}
where
\begin{align}
    \label{eq:gkapp2}
    g_{\xi, \mathbf{k}} = \sqrt{\frac{ \pi G \hbar^{3} \Omega_{k} }{V c^{2}}} \sin^{2}{(\theta_{\mathbf{k}})}\cos{(2 \phi_{\mathbf{k}})},
\end{align}
and $\theta$ is the angle between the polarization of the wave and the alignment of the one-dimensional oscillator and a factor of 2 has been included to account for the two independent polarisation modes.

We again want to study the dynamics of the particle subsystem using its time-evolved reduced density matrix $ {\sigma}(t) = \text{Tr}_{\textsc{f}}[  {\rho}(t) ]$ where $ {\rho}(t)$ denotes the density matrix of the full particle-field system. The density matrix corresponding to the field $ {\sigma}_{\textsc{f}} (t) = \text{Tr}_{\textsc{p}}[  {\rho}(t) ]$. We work in the interaction picture denoted by the tilde on operators
\begin{align}
    \widetilde{b}_{\xi}(t) &= \hat{b}_{\xi}(0) e^{-i \omega t}, \\
    \widetilde{a}_{\mathbf{k}}(t) &= a_{\mathbf{k}}(0) e^{-i \Omega_{\mathbf{k}} t}.
\end{align}
obtained using
\begin{align}
    \widetilde{\mathcal{X}}(t) = e^{\frac{i}{\hbar}\left(H_{\textsc{p}} + H_{\textsc{f}} \right)t} \hat{\mathcal{X}} (0) e^{-\frac{i}{\hbar}\left(H_{\textsc{p}} + H_{\textsc{f}} \right)t},
\end{align}

for any operator $\hat{\mathcal{X}}$.
Using the assumptions (1) - (3), the Liouville equation to second-order in the interaction strength is presented\cite{Cohen}, 
\begin{align}
    \label{eq:evoeq2b}
    \frac{\Delta \widetilde{\sigma}(t)}{\Delta t} = -\frac{1}{\hbar^{2}} \frac{1}{\Delta t} \int^{\infty}_{0} d(t_{1} - t_{2}) \int_{t}^{t + \Delta t} d t_{1} \text{Tr}_{\textsc{f}} [ \widetilde{H}_{\textsc{i}} ( t_{1} ), [ \widetilde{H}_{\textsc{i}} ( t_{2} ) , \widetilde{\sigma} (t) \otimes \widetilde{\sigma}_{\textsc{f}} ]].
\end{align}
This equation can be expressed in terms of the defined quantity $R$
\begin{align}
\nonumber \frac{\Delta \sigma}{\Delta t} &= -\frac{1}{h^{2}} \frac{1}{\Delta t} \int^{\infty}_{0} d(t_{1} - t_{2}) \int_{t}^{t + \Delta t} d t_{1}\\ &\text{Tr}_{\textsc{f}} [ \left(b_{\xi}^{2}(t_{1}) - (b_{\xi}^{\dagger})^{2}(t_{1}) \right) R ( t_{1} ), [ \left((b_{\xi}^{\dagger})^{2}(t_{2}) - b_{\xi}^{2}(t_{2}) \right) R^{\dagger} ( t_{2} ) , \sigma (t) \otimes \sigma_{\textsc{f}} ]] + \text{h.c},
\end{align}
where
\begin{align}
    R(t) &= -\sum_{k} g_{\xi, k} a_{k}(t),
\end{align}
and all operators are in the interaction picture but the tilde has been dropped for convenience.
Evaluating the first commutator.
\begin{equation}
    \label{eq:comm2b}
    \frac{\Delta \sigma}{\Delta t} = -\frac{1}{h^{2}} \frac{1}{\Delta t} \int^{\infty}_{0} d(t_{1} - t_{2}) \int_{t}^{t + \Delta t} d t_{1} \left(\text{Tr}_{\textsc{f}} \left[ \mathcal{O}_{3}, \mathcal{O}_{4} \right] \right)
\end{equation}
where
\begin{align}
\mathcal{O}_{3} &= \left(b_{\xi}^{2}(t_{1}) - (b_{\xi}^{\dagger})^{2}(t_{1}) \right) R(t_{1}), \\
\mathcal{O}_{4} &= \left((b_{\xi}^{\dagger})^{2}(t_{2}) - b_{\xi}^{2}(t_{2}) \right) \sigma(t) \otimes R^{\dagger}(t_{2}) \sigma_{\textsc{f}} - \sigma(t) \left( b_{\xi}^{2}(t_{2}) - (b_{\xi}^{\dagger})^{2}(t_{2}) \right) \otimes \sigma_{\textsc{f}} R^{\dagger}(t_{2}).
\end{align} 
Evaluating the second commutator
\begin{align}
\frac{\Delta \widetilde{\sigma}}{\Delta t} = &-\frac{1}{\hbar^{2}} \frac{1}{\Delta t} \int^{\infty}_{0} d(t_{1} - t_{2}) \int_{t}^{t + \Delta t} d t_{1} \nonumber \\ 
& \bigg(  \left[ (\widetilde{b}_{\xi}^{\dagger 2} \widetilde{b}_{\xi}^{2} \widetilde{\sigma} - \widetilde{b}_{\xi}^{2} \widetilde{\sigma} \widetilde{b}_{\xi}^{\dagger 2} ) e^{ 2 i \omega ( t_{1} - t_{2})} \right] \nonumber \\
+ &  \left[ ( \widetilde{b}_{\xi}^{2} \widetilde{b}_{\xi}^{\dagger 2} \widetilde{\sigma} - \widetilde{b}_{\xi}^{\dagger 2} \widetilde{\sigma} \widetilde{b}_{\xi}^{2} ) e^{ -2 i \omega ( t_{1} - t_{2})}  \right] \nonumber \\
+ &  \left[ ( \widetilde{b}^{\dagger 2} \widetilde{\sigma} \widetilde{b}_{\xi}^{\dagger 2} - \widetilde{b}_{\xi}^{\dagger 4} \widetilde{\sigma} ) e^{ 2 i \omega ( t_{1} + t_{2})} \right] \nonumber \\
+ &  \left[ ( \widetilde{b}_{\xi}^{2} \widetilde{\sigma} \widetilde{b}_{\xi}^{2} - \widetilde{b}_{\xi}^{4} \widetilde{\sigma}  ) e^{- 2 i \omega ( t_{1} + t_{2})}  \right] \bigg) \langle R ( t_{1} ) R^{\dagger} ( t_{2} ) \rangle_{\textsc{f}} \nonumber \\
+ & \bigg(  \left[ ( \widetilde{\sigma} \widetilde{b}^{2} \widetilde{b}_{\xi}^{\dagger 2} - \widetilde{b}_{\xi}^{\dagger 2} \widetilde{\sigma} \widetilde{b}_{\xi}^{2} ) e^{ 2 i \omega ( t_{1} - t_{2})}  \right] \nonumber \\
+ &  \left[ ( \widetilde{\sigma} \widetilde{b}_{\xi}^{\dagger 2} \widetilde{b}_{\xi}^{2} - \widetilde{b}_{\xi}^{2} \widetilde{\sigma} \widetilde{b}_{\xi}^{\dagger 2} ) e^{-2 i \omega ( t_{1} - t_{2})}  \right] \nonumber \\
+ &  \left[ ( \widetilde{b}_{\xi}^{2} \widetilde{\sigma} \widetilde{b}_{\xi}^{2} - \widetilde{\sigma} \widetilde{b}_{\xi}^{4} ) e^{- 2 i \omega ( t_{1} + t_{2})} \right] \nonumber \\
+ &  \left[ ( \widetilde{b}_{\xi}^{\dagger 2} \widetilde{\sigma} \widetilde{b}_{\xi}^{\dagger 2} - \widetilde{\sigma} \widetilde{b}_{\xi}^{\dagger 4}  ) e^{ 2 i \omega ( t_{1} + t_{2})}  \right] \bigg)  \langle R^{\dagger} ( t_{2} ) R ( t_{1} ) \rangle_{\textsc{f}} \nonumber \\
+ & \text{h.c.}
\end{align}
The partial trace over the field gives
\begin{align}
\langle R (t_{1}) R^{\dagger} (t_{2}) \rangle_{\textsc{f}} &= \sum_{\mathbf{k}} |g_{\xi, \mathbf{k}}|^{2} ( \langle n_{\mathbf{k}} \rangle + 1 ) e^{ - i \Omega_{\mathbf{k}} ( t_{1} - t_{2} ) }, \\
\langle R^{\dagger} (t_{2}) R (t_{1}) \rangle_{\textsc{f}} &= \sum_{\mathbf{k}} |g_{\xi, \mathbf{k}}|^{2} \langle n_{\mathbf{k}} \rangle e^{ - i \Omega_{\mathbf{k}} ( t_{1} - t_{2} ) }.
\end{align}
We then have
\begin{align}
\frac{\Delta \widetilde{\sigma}}{\Delta t} = -\frac{1}{\hbar^{2}} \frac{1}{\Delta t} \sum_{k} |g_{\xi, k}|^{2} \int^{\infty}_{0} d(t_{1} - t_{2}) \int_{t}^{t + \Delta t} d t_{1} & \bigg(  \left[ (\widetilde{b}_{\xi}^{\dagger 2} \widetilde{b}_{\xi}^{2} \widetilde{\sigma} - \widetilde{b}_{\xi}^{2} \widetilde{\sigma} \widetilde{b}_{\xi}^{\dagger 2} ) e^{ 2 i \omega ( t_{1} - t_{2})} \right] \nonumber \\
+ &  \left[ ( \widetilde{b}_{\xi}^{2} \widetilde{b}_{\xi}^{\dagger 2} \widetilde{\sigma} - \widetilde{b}_{\xi}^{\dagger 2} \widetilde{\sigma} \widetilde{b}_{\xi}^{2} ) e^{ -2 i \omega ( t_{1} - t_{2})}  \right] \nonumber \\
+ &  \left[ ( \widetilde{b}^{\dagger 2} \widetilde{\sigma} \widetilde{b}_{\xi}^{\dagger 2} - \widetilde{b}_{\xi}^{\dagger 4} \widetilde{\sigma} ) e^{ 2 i \omega ( t_{1} + t_{2})} \right] \nonumber \\
+ &  \left[ ( \widetilde{b}_{\xi}^{2} \widetilde{\sigma} \widetilde{b}_{\xi}^{2} - \widetilde{b}_{\xi}^{4} \widetilde{\sigma}  ) e^{- 2 i \omega ( t_{1} + t_{2})}  \right] \bigg) ( \langle n_{k} \rangle + 1 ) \nonumber \\
+ & \bigg(  \left[ ( \widetilde{\sigma} \widetilde{b}_{\xi}^{2} \widetilde{b}_{\xi}^{\dagger 2} - \widetilde{b}_{\xi}^{\dagger 2} \widetilde{\sigma} \widetilde{b}_{\xi}^{2} ) e^{ 2 i \omega ( t_{1} - t_{2})}  \right] \nonumber \\
+ &  \left[ ( \widetilde{\sigma} \widetilde{b}_{\xi}^{\dagger 2} \widetilde{b}_{\xi}^{2} - \widetilde{b}_{\xi}^{2} \widetilde{\sigma} \widetilde{b}_{\xi}^{\dagger 2} ) e^{-2 i \omega ( t_{1} - t_{2})}  \right] \nonumber \\
+ &  \left[ ( \widetilde{b}_{\xi}^{2} \widetilde{\sigma} \widetilde{b}_{\xi}^{2} - \widetilde{\sigma} \widetilde{b}_{\xi}^{4} ) e^{- 2 i \omega ( t_{1} + t_{2})} \right] \nonumber \\
+ &  \left[ ( \widetilde{b}_{\xi}^{\dagger 2} \widetilde{\sigma} \widetilde{b}_{\xi}^{\dagger 2} - \widetilde{\sigma} \widetilde{b}_{\xi}^{\dagger 4}  ) e^{ 2 i \omega ( t_{1} + t_{2})}  \right] \bigg)  \langle n_{k} \rangle \nonumber \\
+ & \text{h.c.}
\end{align}
Evaluating the integral over $d t_{1}$
\begin{align}
\frac{\Delta \widetilde{\sigma}}{\Delta t} = -\frac{1}{\hbar^{2}} \frac{1}{\Delta t} \sum_{k} |g_{\xi, k}|^{2} \int^{\infty}_{0} d\tau & \bigg(  \left[ (\widetilde{b}_{\xi}^{\dagger 2} \widetilde{b}_{\xi}^{2} \widetilde{\sigma} - \widetilde{b}_{\xi}^{2} \widetilde{\sigma} \widetilde{b}_{\xi}^{\dagger 2} ) e^{ -i \tau (\Omega_{k} - 2 \omega )} \right] \nonumber \\
+ &  \left[ ( \widetilde{b}_{\xi}^{2} \widetilde{b}_{\xi}^{\dagger 2} \widetilde{\sigma} - \widetilde{b}_{\xi}^{\dagger 2} \widetilde{\sigma} \widetilde{b}_{\xi}^{2} ) e^{ -i \tau (\Omega_{k} + 2 \omega )}  \right] \nonumber \\
+ &  \zeta_{1} \left[ ( \widetilde{b}_{\xi}^{\dagger 2} \widetilde{\sigma} \widetilde{b}_{\xi}^{\dagger 2} - \widetilde{b}_{\xi}^{\dagger 4} \widetilde{\sigma} ) e^{ -i \tau (\Omega_{k} + 2 \omega )} \right] \nonumber \\
+ &  \zeta_{2} \left[ ( \widetilde{b}_{\xi}^{2} \widetilde{\sigma} \widetilde{b}_{\xi}^{2} - \widetilde{b}_{\xi}^{4} \widetilde{\sigma} ) e^{-i \tau (\Omega_{k} - 2 \omega )}  \right] \bigg) ( \langle n_{k} \rangle + 1 ) \nonumber \\
+ & \bigg(  \left[ ( \widetilde{\sigma} \widetilde{b}_{\xi}^{2} \widetilde{b}_{\xi}^{\dagger 2} - \widetilde{b}_{\xi}^{\dagger 2} \widetilde{\sigma} \widetilde{b}_{\xi}^{2} ) e^{ -i \tau (\Omega_{k} - 2 \omega )}  \right] \nonumber \\
+ &   \left[ ( \widetilde{\sigma} \widetilde{b}_{\xi}^{\dagger 2} \widetilde{b}_{\xi}^{2} - \widetilde{b}_{\xi}^{2} \widetilde{\sigma} \widetilde{b}_{\xi}^{\dagger 2} ) e^{-i \tau (\Omega_{k} + 2 \omega )}  \right] \nonumber \\
+ &  \zeta_{2} \left[ ( \widetilde{b}_{\xi}^{2} \widetilde{\sigma} \widetilde{b}_{\xi}^{2} - \widetilde{\sigma} \widetilde{b}_{\xi}^{4} ) e^{-i \tau (\Omega_{k} - 2 \omega )} \right] \nonumber \\
+ &  \zeta_{1} \left[ ( \widetilde{b}_{\xi}^{\dagger 2} \widetilde{\sigma} \widetilde{b}_{\xi}^{\dagger 2} - \widetilde{\sigma} \widetilde{b}_{\xi}^{\dagger 4}  ) e^{ -i \tau (\Omega_{k} + 2 \omega )}  \right] \bigg) \langle n_{k} \rangle \nonumber \\
+ & \text{h.c.}
\end{align}
where $\tau = t_{1} - t_{2}$, and the phase factors $\zeta_{1}$ and $\zeta_{2}$ are defined as
\begin{align}
    \zeta_{1} &= \frac{e^{i 4 \omega t}( e^{4 i \omega \Delta t} - 1 )}{4 i \omega \Delta t}, \\
    \zeta_{2} &= \frac{e^{- i 4 \omega t}( 1 - e^{- 4 i \omega \Delta t} )}{4 i \omega \Delta t}.
\end{align}
Further, when $\Delta t \rightarrow 0$, $\zeta_{1} \rightarrow e^{i 4 \omega t}$ and $\zeta_{2} \rightarrow e^{- i 4 \omega t}$ and the master equation becomes
\begin{align}
\frac{\text{d} \widetilde{\sigma}}{\text{d} t} = -\frac{1}{\hbar^{2}} \sum_{k} |g_{\xi, k}|^{2} \int^{\infty}_{0} d\tau & \bigg(  \left[ (\widetilde{b}_{\xi}^{\dagger 2} \widetilde{b_{\xi}}^{2} \widetilde{\sigma} - \widetilde{b}_{\xi}^{2} \widetilde{\sigma} \widetilde{b}_{\xi}^{\dagger 2} ) e^{ -i \tau (\Omega_{k} - 2 \omega )} \right] \nonumber \\
+ &  \left[ ( \widetilde{b}_{\xi}^{2} \widetilde{b}_{\xi}^{\dagger 2} \widetilde{\sigma} - \widetilde{b}_{\xi}^{\dagger 2} \widetilde{\sigma} \widetilde{b}_{\xi}^{2} ) e^{ -i \tau (\Omega_{k} + 2 \omega )}  \right] \nonumber \\
+ &  \zeta_{1} \left[ ( \widetilde{b}_{\xi}^{\dagger 2} \widetilde{\sigma} \widetilde{b}_{\xi}^{\dagger 2} - \widetilde{b}_{\xi}^{\dagger 4} \widetilde{\sigma} ) e^{ -i \tau (\Omega_{k} + 2 \omega )} \right] \nonumber \\
+ &  \zeta_{2} \left[ ( \widetilde{b}_{\xi}^{2} \widetilde{\sigma} \widetilde{b}_{\xi}^{2} - \widetilde{b}_{\xi}^{4} \widetilde{\sigma} ) e^{-i \tau (\Omega_{k} - 2 \omega )}  \right] \bigg) ( \langle n_{k} \rangle + 1 ) \nonumber \\
+ & \bigg(  \left[ ( \widetilde{\sigma} \widetilde{b}_{\xi}^{2} \widetilde{b}_{\xi}^{\dagger 2} - \widetilde{b}_{\xi}^{\dagger 2} \widetilde{\sigma} \widetilde{b}_{\xi}^{2} ) e^{ -i \tau (\Omega_{k} - 2 \omega )}  \right] \nonumber \\
+ &   \left[ ( \widetilde{\sigma} \widetilde{b}_{\xi}^{\dagger 2} \widetilde{b}_{\xi}^{2} - \widetilde{b}_{\xi}^{2} \widetilde{\sigma} \widetilde{b}_{\xi}^{\dagger 2} ) e^{-i \tau (\Omega_{k} + 2 \omega )}  \right] \nonumber \\
+ &  \zeta_{2} \left[ ( \widetilde{b}_{\xi}^{2} \widetilde{\sigma} \widetilde{b}_{\xi}^{2} - \widetilde{\sigma} \widetilde{b}_{\xi}^{4} ) e^{-i \tau (\Omega_{k} - 2 \omega )} \right] \nonumber \\
+ &  \zeta_{1} \left[ ( \widetilde{b}_{\xi}^{\dagger 2} \widetilde{\sigma} \widetilde{b}_{\xi}^{\dagger 2} - \widetilde{\sigma} \widetilde{b}_{\xi}^{\dagger 4}  ) e^{ -i \tau (\Omega_{k} + 2 \omega )}  \right] \bigg) \langle n_{k} \rangle \nonumber \\
+ & \text{h.c.}
\end{align}
The integral over $\tau$ can be found using
\begin{align}
\int_{0}^{\infty}d\tau e^{-i \tau (\Omega_{k} \pm 2 \omega )} &= - \frac{i}{\Omega_{k} \pm 2 \omega } + \pi \delta (\Omega_{k} \pm 2 \omega ),
\end{align}
and defining
\begin{subequations}
\begin{align}
\label{eq:constants1b}
\Gamma_{\pm} &= \sum_{k} \frac{|g_{\xi, k}|^{2}}{\hbar^{2}} 2 \pi \delta \left( \Omega_{k} \pm 2 \omega \right), \\
\label{eq:constants2b}
\Gamma_{\pm}^{'} &= \sum_{k} \frac{|g_{\xi, k}|^{2}}{\hbar^{2}} 2 \pi \langle n_{k} \rangle \delta \left( \Omega_{k} \pm 2 \omega \right), \\
\label{eq:constants3b}
\Delta_{\pm} &= \sum_{k} \frac{|g_{\xi, k}|^{2}}{\hbar^{2}} \frac{1}{2 \omega \pm \Omega_{k}}, \\
\label{eq:constants4b}
\Delta_{\pm}^{'} &= \sum_{k} \frac{|g_{\xi, k}|^{2}}{\hbar^{2}} \frac{1}{2 \omega \pm \Omega_{k}} \langle n_{k} \rangle.
\end{align}
\label{eq:constset2}
\end{subequations}

Noting that $\Gamma_{+} = \Gamma'_{+} = 0$, the subscript is dropped from $\Gamma_{-}$ such that $\Gamma_{-} = \Gamma$.

In order to evaluate the effective rates of damping, gain and detuning in the system, we need to evaluate the summation over $\mathbf{k}$ in (\ref{eq:constants1b}) - (\ref{eq:constants4b}). In a typical setup of our interest, the interaction with the reservoir is mainly due to vacuum fluctuations in an infinite range of field modes. So, it is more relevant to consider an integral over $\mathbf{k}$ in place of the summation, as indicated below 
\begin{equation}
\label{eq:ksum2b}
\sum_{\mathbf{k}} f(\mathbf{k}) \rightarrow \frac{V}{8 \pi^{3}}\int_{\mathbf{k}} f(\mathbf{k}) d^{3} \mathbf{k}.
\end{equation}
Using this along with Eq. (\ref{eq:gkapp2}) in Eq. (\ref{eq:constset2}), we get, for instance,
\begin{align}
    \Gamma &= \frac{ G \hbar}{4 \pi c^{2}} \int \text{d}^{3}\mathbf{k} \; \Omega_{\mathbf{k}} \; \delta \left(\Omega_{\mathbf{k}} - 2 \omega \right) \sin^{4}{(\theta_{\mathbf{k}})}\cos^{2}{(2 \phi_{\mathbf{k}})} , \nonumber \\
    &= \frac{ G \hbar}{4 \pi c^{5}} \int_{0}^{\Omega_{\textsc{max}}} \text{d}\Omega_{k} \Omega^{3}_{k} \delta \left(\Omega_{\mathbf{k}} - 2 \omega \right) \int^{\pi}_{0} \text{d}\theta \sin^{5}{(\theta)} \int_{0}^{2 \pi} \text{d}\phi \cos^{2}{(2 \phi)}, \nonumber \\
    &= \frac{32}{15}\frac{G \hbar \omega^{3}}{c^{5}}.
\end{align}
This exactly matches the decay rate in \cite{Torovs2020}.
\begin{align}
    \Delta_{\pm} &= \frac{G \hbar}{8 \pi^{2} c^{2}} \int \text{d}^{3}\mathbf{k} \frac{\Omega_{\mathbf{k}}}{2 \omega \pm \Omega_{\mathbf{k}}} \sin^{4}{(\theta_{\mathbf{k}})}\cos^{2}{(2 \phi_{\mathbf{k}})}, \nonumber \\
    &= \frac{G \hbar}{8 \pi^{2} c^{5}} \int_{0}^{\Omega_{\textsc{max}}} \text{d} \Omega_{k} \frac{\Omega^{3}_{k}}{2 \omega \pm \Omega_{\mathbf{k}}} \int^{\pi}_{0} \text{d}\theta \sin^{5}{(\theta)} \int_{0}^{2 \pi} \text{d}\phi \cos^{2}{(2 \phi)}, \nonumber \\
    &= \frac{\Gamma}{2 \pi} \left(- \ln{\left| \Omega_{\textsc{max}} \pm 2 \omega  \right|} \pm \frac{\Omega_{\textsc{max}}}{2 \omega} - \frac{\Omega^{2}_{\textsc{max}}}{8 \omega^{2}} \pm \frac{\Omega^{3}_{\textsc{max}}}{24 \omega^{3}} \right).
\end{align}

As we are interested in the influence of the vacuum state of the gravitational field on our particle, we set $\langle n_{\mathbf{k}} \rangle = 0$
\begin{align}
\frac{\text{d} \widetilde{\sigma}}{\text{d} t} = & (  \left[ (\widetilde{b}^{2} \widetilde{\sigma} \widetilde{b}^{\dagger 2} - \widetilde{b}^{\dagger 2} \widetilde{b}^{2} \widetilde{\sigma}  ) \left( i \Delta_{-} + \Gamma \right) \right] \nonumber \\
+ & \left[ ( \widetilde{b}^{\dagger 2} \widetilde{\sigma} \widetilde{b}^{2} - \widetilde{b}^{2} \widetilde{b}^{\dagger 2} \widetilde{\sigma}  ) \left( - i \Delta_{+} \right)  \right] \nonumber \\
+ &  \zeta_{1} \left[ (\widetilde{b}^{\dagger 4} \widetilde{\sigma} - \widetilde{b}^{\dagger 2} \widetilde{\sigma} \widetilde{b}^{\dagger 2}  ) \left( - i \Delta_{+} \right) \right] \nonumber \\
+ &  \zeta_{2} \left[ ( \widetilde{b}^{4} \widetilde{\sigma} - \widetilde{b}^{2} \widetilde{\sigma} \widetilde{b}^{2} ) \left( i \Delta_{-} + \Gamma \right) )  \right] ) \nonumber \\
+ & h.c.
\end{align}

Moving back into the Schr{\"o}dinger picture
\begin{align}
\frac{\text{d} \sigma}{\text{d} t} = -&i \omega \hat{b}_{\xi}^{\dagger} \hat{b}_{\xi} \sigma \nonumber \\
+& ( \left[ (\hat{b}_{\xi}^{2} \sigma \hat{b}_{\xi}^{\dagger 2} - \hat{b}_{\xi}^{\dagger 2} \hat{b}_{\xi}^{2} \sigma  ) \left( i \Delta_{-} + \Gamma \right) \right] \nonumber \\
+ &  \left[ ( \hat{b}_{\xi}^{\dagger 2} \sigma \hat{b}_{\xi}^{2} - \hat{b}_{\xi}^{2} \hat{b}_{\xi}^{\dagger 2} \sigma  ) \left( - i \Delta_{+} \right)  \right] \nonumber \\
+ &  \left[ (\hat{b}_{\xi}^{\dagger 4} \sigma - \hat{b}_{\xi}^{\dagger 2} \sigma \hat{b}_{\xi}^{\dagger 2}  ) \left( - i \Delta_{+} \right) \right] \nonumber \\
+ &  \left[ ( \hat{b}_{\xi}^{4} \sigma - \hat{b}_{\xi}^{2} \sigma \hat{b}_{\xi}^{2} ) \left( i \Delta_{-} + \Gamma \right) )  \right] ) \nonumber \\
+ & h.c.
\end{align}

Including the Hermitian conjugate and rearranging, the final master equation is then
\begin{align}
\label{eq:mastereqd2}
\frac{\text{d} \sigma}{\text{d} t} &= -i \left(\omega - \Delta_{-} -  3\Delta_{+} \right)  \left[ \hat{b}_{\xi}^{\dagger} \hat{b}_{\xi}, \sigma \right] \nonumber \\
&+ i \left( \Delta_{+} -  \Delta_{-} \right) \left[ \left( \hat{b}_{\xi}^{\dagger} \hat{b}_{\xi} \right)^{2}, \sigma  \right] \nonumber \\
&+ \Gamma \left( \hat{b}_{\xi}^{2} \sigma \hat{b}_{\xi}^{\dagger 2} - \frac{1}{2} \hat{b}_{\xi}^{\dagger 2} \hat{b}_{\xi}^{2} \sigma - \frac{1}{2} \sigma \hat{b}_{\xi}^{\dagger 2} \hat{b}_{\xi}^{2} \right) \nonumber \\
&+  i \Delta_{+} \left( \hat{b}_{\xi}^{\dagger 2} \sigma \hat{b}_{\xi}^{\dagger 2} - \hat{b}_{\xi}^{\dagger 4} \sigma + \sigma \hat{b}_{\xi}^{4} - \hat{b}_{\xi}^{2} \sigma \hat{b}_{\xi}^{2} \right) \nonumber \\
&+  \left(i \Delta_{-} + \frac{\Gamma}{2} \right)\left( \hat{b}_{\xi}^{4} \sigma - \hat{b}_{\xi}^{2} \sigma \hat{b}_{\xi}^{2} \right) \nonumber \\
&+  \left(i \Delta_{-} - \frac{\Gamma}{2} \right) \left( \hat{b}_{\xi}^{\dagger 2} \sigma \hat{b}_{\xi}^{\dagger 2} - \sigma \hat{b}_{\xi}^{\dagger 4} \right).
\end{align}
In the limit of the rotating-wave approximation,
\begin{align}
\label{eq:mastereqdrwa2}
\frac{\text{d} \sigma}{\text{d} t} &= -i \left(\omega - \Delta_{-}  \right)  \left[ \hat{b}_{\xi}^{\dagger} \hat{b}_{\xi}, \sigma \right] - i \Delta_{-} \left[ \left( \hat{b}_{\xi}^{\dagger} \hat{b}_{\xi} \right)^{2}, \sigma  \right] \nonumber \\
&+ \Gamma \left( \hat{b}_{\xi}^{2} \sigma \hat{b}_{\xi}^{\dagger 2} - \frac{1}{2} \hat{b}_{\xi}^{\dagger 2} \hat{b}_{\xi}^{2} \sigma - \frac{1}{2} \sigma \hat{b}_{\xi}^{\dagger 2} \hat{b}_{\xi}^{2} \right).
\end{align}

We now present the master equation in terms of $\xi$ and $p_{\xi}$, both in the limit of the rotating-wave approximation and outside this limit. In the rotating-wave limit
\begin{align}
\frac{\text{d}\sigma}{\text{d}t} &= -i(\omega - \Delta_{-}) \left(\frac{m \omega}{2 \hbar} \right) \Big( [\xi^{2}, \sigma] + \frac{i}{m \omega} [(\xi p_{\xi} - p_{\xi} \xi), \sigma] + \frac{1}{m^{2} \omega^{2}} [p^{2}_{\xi}, \sigma] \Big) \nonumber \\
&-i \Delta_{-} \left(\frac{m \omega}{2 \hbar} \right)^{2} \Big( [\xi^{4}, \sigma] + \frac{i}{m \omega} [(\xi^{3}p_{\xi}-\xi^{2}p_{\xi}\xi + \xi p_{\xi} \xi^{2} - p_{\xi} \xi^{3}),\sigma] \Big) \nonumber \\
&+\frac{1}{m^{2} \omega^{2}} [(\xi^{2} p_{\xi}^{2} + p_{\xi}^{2} \xi^{2} - (\xi p_{\xi})^{2} + \xi p_{\xi}^{2} \xi + p_{\xi} \xi^{2} p_{\xi} - (p_{\xi} \xi)^{2}), \sigma] \nonumber \\
&+\frac{i}{m^{3} \omega^{3}} [(\xi p_{\xi}^{3} - p_{\xi} \xi p_{\xi}^{2} + p_{\xi}^{2} \xi p_{\xi} - p_{\xi}^{3} \xi), \sigma] + \frac{1}{m^{4} \omega^{4}} [p_{\xi}^{4}, \sigma] \Big) \nonumber \\
&+\Gamma \left(\frac{m \omega}{2 \hbar} \right)^{2} \Big( \mathcal{L} [\hat{\xi}^{2}, \hat{\xi}^{2}] + \frac{i}{m \omega} \mathcal{L} [\{\hat{\xi}, \hat{p}_{\xi}\}, \hat{\xi}^{2}] - \frac{i}{m \omega} \mathcal{L} [ \hat{\xi}^{2}, \{\hat{\xi}, \hat{p}_{\xi}\}] \nonumber \\
&+ \frac{1}{m^{2} \omega^{2}} \mathcal{L} [\hat{\xi} \hat{p}_{\xi}, \{\hat{\xi}, \hat{p}_{\xi} \}] + \frac{1}{m^{2} \omega^{2}} \mathcal{L} [\{\hat{\xi}, \hat{p}_{\xi} \}, \hat{\xi} \hat{p}_{\xi} ] - \frac{1}{m^{2} \omega^{2}} \mathcal{L} [\hat{\xi}^{2}, \hat{p}^{2}_{\xi}] \nonumber\\
&- \frac{1}{m^{2} \omega^{2}} \mathcal{L} [ \hat{p}^{2}_{\xi}, \hat{\xi}^{2}] + \frac{i}{m^{3} \omega^{3}} \mathcal{L} [\hat{p}^{2}_{\xi},\{ \hat{\xi},  \hat{p}_{\xi} \}] -  \frac{i}{m^{3} \omega^{3}} \mathcal{L} [\{\hat{\xi}, \hat{p}_{\xi} \}, \hat{p}_{\xi}^{2} ] \nonumber \\
&+ \frac{1}{m^{4} \omega^{4}} \mathcal{L} [\hat{p}_{\xi}^{2}, \hat{p}_{\xi}^{2}] \Big).
\end{align}
The full master equation is
\begin{align}
\frac{\text{d}\sigma}{\text{d}t} &= -i(\omega - \Delta_{-}) \left(\frac{m \omega}{2 \hbar} \right) \Big( [\xi^{2}, \sigma] + \frac{i}{m \omega} [(\xi p_{\xi} - p_{\xi} \xi), \sigma] + \frac{1}{m^{2} \omega^{2}} [p^{2}_{\xi}, \sigma] \Big) \nonumber \\
&-i \Delta_{-} \left(\frac{m \omega}{2 \hbar} \right)^{2} \Big( [\xi^{4}, \sigma] + \frac{i}{m \omega} [(\xi^{3}p_{\xi}-\xi^{2}p_{\xi}\xi + \xi p_{\xi} \xi^{2} - p_{\xi} \xi^{3}),\sigma] \Big) \nonumber \\
&+\frac{1}{m^{2} \omega^{2}} [(\xi^{2} p_{\xi}^{2} + p_{\xi}^{2} \xi^{2} - (\xi p_{\xi})^{2} + \xi p_{\xi}^{2} \xi + p_{\xi} \xi^{2} p_{\xi} - (p_{\xi} \xi)^{2}), \sigma] \nonumber \\
&+\frac{i}{m^{3} \omega^{3}} [(\xi p_{\xi}^{3} - p_{\xi} \xi p_{\xi}^{2} + p_{\xi}^{2} \xi p_{\xi} - p_{\xi}^{3} \xi), \sigma] + \frac{1}{m^{4} \omega^{4}} [p_{\xi}^{4}, \sigma] \Big) \nonumber \\
&+\Gamma \left(\frac{m \omega}{2 \hbar} \right)^{2} \Big( \mathcal{L} [\hat{\xi}^{2}, \hat{\xi}^{2}] + \frac{i}{m \omega} \mathcal{L} [\{\hat{\xi}, \hat{p}_{\xi}\}, \hat{\xi}^{2}] - \frac{i}{m \omega} \mathcal{L} [ \hat{\xi}^{2}, \{\hat{\xi}, \hat{p}_{\xi}\}] \nonumber \\
&+ \frac{1}{m^{2} \omega^{2}} \mathcal{L} [\hat{\xi} \hat{p}_{\xi}, \{\hat{\xi}, \hat{p}_{\xi} \}] + \frac{1}{m^{2} \omega^{2}} \mathcal{L} [\{\hat{\xi}, \hat{p}_{\xi} \}, \hat{\xi} \hat{p}_{\xi} ] - \frac{1}{m^{2} \omega^{2}} \mathcal{L} [\hat{\xi}^{2}, \hat{p}^{2}_{\xi}] \nonumber\\
&- \frac{1}{m^{2} \omega^{2}} \mathcal{L} [ \hat{p}^{2}_{\xi}, \hat{\xi}^{2}] + \frac{i}{m^{3} \omega^{3}} \mathcal{L} [\hat{p}^{2}_{\xi},\{ \hat{\xi},  \hat{p}_{\xi} \}] -  \frac{i}{m^{3} \omega^{3}} \mathcal{L} [\{\hat{\xi}, \hat{p}_{\xi} \}, \hat{p}_{\xi}^{2} ] + \frac{1}{m^{4} \omega^{4}} \mathcal{L} [\hat{p}_{\xi}^{2}, \hat{p}_{\xi}^{2}] \Big) \nonumber \\
&+(\Delta_{-} - \Delta_{+}) \left(\frac{m \omega}{2 \hbar^{2}} \right) \Big(\xi^{2} \sigma \{\xi,p_{\xi} \} + \{\xi,p_{\xi} \} \sigma \xi^{2} - \{\xi,p_{\xi} \} \sigma p^{2}_{\xi} - p^{2}_{\xi} \sigma \{\xi,p_{\xi} \} \Big) \nonumber \\
&+ i ( \Delta_{-} - \Delta_{+} ) \left( \frac{m \omega}{2 \hbar} \right)^{2} \Big( [\xi^{4}, \sigma] + \frac{i}{m \omega} \{ (\xi^{3} p_{\xi} + \xi^{2} p_{\xi} \xi + \xi p_{\xi} \xi^{2} + p_{\xi} \xi^{3}), \sigma \} \nonumber \\
-\frac{1}{m^{2} \omega^{2}}  \Big).
\end{align}

\section{ \label{app:timeind} Time-independent perturbation theory for the harmonic potential case}

\subsection{Time-Independent Perturbation Theory}

In the realm of quantum mechanics, exact solutions to the Schr$\ddot{\text{o}}$dinger equation are available only for a few systems. Many physical problems don't fall into this category, requiring physicists to deploy approximation methods. One such powerful method is the time-independent perturbation theory.
Consider a quantum system governed by a Hamiltonian $H_0$, the solutions to which we know. If we add a small perturbation $\gamma \Phi$ to it, the solutions of the total Hamiltonian $H = H_0 + \gamma \Phi$ can be expressed as a perturbation series, which builds upon the known solutions of $H_0$.

The Schr$\ddot{\text{o}}$dinger equation for the unperturbed system is 
\begin{equation}
H_0 | \psi_n^{(0)} \rangle = E_n^{(0)} | \psi_n^{(0)} \rangle ,  \qquad n = 1, 2, 3, \cdots ,
\end{equation}
where \( | \psi_n^{(0)} \rangle \) and \( E_n^{(0)} \) are the unperturbed system's n-th eigenstate and eigenenergy , respectively.

The total Hamiltonian is \( H = H_0 + \gamma \Phi \), where \( \gamma \) is a small parameter indicating the strength of the perturbation. The perturbed Schr$\ddot{\text{o}}$dinger equation is 
\begin{equation}
H |\psi_n \rangle = (H_0 + \gamma \Phi) |\psi_n \rangle = E_n |\psi_n \rangle
\end{equation}
Using perturbation theory, \( E_n \) may be expanded in a power series of \( \lambda \). The first-order correction is
\begin{eqnarray}
    E_n^{(1)} = \langle \psi_n^{(0)} | \Phi | \psi_n^{(0)} \rangle  ,
\end{eqnarray}
and the second-order energy correction is 
\begin{equation}
E_n^{(2)} = \sum_{m \neq n} \frac{| \langle \psi_m^{(0)} | \Phi | \psi_n^{(0)} \rangle |^2}{E_n^{(0)} - E_m^{(0)}} .
\end{equation}
For our study, we can see that we do not need to consider the higher-order terms.

\subsection{From single mode cases to multi-mode cases}

For this work, we may first consider the single mode case and choose the harmonic potential $\Phi(\xi)$ as $\frac{1}{2}m \omega^2   \xi^2$. Then the Hamiltonian may be written as
\begin{eqnarray}
    H = \frac{p_{\xi}^{2}}{2 m} + \frac{1}{2}m \omega^2   \xi^2 + \frac{\xi p_{\xi} + p_{\xi} \xi}{4 \rho_{\textsc{f}}} p_{\mathbf{k}} + \frac{V p_{\mathbf{k}}^{2}}{2 \rho_{\textsc{f}}} + \frac{\rho_{\textsc{f}}}{2V} \Omega_{\mathbf{k}}^{2} q_{\mathbf{k}}^{2}.
\end{eqnarray}
We may define the creation and annihilation operators of the particle as
\begin{eqnarray}
    b = \sqrt{\frac{m\omega}{2\hbar}} (\xi + \frac{i p_\xi}{m\omega}) \;\;\;, \;\;\;
    b^\dagger = \sqrt{\frac{m\omega}{2\hbar}} (\xi - \frac{i p_\xi}{m\omega}) .\nonumber \\
\end{eqnarray}
From  $[q_\xi ,p_\xi]=i \hbar \phi_\text{P}$, we have $[b,b^\dagger]=\phi_\text{P}$.
The position and momentum operator may be expressed as
\begin{eqnarray}
   \xi = \sqrt{\frac{\hbar }{2m\omega}} (b + b^\dagger) \;\;\;, \;\;\;
    p_\xi = -i\sqrt{\frac{m\omega \hbar}{2}} (b - b^\dagger) .\nonumber \\
\end{eqnarray}
Similarly, the creation and annihilation operators for the single mode of the gravitational field may be written as
\begin{eqnarray}
    a = \sqrt{\frac{\frac{\rho_{\textsc{f}}}{V}\Omega_k}{2\hbar}} (q_k + \frac{i p_k}{\frac{\rho_{\textsc{f}}}{V}\Omega_k}) \;\;\;, \;\;\;
a^\dagger = \sqrt{\frac{\frac{\rho_{\textsc{f}}}{V}\Omega_k}{2\hbar}} (q_k - \frac{i p_k}{\frac{\rho_{\textsc{f}}}{V}\Omega_k}) . 
\end{eqnarray}
Here $\frac{\rho_{\textsc{f}}}{V}$ equal to the mass of a harmonic oscillator.
Recall that $[q_k ,p_k]=i \hbar \phi_\text{F}$, thus we have $[a,a^\dagger]=\phi_\text{F}$.
The position and momentum operator may be expressed as
\begin{eqnarray}
   q_k = \sqrt{\frac{\hbar }{2\frac{\rho_{\textsc{f}}}{V}\Omega_k}} (a + a^\dagger) \;\;\;, \;\;\;
    p_k = -i\sqrt{\frac{\hbar \frac{\rho_{\textsc{f}}}{V}\Omega_k}{2}} (a - a^\dagger) .
\end{eqnarray}
With these operators, the Hamiltonian may be written as
\begin{eqnarray}
    H = \hbar \Omega_k (a^\dagger a + \frac{1}{2} \phi_\text{F})  + \hbar \omega (b^\dagger b + \frac{1}{2} \phi_\text{P} ) - \sqrt{\frac{\hbar \Omega_k}{32 V\rho_{\textsc{f}}}} \hbar  (a - a^\dagger) (b b - b^\dagger b^\dagger) .
    \label{NApproHamil3}
\end{eqnarray}
Recall that $\rho_{\textsc{f}} = \frac{c^{2} }{8 \pi G}$, Eq.~(\ref{NApproHamil3}) may be written as
\begin{eqnarray}
    H = \hbar \Omega_k (a^\dagger a + \frac{1}{2} \phi_\text{F})  + \hbar \omega (b^\dagger b + \frac{1}{2} \phi_\text{P} ) - \hbar \Omega_k \sqrt{\frac{ \pi G \hbar }{4c^2 \Omega_k V}} (a - a^\dagger) (b b - b^\dagger b^\dagger) .
    \label{NApproHamil4}
\end{eqnarray}
We may label $- \sqrt{\frac{ \pi G \hbar }{4c^2 \Omega_k V}}$ as $\gamma$, which is the parameter labelling the strength of the perturbation. 
Thus, we have
\begin{eqnarray}
    H_0 = \hbar \Omega_k (a^\dagger a + \frac{1}{2} \phi_\text{F})  + \hbar \omega (b^\dagger b + \frac{1}{2} \phi_\text{P} ) \;\;\; \text{and} \;\;\; \Phi = \hbar \Omega_k  (a - a^\dagger) (b b - b^\dagger b^\dagger)  .
    \label{OurHamil}
\end{eqnarray}
If we roughly take the frequency to be around $100$Hz range, the $\gamma = - \sqrt{\frac{ \pi G \hbar }{4c^2 \Omega_k V}} \approx  1 \times 10^{-42}$ and we have assumed $V=\lambda^3$. Thus, we can see that this is a tiny perturbation.

The eigenvalues and eigenstates of two harmonic oscillators without perturbation equal
\begin{eqnarray}
   E_n^{(0)} = (n_b + \frac{1}{2}) \hbar \omega + (n_a + \frac{1}{2}) \hbar \Omega_k   \;\;\; \text{and} \;\;\;  | \psi_n^{(0)} \rangle &= | n_a, n_b^{(0)} \rangle
\end{eqnarray}
Inserting our Hamiltonian and perturbation in Eq.~(\ref{OurHamil}) into the time-independent perturbation theory, 
the first-order correction for the eigenvalue equals
\begin{eqnarray}
E_n^{(1)} = \gamma \langle  n_a, n_b^{(0)} | \hbar \Omega_k  (  a -   a^\dagger) (  b   b -   b^\dagger   b^\dagger) |  n_a, n_b^{(0)} \rangle =0 . 
\end{eqnarray}
Similarly, the second-order correction to the eigenvalue may be calculated as
\begin{eqnarray}
    E_n^{(2)} &=& \gamma^2 \sum_{m \neq n} \frac{| \langle \psi_m^{(0)} | \Phi | \psi_n^{(0)} \rangle |^2}{E_n^{(0)} - E_m^{(0)}} 
    = \gamma^2 \sum_{m \neq n} \frac{| \langle  m_a, m_b^{(0)} | \hbar \Omega_k  (  a -   a^\dagger) (  b   b -   b^\dagger   b^\dagger) |  n_a, n_b^{(0)} \rangle |^2}{E_n^{(0)} - E_m^{(0)}}  \nonumber \\
    &=& \gamma^2 \sum_{m \neq n} \frac{|\langle  m_a, m_b^{(0)} | \hbar \Omega_k  (  a   b   b -   a   b^\dagger   b^\dagger -   a^\dagger   b   b +   a^\dagger   b^\dagger   b^\dagger) |  n_a, n_b^{(0)} \rangle |^2}{(n_b + \frac{1}{2}) \hbar \omega + (n_a + \frac{1}{2}) \hbar \Omega_k - E_m^{(0)}}    \nonumber \\
    &=& \gamma^2 \frac{ \hbar^2 \Omega_k^2  n_a n_b (n_b-1) }{2 \hbar \omega +  \hbar \Omega_k } + \gamma^2 \frac{ \hbar^2 \Omega_k^2  n_a (n_b+1)(n_b+2) }{-2 \hbar \omega +  \hbar \Omega_k } + \gamma^2 \frac{ \hbar^2 \Omega_k^2  (n_a+1) n_b (n_b-1) }{2 \hbar \omega -  \hbar \Omega_k }  + \gamma^2 \frac{ \hbar^2 \Omega_k^2  (n_a+1) (n_b+1) (n_b+2) }{-2 \hbar \omega -  \hbar \Omega_k }   \nonumber \\
    &=&  - \gamma^2 \hbar^2 \Omega_k^2 \frac{ 4 n_a n_b + 2n_a + n_b^2 + 3 n_b +2 }{2 \hbar \omega +  \hbar \Omega_k } + \gamma^2 \hbar^2 \Omega_k^2  \frac{  4 n_a n_b + 2n_a - n_b^2 + n_b }{-2 \hbar \omega +  \hbar \Omega_k } .
\end{eqnarray}
Thus, the energy until the second-order correction equals
\begin{eqnarray}
    E_n^{(0)} &=& (n_b + \frac{1}{2}) \hbar \omega + (n_a + \frac{1}{2}) \hbar \Omega_k  \nonumber \\
    E_n^{(1)} &=& 0  \nonumber \\
    E_n^{(2)} &=& - \frac{ \pi G \hbar^2 \Omega_k}{4c^2 V}  \Bigl( \frac{ 4 n_a n_b + 2n_a + n_b^2 + 3 n_b +2 }{2 \omega +  \Omega_k } + \frac{ 4 n_a n_b + 2n_a - n_b^2 + n_b }{2 \omega - \Omega_k } \Bigr)  \nonumber \\
    &\vdots&
\end{eqnarray}
where we have used $\gamma=- \sqrt{\frac{ \pi G \hbar }{4c^2 \Omega_k V}}$.
The frequency shift may be calculated as the energy shift over $\hbar$, which equals
\begin{eqnarray}
    \Delta_{\Omega_k} = \frac{E_n^{(2)}}{\hbar} = - \frac{ \pi G \hbar \Omega_k}{4c^2 V}  \Bigl( \frac{ 4 n_a n_b + 2n_a + n_b^2 + 3 n_b +2 }{2 \omega +  \Omega_k } + \frac{  4 n_a n_b + 2n_a - n_b^2 + n_b }{2 \omega - \Omega_k } \Bigr)
    \label{frequency1}
\end{eqnarray}
For vacuum spacetime ($n_a=0$), the frequency shift reduces into
\begin{eqnarray}
    \Delta_{\Omega_k}=\frac{E_n^{(2)}}{\hbar} = - \frac{ \pi G \hbar \Omega_k}{4c^2 V}  \Bigl( \frac{n_b^2 + 3 n_b +2 }{2 \omega +  \Omega_k } + \frac{- n_b^2 + n_b }{2 \omega - \Omega_k } \Bigr)
    \label{frequency2}
\end{eqnarray}
We can further construct the multi-mode results from the above single mode case. The multi-mode case may be calculated as
\begin{eqnarray}
    \Delta &=& \sum_{\mathbf{k}, s} \Delta_{\Omega_k} \sin^{4}{(\theta_{\mathbf{k}})} \cos^{2}{(\phi_\mathbf{k})} = - \sum_{\mathbf{k}} \frac{ 4 \pi G \hbar \Omega_k}{4c^2 V}  \Bigl( \frac{n_b^2 + 3 n_b +2 }{2 \omega +  \Omega_k } + \frac{- n_b^2 + n_b }{2 \omega - \Omega_k } \Bigr) \sin^{4}{(\theta_{\mathbf{k}})} \cos^{2}{(\phi_\mathbf{k})} \nonumber \\
    &=& - \frac{1}{(2 \pi)^3} \int d^3 \mathbf{k} \frac{  \pi G \hbar \Omega_k}{c^2}  \Bigl( \frac{ n_b^2 + 3 n_b +2 }{2 \omega +  \Omega_k } + \frac{- n_b^2 + n_b }{2 \omega - \Omega_k } \Bigr) \sin^{4}{(\theta_{\mathbf{k}})} \cos^{2}{(\phi_\mathbf{k})} \nonumber \\
    &=& - \frac{1}{(2 \pi)^3} \int_0^{k_{\textsc{max}}} k^2 dk \int_0^{2\pi} d\phi_\mathbf{k} \int_0^\pi d\theta \sin^5{(\theta_{\mathbf{k}})} \cos^2(\phi_\mathbf{k}) \frac{ \pi G \hbar \Omega_k}{c^2}  \Bigl( \frac{n_b^2 + 3 n_b +2 }{2 \omega +  \Omega_k } + \frac{- n_b^2 + n_b }{2 \omega - \Omega_k } \Bigr) \nonumber \\
    &=& - \frac{2G \hbar }{15\pi c^2} \int_0^{k_{\textsc{max}}} dk \, k^2 \Omega_k  \Bigl( \frac{ n_b^2 + 3 n_b +2 }{2 \omega +  \Omega_k } + \frac{- n_b^2 + n_b }{2 \omega - \Omega_k } \Bigr) ,
    \label{shift1}
\end{eqnarray}
where the factor of 4 in the first line arises from the squaring of the two polarizations in the second-order energy perturbation; and $\frac{1}{V} \sum_{\mathbf{k}} = \frac{1}{(2 \pi)^3} \int d^3 \mathbf{k}$ is used in moving from the first to the second line. If we apply $\Omega_k=ck$ to Eq.~(\ref{shift1}), we further have
\begin{eqnarray}
    \Delta &=& - \frac{2G \hbar }{15\pi c^5} \int_0^{\Omega_{\textsc{max}}} d\Omega_k \, \Omega_k^3  \Bigl( \frac{ n_b^2 + 3 n_b +2 }{2 \omega +  \Omega_k } + \frac{- n_b^2 + n_b }{2 \omega - \Omega_k } \Bigr) \nonumber \\
    &=& - \Bigl( (n_b^2 + 3 n_b +2 ) \Delta_+ + (- n_b^2 + n_b ) \Delta_- \Bigr) ,
    \label{shift2}
\end{eqnarray}
where $\Delta_\pm = \frac{2G \hbar}{15 \pi c^5} \int_0^{\Omega_{\textsc{max}}} d\Omega_k \, \frac{\Omega_k^3}{2 \omega \pm \Omega_k}$ is used in moving from the first to the second line.

Since the constant term $2 \Delta_-$ is an overall constant shift for all different number state, it will not contribute to the experiment results and can be ignored. Thus, the physical frequency shift may be written as
\begin{eqnarray}
    \Delta = - \Bigl( (n_b^2 + 3 n_b ) \Delta_+ + (- n_b^2 + n_b ) \Delta_- \Bigr) .
\end{eqnarray}
Equivalently,
\begin{eqnarray}
    \Delta = (\Delta_- - \Delta_+ ) n_b^2 - ( 3\Delta_+ + \Delta_- ) n_b  .
    \label{shift3}
\end{eqnarray}
We will see that these results agree with the frequency shift indicated by the master equation obtained from the open quantum system method.

\section{\label{app:validity} Validity of Master Equation}

The two conditions that must be satisfied to ensure the validity of the time-evolved state $\hat{\sigma}(t)$ (obtained by solving the master equation Eq. (\ref{eq:mastereqd2}) with the renormalized constants) are as follows.
\begin{enumerate}
    \item The trace of $\hat{\sigma}$ is preserved.
    \item $\hat{\sigma}$ is a completely positive matrix.
\end{enumerate}
The trace of $\hat{\sigma}$ is easily checked by taking the trace of both sides of the master equation (Eq. \ref{eq:mastereqd2})
\begin{align}
    \text{Tr}\left[\frac{\text{d} \sigma}{\text{d} t}\right] &= - i \left( \omega - \Delta_{-} -3 \Delta_{+} \right)  \text{Tr}\left[ \left[ b^{\dagger} b , \hat{\sigma} \right] \right] \nonumber \\
    &+i \left( \Delta_{+} - \Delta_{-} \right) \text{Tr}\left[\left[(b^{\dagger} b)^{2}, \sigma \right] \right] \nonumber \\
    &+ \Gamma \left( \text{Tr}\left[\hat{b}^{2} \hat{\sigma} \hat{b}^{\dagger 2} \right] - \frac{1}{2} \text{Tr}\left[\hat{b}^{\dagger 2} \hat{b}^{2} \hat{\sigma} \right] - \frac{1}{2}\text{Tr}\left[\hat{\sigma} \hat{b}^{\dagger 2} \hat{b}^{2} \right] \right) \nonumber \\
    &+ i \Delta_{+} \left( \text{Tr}\left[\hat{b}^{\dagger 2} \hat{\sigma} \hat{b}^{\dagger 2} \right] - \text{Tr}\left[\hat{b}^{\dagger 4} \hat{\sigma} \right] - \text{Tr}\left[\hat{b}^{2} \hat{\sigma} \hat{b}^{2}\right] + \text{Tr}\left[\hat{\sigma} \hat{b}^{4} \right] \right) \nonumber \\
    &+ \left( \frac{1}{2}\Gamma + i \Delta_{-} \right) \left( \text{Tr}\left[ \hat{b}^{4} \hat{\sigma} \right] - \text{Tr}\left[ \hat{b}^{2} \hat{\sigma} \hat{b}^{2} \right] \right) \nonumber \\
    &+ \left( \frac{1}{2}\Gamma -  i \Delta_{-} \right) \left(\text{Tr}\left[ \hat{b}^{\dagger 2} \hat{\sigma} \hat{b}^{\dagger 2} \right] - \text{Tr}\left[\hat{\sigma} \hat{b}^{\dagger 4}\right] \right) 
\end{align}
Using the invariance of cyclic permutations of the trace it can be seen
\begin{align}
    \text{Tr}\left[\frac{\text{d} \hat{\sigma}}{\text{d}t} \right] = 0,
\end{align}
and hence the trace is preserved.

The second condition requires the density matrix to be a positive-definite matrix, which is not guaranteed in general as the master equation Eq. (\ref{eq:mastereqd2}) is not in Lindblad form. However, we are most intersted in matter states which are initially thermal. We therefore consider an initial density matrix which is diagonal
\begin{equation}
\label{eq:sigmainit}
\hat{\sigma}(0) = \begin{pmatrix}
\sigma_{0} & 0 & 0 & \dots &\\
0 & \sigma_{1} & 0 & \dots &\\
\vdots & \vdots & \ddots & &\\
0 & 0 &  & \sigma_{n}
\end{pmatrix}.
\end{equation}
where
\begin{align}
    \label{eq:thermal}
    \sigma_{n} = \frac{e^{-n \beta \hbar \omega}}{1 - e^{-\beta \hbar \omega}},
\end{align}
and $\beta = 1 / k_{\textsc{b}} T$ with $T$ as the temperature of the inital state.
In the Fock basis, the matrix representation of $\hat{b}$ and $\hat{b}^{\dagger}$
is

\begin{align}
\label{eq:b}
    \hat{b} &= \begin{pmatrix}
0 & \sqrt{1} & 0 & \dots & 0 & \dots &\\
0 & 0 & \sqrt{2} & \dots & 0 & \dots &\\
\vdots & \vdots & \vdots & \ddots & \vdots & &\\
0 & 0 & 0 & & \sqrt{n+1} & \dots &
\end{pmatrix}, \\
\label{eq:bd}
\hat{b}^{\dagger} &= \begin{pmatrix}
0 & 0 &  \dots & 0 & \dots &\\
\sqrt{1} & 0 &  \dots & 0 & \dots &\\
0 & \sqrt{2} &  \dots & 0 & \dots&\\
\vdots & \vdots & \ddots & \vdots &  &\\
0 & 0 &  & \sqrt{n} & \dots & 
\end{pmatrix},
\end{align}
where $n$ is the row number.
The master equation Eq. (\ref{eq:mastereqd2}), can be expressed in terms of the superoperator $\mathcal{L}$ such that
\begin{equation}
    \dot{\sigma}(t) = \mathcal{L}\left[\sigma(t)\right].
\end{equation}
The solution to the master equation is then
\begin{equation}
    \sigma(t) = e^{t \mathcal{L}} \sigma(0).
\end{equation}
We assume that to leading order $\mathcal{L}\left[\sigma (t) \right] \sim \Gamma$. (The leading term in the master equation is in fact to order $\sim \omega$. However, as $\omega$ appears in the unitary term of the master equation, it is not present in the time dependence of the density matrix.) Then if $ t \ll \Gamma^{-1}$, we can expand the above equation to
\begin{equation}
    \sigma(t) = \sigma(0) + \mathcal{L}\left[ \sigma(0) \right] t + \dots
\end{equation}
Inserting Eqs.~\ref{eq:sigmainit}, \ref{eq:b} and \ref{eq:bd} into the master equation results in
\begin{equation}
\label{eq:sigmasol}
    \sigma(t) = \begin{pmatrix}
x_{0} & 0 & 0 & 0 & y_{0} & 0& 0& \dots&0&\dots \\
0 & x_{1} & 0 & 0 & 0 & y_{1} &0&\dots&0&\dots \\
0 & 0 & x_{2} & 0 & 0 & 0 & y_{2}& \dots &0&\dots \\
0 & 0 & 0 & x_{3} & 0 & 0 & 0 &  \ddots & 0&\dots \\
y^{*}_{0} & 0 & 0 & 0 & x_{4} &0&0&\dots&y_{n-4}&\dots\\
0 & y^{*}_{1} & 0 & 0 & 0 &x_{5}&0&\dots&0&\dots\\
0 & 0 & y^{*}_{2} & 0 & 0 &0&x_{6} &\dots&0&\dots\\
\vdots & \vdots & \vdots & \ddots & \vdots &\vdots&\vdots&\ddots&0&\dots\\
0 & 0 & 0 & 0 & y^{*}_{n-4} &0&0 &0&x_{n}&\dots\\
\vdots & \vdots &\vdots &\vdots &\vdots &\vdots &\vdots &\vdots &\vdots & \ddots\\
\end{pmatrix},
\end{equation}
where
\begin{align}
    \label{eq:x}
    x_{n} &= \sigma_{n} + \Gamma \Big( (n+1) (n+2) \sigma_{n+2} - n(n-1)\sigma_{n} \Big) t, \\
    y_{n} &= \sqrt{(n+1)(n+2)(n+3)(n+4)} \Big( i \Delta_{+} \big(\sigma_{n} - \sigma_{n+2} \big) + \left(i \Delta_{-} + \frac{\Gamma}{2} \right) \big(\sigma_{n+4} - \sigma_{n+2} \big) \Big) t.
\end{align}
While proving the infinite dimensional \ref{eq:sigmasol} is positive-definite is challenging, in practice a thermal state can be approximated with a finite $n$-dimensional matrix so long as $n$ is sufficiently large. We therefore deine the following finite matrix
$S_{n}$,
\begin{equation}
    S_{n}(t) = \begin{pmatrix}
x_{0} & 0 & 0 & 0 & y_{0} & 0& 0& \dots&0 \\
0 & x_{1} & 0 & 0 & 0 & y_{1} &0&\dots&0 \\
0 & 0 & x_{2} & 0 & 0 & 0 & y_{2}& \dots &0 \\
0 & 0 & 0 & x_{3} & 0 & 0 & 0 &  \ddots & 0 \\
y^{*}_{0} & 0 & 0 & 0 & x_{4} &0&0&\dots&y_{n-4}\\
0 & y^{*}_{1} & 0 & 0 & 0 &x_{5}&0&\dots&0&\\
0 & 0 & y^{*}_{2} & 0 & 0 &0&x_{6} &\dots&0\\
\vdots & \vdots & \vdots & \ddots & \vdots &\vdots&\vdots&\ddots&0\\
0 & 0 & 0 & 0 & y^{*}_{n-4} &0&0 &0&x_{n}\\
\end{pmatrix},
\end{equation}
The determinant of the first five $S_{n}$ are
\begin{align}
    \text{det}[S_{0}] &= x_{0}, \\
    \text{det}[S_{1}] &= x_{0} x_{1}, \\
    \text{det}[S_{2}] &= x_{0} x_{1} x_{2}, \\
    \text{det}[S_{3}] &= x_{0} x_{1} x_{2} x_{3}, \\
    \text{det}[S_{4}] &= x_{0} x_{1} x_{2} x_{3} x_{4} + |y_{0}|^{2} x_{1} x_{2} x_{3}. \\
\end{align}
The determinant of $S_{n}$ can be calculated as follows
\begin{align}
    \text{det}\left[S_{n}(t)\right] &= \begin{vmatrix}
x_{0} & 0 & 0 & 0 & y_{0} & 0& 0& \dots&0 \\
0 & x_{1} & 0 & 0 & 0 & y_{1} &0&\dots&0 \\
0 & 0 & x_{2} & 0 & 0 & 0 & y_{2}& \dots &0 \\
0 & 0 & 0 & x_{3} & 0 & 0 & 0 &  \ddots & 0 \\
y^{*}_{0} & 0 & 0 & 0 & x_{4} &0&0&\dots&y_{n-4}\\
0 & y^{*}_{1} & 0 & 0 & 0 &x_{5}&0&\dots&0&\\
0 & 0 & y^{*}_{2} & 0 & 0 &0&x_{6} &\dots&0\\
\vdots & \vdots & \vdots & \ddots & \vdots &\vdots&\vdots&\ddots&0\\
0 & 0 & 0 & 0 & y^{*}_{n-4} &0&0 &0&x_{n}\\
\end{vmatrix}, \nonumber \\
\label{eq:det1}
&= x_{0} \; \text{minor}_{n, n} \left[S_{n}(t)\right] + y^{*}_{n-4} \; \text{minor}_{n, n-4} \left[S_{n}(t)\right],
\end{align}
where $\text{minor}_{(i, j)} \left[S_{n}(t)\right]$ is the minor of $S_{n}(t)$ omitting the $i$th row and $j$th column. Evaluating the second term in Eq.~\ref{eq:det1} results in
\begin{align}
    \text{det}\left[S_{n}(t)\right] &= x_{0} \; \text{minor}_{(n, n)} \left[S_{n}(t)\right] + |y_{n-4}|^{2} \; \text{minor}_{(n, n-4),(n-4,n)} \left[S_{n}(t)\right], \nonumber \\
    \label{eq:det2}
    &= x_{0} \; \text{minor}_{(n, n)} \left[S_{n}(t)\right] + x_{0} |y_{n-4}|^{2} \; \text{minor}_{(n, n-4),(n-4,n),(0,0)} \left[S_{n}(t)\right].
\end{align}
It can be seen that the two minors of Eq.~(\ref{eq:det2}) are determinants of matrices of the same form as $S_{n}(t)$
\begin{align}
    \label{eq:det3}
    \text{det}\left[S_{n}(t)\right] &= x_{0}
    \begin{vmatrix}
 x_{0} & 0 & 0 & 0 & y_{0} &0&\dots&0 \\
 0 & x_{1} & 0 & 0 & 0 & y_{1}& \dots &0 \\
 0 & 0 & x_{2} & 0 & 0 & 0 &  \ddots & 0 \\
 0 & 0 & 0 & x_{3} &0&0&\dots&y_{n-5}\\
 y^{*}_{0} & 0 & 0 & 0 &x_{4}&0&\dots&0&\\
 0 & y^{*}_{1} & 0 & 0 &0&x_{5} &\dots&0\\
 \vdots & \vdots & \ddots & \vdots &\vdots&\vdots&\ddots&0\\
 0 & 0 & 0 & y^{*}_{n-5} &0&0 &0&x_{n-1}\\
\end{vmatrix} + x_{0} |y_{n-4}|^{2}
\begin{vmatrix}
 x_{1} & 0 & 0 & 0 & y_{1} &0&\dots&0 \\
 0 & x_{2} & 0 & 0 & 0 & y_{2}& \dots &0 \\
 0 & 0 & x_{3} & 0 & 0 & 0 &  \ddots & 0 \\
 0 & 0 & 0 & x_{4} &0&0&\dots&y_{n-5}\\
 y^{*}_{1} & 0 & 0 & 0 &x_{5}&0&\dots&0&\\
 0 & y^{*}_{2} & 0 & 0 &0&x_{6} &\dots&0\\
 \vdots & \vdots & \ddots & \vdots &\vdots&\vdots&\ddots&0\\
 0 & 0 & 0 & y^{*}_{n-5} &0&0 &0&x_{n-1}\\
\end{vmatrix}.
\end{align}
Finding the determinants in Eq.~\ref{eq:det3} can be found by repeating the above steps to again obtain determinants of matrices that are of the same form. Therefore the determinant of $S_{n}(t)$ is the summation of terms that contain only products of $x_{i}$ and $|y_{j}|^{2}$ for any $i,j$. The determinant of $S_{n}(t)$, and all its leading principal minors are therefore non-negative so long as all $x_{i}$ are non-negative. Substituting Eq.~\ref{eq:thermal} into Eq.~\ref{eq:x} results in the condition
\begin{align}
    \frac{e^{-n \beta \hbar \omega}}{1 - e^{-\beta \hbar \omega}} + \Gamma \left((n+1)(n+2) \frac{e^{-n \beta \hbar \omega} e^{-2 \beta \hbar \omega}}{1 - e^{-\beta \hbar \omega}} - n(n-1) \frac{e^{-n \beta \hbar \omega}}{1 - e^{-\beta \hbar \omega}} \right)t &\geq 0, \nonumber \\
    1 + \Gamma \left( (n+1)(n+2) e^{-2 \beta \hbar \omega} - n(n-1) \right) t &\geq 0, \nonumber \\
    \frac{1}{n(n-1) - (n+1)(n+2)e^{-2\beta \hbar \omega}} & \leq  \Gamma t .
\end{align}
The minimum value of $t$ is the initial time $t=0$. The requirement for $S_{n}(t)$ to be postive-semidefinite then becomes
\begin{align}
    (e^{-2\beta \hbar \omega} -1)n^{2} + (3e^{-2\beta \hbar \omega} + 1)n + 2e^{-2\beta \hbar \omega} &\geq 0.
\end{align}
This means we require
\begin{align}
    n \leq \frac{1 + 3e^{-2\beta \hbar \omega} + \sqrt{1 + e^{-4\beta \hbar \omega}} + 14e^{-2\beta \hbar \omega}}{2(1 - e^{-2\beta \hbar \omega})}.
\end{align}
Therefore any temperature $T$ of our initial state sets a maximum allowed size of matrix $n$ that we can use to model the system. Our master equation therefore corresponds to a CPTP map so long as the following condtions are met:
\begin{itemize}
    \item The state is initially a thermal state,
    \item We consider timescales such that $t\ll \Gamma^{-1}$,
    \item We can model the state with a density matrix that is finite and has dimension $n$ where
    \begin{equation}
        n \leq \frac{1 + 3e^{-2\beta \hbar \omega} + \sqrt{1 + e^{-4\beta \hbar \omega}} + 14e^{-2\beta \hbar \omega}}{2(1 - e^{-2\beta \hbar \omega})}.
    \end{equation}
\end{itemize}

\end{document}